\documentclass[journal=rsm]{CUP-JNL-DTM}%
\makeatletter
\def\ps@titlepage{%
    \let\@oddhead\@empty
    \let\@evenhead\@empty
    \def\@oddfoot{\hfil\thepage\hfil}%
    \let\@evenfoot\@oddfoot}
\makeatother

\makeatletter
\def\ps@headings{%
    \let\@oddfoot\@empty
    \let\@evenfoot\@empty
    \def\@evenhead{\headerfont\thepage\hfill}%
    \def\@oddhead{\headerfont\hfill\thepage}%
    \let\@mkboth\@gobbletwo}
\makeatother

\let\cite\parencite

\usepackage{parskip}
\usepackage{graphicx}
\usepackage{multicol,multirow}
\usepackage{amsmath,amssymb,amsfonts}
\usepackage{mathrsfs}
\usepackage{amsthm}
\usepackage{rotating}
\usepackage{appendix}
\usepackage{ifpdf}
\usepackage[T1]{fontenc}
\usepackage{newtxtext}
\usepackage{newtxmath}
\usepackage{textcomp}
\usepackage{xcolor}
\usepackage{lipsum}
\usepackage[colorlinks,allcolors=blue]{hyperref}

\theoremstyle{definition}

\numberwithin{equation}{section}

\usepackage{amsmath}
\usepackage{amssymb}
\usepackage{amssymb}
\usepackage{mathtools}

\allowdisplaybreaks

\usepackage{booktabs}
\usepackage{longtable}
\usepackage{graphicx}
\usepackage{float}
\usepackage{multirow}
\usepackage[export]{adjustbox}
\usepackage{graphicx}
\usepackage{subcaption} 
\usepackage{threeparttable}

\usepackage{ulem}
\usepackage{url}
\usepackage{hyperref}

\usepackage{titlesec}
\titlespacing{\section}{0pt}{14pt plus 4pt minus 2pt}{4pt plus 2pt minus 2pt}
\titlespacing{\subsection}{0pt}{12pt plus 4pt minus 2pt}{2pt plus 2pt minus 2pt}
\titlespacing{\subsubsection}{0pt}{10pt plus 4pt minus 2pt}{2pt plus 2pt minus 2pt}

\usepackage{enumerate}
\usepackage[shortlabels]{enumitem}

\newcommand{\bbeta}{\boldsymbol{\beta}}

\newcommand{\E}{\mathrm{E}}
\renewcommand{\P}{\mathrm{P}}
\newcommand{\Var}{\mathrm{Var}}
\newcommand{\SE}{\mathrm{SE}}

\jname{Research Synthesis Methods}
\articletype{Research Articles}
\jyear{2026}

\begin{document}

\begin{Frontmatter}

\title[Article Title]{Forecasting Replication Study Results from Meta-Analyses}

\author[1]{Yi Li}
\author[2]{Alvin Wei Ming Tan}
\author[3]{Francesco Margoni}
\author[2]{Michael C. Frank}
\author[4]{Maya B. Mathur}

\authormark{Yi Li \textit{et al}.}

\address[1]{\orgdiv{Department of Epidemiology, Biostatistics and Occupational Health}, \orgname{McGill University}, \orgaddress{\city{Montreal}, \postcode{H3C 0X6}, \state{Quebec},  \country{Canada}.}
\email{yi.li10@mail.mcgill.ca}}

\address[2]{\orgdiv{Department of Psychology}, \orgname{Stanford University}, \orgaddress{\city{Stanford}, \postcode{94305}, \state{California},  \country{USA}}}

\address[3]{\orgdiv{Department of Psychology}, \orgname{University of Stavanger}, \orgaddress{\city{Stavanger}, \postcode{4036}, \state{Rogaland},  \country{Norway}}}

\address[4]{\orgdiv{Department of Pediatrics}, \orgname{Stanford University},
\orgaddress{\city{Stanford}, \postcode{94305}, \state{California}, \country{USA}}
}

\authormark{Li et al.}



\abstract{
Multisite replication projects often rely on published meta-analyses to plan their sample sizes, but meta-analyses tend to overestimate population effects because of publication bias and differences in study characteristics. We developed methods to forecast site-level estimates of a not-yet-conducted replication from a meta-analysis. Extending calibrated estimates from meta-regression, we constructed nonparametric prediction intervals that condition on each site’s moderator levels and correct for publication bias through inverse-probability-of-publication weighting under an assumed severity. These intervals achieve nominal coverage asymptotically when population effects are normally distributed, with a bounded coverage error otherwise. We compared four forecasting models (intercept-only meta-analysis or meta-regression, each with and without publication bias correction) in simulation study (81 scenarios and blinded retrospective forecasts of two multisite replications in infant cognition: ManyBabies 1 (104 sites) and ManyBabies 4 (33 sites). Correcting for publication bias improved interval calibration, with gains that increased with bias severity. Meta-regression improved forecasts only when measured moderators explained effect variation and replication sites differed meaningfully on them. In practice, publication bias correction should be the default when forecasting replications from meta-analyses, and replication projects that deliberately vary measured moderators across sites would enable more informative forecasts.

}

\end{Frontmatter}

\newpage
\section*{Highlights}
\subsection*{What is already known}
$\bullet$ In psychology, meta-analytic estimates are on average nearly three times larger than the matching multisite replication estimates. This gap is largely attributed to publication bias.

$\bullet$ There is no consensus on how to judge replication success. Proposed criteria include significance agreement, combined p-value tests, and checking whether the replication estimate falls within a prediction interval. Most of these compare one original study with one replication and ignore heterogeneity across multiple replication sites.

$\bullet$ Existing prediction intervals for meta-analysis cannot handle study-level moderators and publication bias at the same time, hence cannot produce site-specific forecasts for a planned replication.

\subsection*{What is new}
$\bullet$ We propose nonparametric prediction intervals for each replication site with an asymptotic coverage guarantee. They are built from meta-regression calibrated estimates that condition on the replication site’s moderators and correct for publication bias in the meta-analysis study with inverse-probability-of-publication weighting.

$\bullet$ In a simulation study and in two ManyBabies replications, publication bias correction drove most of the improvement in forecasts. Meta-regression helped only when the measured moderators explained effect variation and the replication sites varied on them.

$\bullet$ No single model was both the most accurate and the best calibrated, revealing a tradeoff between point accuracy and interval calibration.

\subsection*{Potential impact for readers outside the authors’ field}
$\bullet$ Before a multisite replication begins, the method turns an existing meta-analysis into a point forecast and a 95\% prediction interval for each site. Researchers can use these to plan sample sizes, choose site conditions, and judge whether an observed result is surprising. The method applies in any field where a published meta-analysis exists for the effect a multisite replication will test.

$\bullet$ When forecasting from a meta-analysis, correct for publication bias by default. Use meta-regression only when the moderators explain effect variation in the meta-analysis and the replication sites differ on them.

$\bullet$ Replication projects can make forecasts more informative by varying prespecified moderators across sites. A fully standardized protocol leaves no moderation variation for meta-regression, which limits forecasting and assessment of whether the effect generalizes across study characteristics.


\newpage
\section{Introduction}
\label{sec:introduction}
Multisite replication projects, in which many independent laboratories follow a shared protocol to re-verify a published finding, have become a central tool for assessing the credibility of published research. However, these projects are costly and time consuming. When one is being planned, the best available evidence about the effect size usually comes from a meta-analysis of the published literature. The meta-analysis literature, however, suffers from publication bias and spans heterogeneous study designs: a study found that across 15 psychology topics, meta-analytic estimates were on average nearly three times larger than the corresponding multisite replication estimates \cite{kvarven2020comparing}. This then raises a practical question for researchers planning a replication: given an existing meta-analysis, what effect should they expect at each replication site, and how much should that effect plausibly vary across sites?

In this paper, we develop and evaluate methods for forecasting the results of not-yet-conducted replication studies from published meta-analyses. For each replication site, our approach yields a point prediction and a 95\% prediction interval that adjust for a hypothesized degree of publication bias (the estimated selection ratio depending on the discipline\cite{mathur2021estimating}) and condition on the replication study site's own study-level characteristics through meta-regression. The selection ratio quantifies the severity of publication bias as the number of times more likely affirmative results are to be published than nonaffirmative results. We compare four forecasting models, crossing intercept-only meta-analysis versus meta-regression with and without publication bias correction, in a simulation study and in two multisite replications of infant cognition findings, ManyBabies 1\cite{manybabies2020quantifying} and ManyBabies 4\cite{lucca2025infants}. Finally, we discuss the limitations of the forecasting models and the conditions under which each is likely to be informative. In particular, we consider how the different purposes of meta-analyses and multisite replications bear on forecasting: meta-analyses synthesize studies whose populations and procedures vary, whereas multisite replications typically standardize their protocols to estimate a single effect precisely. We also examine the role of measured and unmeasured moderators in explaining differences between meta-analytic and replication results, and what these considerations imply for the design of future replication projects in psychology and beyond.

\subsection{Why Forecast Replication Results from Meta-Analyses?}
Large-scale efforts to measure the reliability of published research have shown that a single study is rarely decisive. The \href{https://www.cos.io/score}{Systematizing Confidence in Open Research and Evidence (SCORE)} program assessed the credibility of 3,900 claims from the social and behavioral sciences and empirically tested a subset of them for reproducibility, robustness, and replicability \cite{aczel2026investigating, miske2026investigating, tyner2026investigating}. Most claims could be reproduced when the original data and code were available \cite{miske2026investigating}, but only about a third proved robust across independent reanalyses \cite{aczel2026investigating}, and half of well-powered replication attempts reproduced the original study's statistically significant result, typically with substantially smaller effect sizes \cite{tyner2026investigating}. These findings caution against treating a single published analysis, or even a single replication of it, as dispositive evidence for or against a claim.

One solution has been the multisite, or ``many-to-one,'' replication design, in which a single original study is replicated across many independent laboratories following a standardized, preregistered protocol, often developed together with the original authors \cite{simons2014introduction}. ``Many Labs'' projects extend this design by replicating several original studies simultaneously across the same set of sites \cite{ebersole2016many, klein2014data}. Because their results are reported regardless of outcome, multisite replications provide large-sample estimates that are largely free of publication bias, and they have become a benchmark against which the published literature is judged. 

The discrepancy between estimates of meta-analysis studies and large preregistered replications is consistent in direction rather than random \cite{kvarven2020comparing}. This suggests that a meta-analysis, once corrected for selective publication and conditioned on the characteristics of each replication site, may still carry useful predictive information about what a planned replication study will find.
A forecast made before data collection would serve several purposes: it provides a prespecified benchmark for judging whether a replication result is surprising, informs sample size calculations and the choice of study conditions, and clarifies whether publication bias or differences in study design could drive any discrepancy with the replication.

Developmental psychology offers a natural test case. Research with infants faces discipline-specific replicability challenges, notably small samples and noisy measurement \cite{frank2017collaborative}. The ManyBabies consortium conducts multisite, preregistered replications of influential findings in infant cognition and selects effects for replication in part on the basis of existing meta-analyses \cite{frank2017collaborative}. As a result, both projects we study in this paper were paired with a meta-analysis of the same effect. The two projects also differed in a way that matters for forecasting: ManyBabies 1 sites varied on several study characteristics that were also measured and recorded in the corresponding meta-analysis \cite{manybabies2020quantifying}, whereas ManyBabies 4 standardized its protocol and stimuli across all study sites \cite{lucca2025infants}. We next consider two mechanisms that may account for discrepancies between meta-analytic and multisite replication estimates.

\subsection{First Source of Discrepancy: Publication Bias in Meta-Analyses}
Meta-analyses are widely used to synthesize evidence across studies, but their validity can be compromised at two levels: within individual studies, through confounding and measurement error, and across studies, through the selective publication of results. In this paper, we focus on across-study level, the most direct explanation for the systematic gap between meta-analytic and replication estimates.

Kvarven et al.\cite{kvarven2020comparing} compared 15 psychology topics for which both a meta-analysis and a preregistered multi-laboratory replication were available. In 12 of the 15 pairs, the meta-analytic effect size was significantly larger than the replication effect size, and on average the meta-analytic estimates were nearly three times as large. This inflation was consistent with publication bias and selective reporting in the primary literature. However, applying three established publication bias correction methods (trim-and-fill, PET-PEESE, and the three-parameter selection model) to the meta-analyses did not meaningfully improve their agreement with the replication estimates. 

One possible reason these corrections failed is that each must infer the severity of selection from the same, often small, set of published studies it is trying to correct, and such estimates can be unstable. An alternative approach treats the severity of publication bias as a sensitivity parameter rather than estimating it from the meta-analysis itself, and corrects the estimate under an assumed degree of selection favoring statistically significant results in the expected direction\cite{mathur2020sensitivity}. This assumed degree of selection can be informed by evidence pooled across many meta-analyses. In a meta-meta-analysis spanning medicine and psychology, Mathur and VanderWeele\cite{mathur2021estimating} found that publication bias was generally mild, with one notable exception: selective publication was considerably more severe among meta-analyses in developmental psychology. Although this finding was based on only five meta-analyses, the authors suggested that the small samples typical of infant research may contribute to more severe selection in this field.

\subsection{Second Source of Discrepancy: Differences in Study Characteristics}
Even under a shared protocol, replication results vary across sites: eight of sixteen effects in the first Many Labs project showed statistically significant heterogeneity\cite{klein2014data}. Published studies in a meta-analysis typically vary even more, in participant populations, stimuli, and procedures. A single pooled estimate can mask this variation, and a forecast that ignores it implicitly assumes that every replication site resembles an ``average'' published study.

Meta-regression\cite{sutton2000methods} addresses this by modeling the mean effect as a function of study-level characteristics, or moderators, such as participant age or experimental method. Mathur and VanderWeele\cite{mathur2019new, mathur2021meta} extended this approach with calibrated estimates, which shrink each study's observed estimate toward its model-predicted mean in proportion to the study's imprecision. The resulting set of calibrated estimates approximates the distribution of true effects at a given moderator level, not just its mean. They used this distribution to quantify, for example, the percentage of true effects that are meaningfully strong at a given moderator level. These metrics describe heterogeneity in existing evidence but were not designed to project the results of future studies.

\subsection{Prediction Intervals and Evaluating Replication Success}
A confidence interval for a meta-analytic mean describes uncertainty about the average effect. A prediction interval instead gives the range within which the estimate from a new study is expected to fall, accounting for both between-study heterogeneity and the new study's own sampling error. For forecasting a replication site's result, the prediction interval is the relevant quantity. Wang and Lee\cite{wang2019simple} proposed a nonparametric prediction interval for standard meta-analysis based on calibrated estimates which makes the variance of the empirical distribution equal the true variance asymptotically. However, their approach, like most existing prediction interval methods\cite{burr2005bayesian, ohlssen2007flexible, lee2008flexible, jpt2009re, graham2012robust, kontopantelis2012performance, inthout2016plea, yamaguchi2017random}, does not accommodate study-level moderators and publication bias simultaneously.

Forecasting differs from a related body of literature that evaluates replication success after both the original and replication results are available. Common practice relies on ``significance agreement,'' a criterion whose expected value depends on the statistical power of both studies. More principled criteria compare a single original study with a single replication. Held et al.\cite{held2022ReplicationSuccess} proposed a sum-of-$p$-values test that avoids requiring independent significance in both studies\cite{held2024assessment}, a test of whether the replication estimate falls within the original study's prediction interval\cite{cochran1954combination}, and the skeptical $p$-value, which penalizes shrinkage from the original to the replication estimate\cite{held2020sceptical}. Inverting Edgington's combined test\cite{edgington1972additive, edgington1972normal} additionally yields a combined point estimate and confidence interval with favorable coverage properties\cite{held2025comparison}. These pairwise criteria, however, do not account for heterogeneity among multiple replications. 

Mathur and VanderWeele\cite{mathur2020new} addressed this evaluation metric for multisite designs by assessing whether a single estimate is statistically consistent with a heterogeneous distribution of effects estimated from many others, rather than with a single pooled estimate. Because this framework compares one estimate against a distribution, it can be extended naturally to comparing a replication result with a meta-analysis. 

\subsection{Contributions and Preview of Findings}
This paper makes three contributions. First, we construct nonparametric prediction intervals for replication studies by extending the meta-regression calibrated estimates of Mathur and VanderWeele\cite{mathur2021meta} and the prediction intervals of Wang and Lee\cite{wang2019simple}. The intervals condition on each replication site's moderator levels and incorporate a weighting-based correction for publication bias. We show that these intervals achieve nominal coverage asymptotically when the population effects are normally distributed, and approximately nominal coverage otherwise, with a bound on the coverage error. Second, we evaluate four forecasting models, together with kernel density estimate (KDE)-based and standard parametric prediction intervals, in a simulation study that varies the severity of publication bias, between-study heterogeneity, moderator strength, and the number of meta-analyzed studies. Third, we apply the models to two ManyBabies projects whose sites differ sharply in how much moderators varied.

We highlight three main findings. First, correcting for publication bias consistently reduced the tendency of meta-analyses to overpredict replication results. Second, meta-regression improved forecasts only when the measured moderators explained variation in the population effects: in simulation, the corrected meta-regression gave the most accurate forecasts whenever moderation was present, but the corrected meta-analysis was more accurate without moderation, and in the ManyBabies applications meta-regression offered little improvement. Third, correcting for publication bias improved the calibration of the prediction intervals as well as the location of the forecasts, provided the inverse probability of publication weight (IPPW) weights were also applied to the KDE, but it did not fully restore nominal coverage. Under strong publication bias in the simulation, the corrected models' KDE intervals covered substantially more replication estimates than the uncorrected models', and in ManyBabies~4, where the shared protocol fixed most moderators across sites, no model reached nominal coverage, suggesting that unmeasured factors drove much of the variation across sites. This pattern has a design implication: replication projects that deliberately vary prespecified moderators across sites would allow more informative forecasts and tests of generalizability.

\subsection{Organization of the Paper}
Section~\ref{sec:mbintro} describes the ManyBabies~1 and ManyBabies~4 replication projects and their corresponding meta-analyses. Section~\ref{sec:methods} introduces notation, calibrated estimates, publication bias correction, and the proposed methods for prediction intervals. Section~\ref{sec:simulation} presents a simulation study, and Section~\ref{sec:mbapplication} applies the four forecasting models to the ManyBabies data. Section~\ref{sec:discussion} discusses the trade-off between accuracy and calibration, the role of unmeasured moderators, and implications for the design of future replication projects.


\section{Motivating Study: ManyBabies}
\label{sec:mbintro}

Developmental psychology research with infant participants faces discipline-specific replicability challenges, notably small samples and a reliance on indirect behavioral measures, including both looking- and action-based measures (e.g. looking time, eye tracking, manual choice, head-turn procedure, imitation). ManyBabies is a large-scale, multi-laboratory collaborative initiative that conducts multisite, preregistered replications of landmark findings in developmental psychology\cite{frank2017collaborative}. Studies are selected for replication based in part on evidence from existing meta-analyses.

\subsection{ManyBabies 1: Infant-Directed Speech Preference}

Infant-directed speech (IDS) is the characteristic speech caregivers use with infants across cultures. Compared with adult-directed speech (ADS), IDS is typically higher pitched, has greater pitch excursions, and is characterized by shorter utterances, among other differences \cite{fernald1989intonation}. A substantial literature has demonstrated that infants prefer IDS over ADS \cite{cooper1997development, cooper1994developmental, fernald1985four, hayashi2001developmental, kitamura2009age, newman2006changes, pegg1992preference, santesso2007frontal, singh2002infants, werker1989infant}, with looking time as a typical dependent measure.

A meta-analysis of 34 traditionally published studies including 840 infants estimated an uncorrected point estimate of 0.67 on Cohen's d scale (95\% CI: [0.57, 0.76]) for the IDS preference in looking time \cite{dunst2012preference}. In the ManyBabies 1 replication project \cite{byers2020building, frank2017collaborative, 
klein2014data, manybabies2020quantifying}, the pooled estimate from 67 labs including 2,329 infants was approximately half as large (0.35; 95\% CI: [0.29, 0.42]) \cite{zettersten2024evidence}. The moderated model found that infant's age and experimental method (single screen central fixation, head-turn preference procedure, or eye tracking) was predictive of the IDS preference in ManyBabies 1 only, but not in the meta-analysis\cite{zettersten2024evidence}.

A recent synthesis \cite{zettersten2024evidence} re-analyzed the original meta-analysis \cite{dunst2012preference} and the  ManyBabies-Consortium\cite{manybabies2020quantifying} data, concluding that the IDS preference generalizes across a broad range of participants, ages, experimental methods, and stimuli, albeit with substantial remaining uncertainty about how moderators influence the effect magnitude. This work also highlights the key distinctions between meta-analyses and multi-lab replications: meta-analyses estimate effects across heterogeneous design choices and populations in the extant literature, while multi-lab replications offer large-scale, high-precision estimation of key effects within similar implementations, free from publication bias.

\subsection{ManyBabies 4: Social Evaluation}
Hamlin, Wynn, and Bloom \cite{hamlin2007social} conducted a study among preverbal infants to determine if they are able to evaluate people around them as friends or foes. The study used two paradigms: (1) a choice paradigm\cite{choi2023understanding}, where infants indicated preferences through reaching behavior, and (2) a violation-of-expectation paradigm \cite{margoni2024violation}, in which infants' looking times are measured on the premise that infants look longer at events that violate their expectations. In the latter paradigm, infants who had watched a climber being helped by one character and hindered by another looked longer when the climber subsequently approached the hinderer than when it approached the helper, suggesting that infants expected the climber to prefer its helper \cite{hamlin2007social, lee2015development}.

This initial paper has motivated numerous studies to explore infants' sociomoral preferences and their expectations regarding scenarios involving morally good and bad agents \cite{bloom2016develops, buon2014friend, chae2018negativity, kiley2010three, hamlin2011young, scola2015further, van2016evidence, holvoet2016infants, loheide2021toddlers, shimizu2021cultural}. Although many studies have reported significant preferences for helpful or fair agents over hindering or unfair agents, several subsequent published studies\cite{scarf2012social, schlingloff202015, salvadori2015probing, cowell2015precursors, nighbor2017stability, vaporova2020social, shimizu2021cultural} and  unpublished raw data and manuscripts \cite{abramson2016preferences, hamlin2016box, woo2016older} failed to replicate Hamlin et al.’s findings\cite{hamlin2007social}. 

These failed attempts motivated Margoni and Surian’s meta-analysis of 26 studies\cite{margoni2018infants}. Margoni and Surian\cite{margoni2018infants} included only the studies with the manual task outcome, i.e. the choice paradigm, in Hamlin et al.\cite{hamlin2007social}, and their results suggested that about two thirds of infants preferred prosocial agents to antisocial agents. The estimated effect was not moderated by age, the type of dependent variable (selective reaching or helping), or the modality of stimulus presentation (cartoons or real events). Furthermore, there was evidence of publication bias, with the point estimate in the published studies greater than the unpublished ones. 

Tan's\cite{tan2024prosocial} overview of the recently published studies in this field including 65 papers and 213 point estimates\cite{margoni2018infants} meta-analyzed studies that examined infants’ social moral preferences with manual tasks and infants’ looking time (e.g., violation of expectation regarding
socially relevant events, preferential looking between the two agents,
anticipatory looking for expectation of event completion). For the manual task outcome, Margoni and Surian found that scenario type (help/hinder, fair/unfair, give/take, other) and presentation modality (movie, live show) were two significant moderators of the point estimate, but the lab of origin, age, dependent variable (reaching, helping), modality of stimuli presentation (real events, cartoons), and choice object (shapes, puppets, people, experimenters) were not, and there was evidence of publication bias, with the point estimate in the published studies greater than the unpublished ones\cite{margoni2018infants}. For the looking time outcome, the only moderator was the method of measurement (i.e., the type of paradigm used in the task: violation of expectation, preferential looking, anticipatory looking).

The recently published ManyBabies 4\cite{lucca2025infants} attempted to obtain a more precise estimate of the true point estimate for infants' preference for helpers over hinderers by standardizing methodology across 37 labs worldwide: each lab followed the same protocol and used the same video stimuli. Consequently, most of the moderators recorded in the meta-analytic literature took a single, constant value across the 37 replication studies. The study found no evidence that infants prefer helpers over hinderers (49.34\% of infants preferred helpers over hinderers). The exploratory analyses examined several moderators, including attention to video events, clarity of action choices, and experiment-, child-, and lab-level factors. None of the interaction effects were significant.

ManyBabies 1 and ManyBabies 4 differ sharply in how much their replication sites varied on study-level covariates: ManyBabies 1 sites varied meaningfully in infant age, test language, and experimental method, whereas ManyBabies 4 standardized protocol and stimuli across all 37 sites. As we show in Section~\ref{sec:mbapplication}, this contrast affects how informative ManyBabies 1 and ManyBabies 4 are for comparing the performance of the four forecasting models.

\section{Methods}
\label{sec:methods}
The ManyBabies studies described in Section~\ref{sec:mbintro} share a common structure: each is a multisite replication of a key finding in infant cognition, motivated by a meta-analysis of the published literature on that finding. Each replication site contributes its own estimate of the effect.
Our goal is to use the meta-analysis dataset to forecast each site's estimate before the replication is conducted. Both ManyBabies examples also illustrate the two sources of discrepancy introduced in Section~\ref{sec:introduction}: the published literature may be affected by publication bias, and the replication sites may differ from the meta-analyzed studies in their moderator distribution. In this section, we introduce and develop forecasting models that address both. Section~\ref{sec:notation} introduces notation and the intercept-only and the meta-regression models. We then describe calibrated estimates, which adjust each meta-analyzed study's estimate to a given moderator level (Section~\ref{sec:calibrated}), and a weighting approach that corrects for publication bias (Section~\ref{sec:pubbias_correction}). Section~\ref{sec:prediction} combines these components to construct prediction intervals for the estimate at each replication site, states the assumptions under which the intervals are valid, and establishes their asymptotic coverage. Section~\ref{sec:4models} introduces four forecasting models, and Section~\ref{sec:metrics} introduces the evaluation metrics.

\subsection{Notation}
\label{sec:notation}

Let $k$ denote the total number of studies in the meta-analysis dataset. For the $i$th meta-analysis study ($i = 1, \ldots, k$), let $\theta_i$, $\hat{\theta}_i$, and $\hat{\sigma}_i^2$ denote the true population
effect, point estimate, and estimated within-study variance, respectively.

Denote $Z_i$ as the $1 \times (p+1)$ row vector of study-level covariates for study $i$, with the first entry fixed at $1$ ($z_{i1}=1$ for all $i$) to accommodate the population-level intercept. Denote $z_i$ as the vector of realized values of covariates observed for study $i$, with scalar entries $z_{i1},\dots,z_{ip}$.

Under a linear meta-regression model,
\[
\E(\theta_i \mid Z_i) = \beta_0 + Z_i \boldsymbol{\beta}_1,
\]
the full $k \times (p+1)$ moderator matrix is
\[
\boldsymbol{Z} = \begin{bmatrix}
    1 & z_{11} & \cdots & z_{1p} \\
    \vdots & \vdots & \ddots & \vdots \\
    1 & z_{k1} & \cdots & z_{kp}
\end{bmatrix},
\]
with coefficient vector
$\boldsymbol{\beta} = (\beta_0, \beta_1, \ldots, \beta_p)^T$, so that
$\boldsymbol{\Theta} = \boldsymbol{Z}\boldsymbol{\beta}$, where
$\boldsymbol{\Theta} = (\theta_1, \ldots, \theta_k)^T$. We assume the residual variance of the population effects conditional on the covariates, $\mathrm{Var}(\theta_i \mid Z_i)$, is a constant $\tau_c^2$; this is the residual heterogeneity conditional on the meta-regression covariates, as distinct from the marginal $\tau^2$ of a standard intercept-only meta-analysis model.

When the model contains only an intercept, $\beta_0$ and $\tau_c^2$ reduce to the mean and between-study variance of the population effects, denoted $\mu$ and $\tau^2$, respectively, with estimates $\hat{\mu}$ and $\hat{\tau}^2$. More generally, let $\hat\beta_0$, $\hat{{\beta}}_1$, and $\hat\tau_c^2$ denote the meta-regression estimates of $\beta_0$, $\beta_1$, and $\tau_c^2$. In practice, we estimate these parameters using robust variance estimation\cite{hedges2010robust, tipton2015small}, a semiparametric approach analogous to generalized estimating equations that yields consistent coefficient estimates without distributional assumptions on the true population effect of study $i$ \cite{mathur2021meta}.

For replication studies, we use the index $j = 1, \ldots, n$, and denote the
point estimate and estimated variance of the $j$th replication study as
$\hat{\theta}_j^{\mathrm{rep}}$ and $\hat{\sigma}_j^{2,\mathrm{rep}}$,
respectively, and the point forecast as $\hat{\theta}_j^{\mathrm{pred}}$, with
estimated variance $\widehat{\mathrm{Var}}\bigl(\hat{\theta}_j^{\mathrm{pred}}\bigr)$.
The realized value of the $j$th replication study's covariate vector is $z_j$.
For the meta-analysis of $k$ studies, $\hat{\mu}$ denotes the pooled point
estimate from an intercept-only model; $\hat{\beta}$ and $\widehat{V}_{R}$
denote the estimated coefficient vector of a meta-regression and its robust
covariance matrix, respectively; and $\hat{\tau}^2$ denotes the estimated
between-study variance, which is the residual
between-study variance for meta-regression models.

\subsubsection{Calibrated Estimate}
\label{sec:calibrated}
When the model contains moderators,
following  Mathur and VanderWeele \cite{mathur2021meta}, define the \textit{calibrated estimate} of the $i$th meta-analysis study, shifted to a specific covariate level $Z = z^\ast$, as
\begin{equation}
\tilde\theta_i(Z=z^\ast) = \hat\beta_0 + z^\ast\hat{\boldsymbol{\beta}}_1 + \sqrt{\frac{\hat\tau_c^2}{\hat\tau_c^2 + \hat\sigma_i^2}}\left[\hat\theta_i - \left(\hat\beta_0 + z_i\hat{\boldsymbol{\beta}}_1\right)\right].
\label{eq:calibrated}
\end{equation}
The shrinkage factor $\sqrt{\hat\tau_c^2/(\hat\tau_c^2+\hat\sigma_i^2)}$ pulls each study's residual toward zero in proportion to the study's imprecision: relatively imprecise estimates (large $\hat\sigma_i^2$) are shrunk strongly toward the meta-regression mean $\hat\beta_0 + z^\ast\hat{\boldsymbol{\beta}}_1$, while precise estimates remain close to their reported values \cite{mathur2021meta, wang2019simple}.
When there are no moderators, equation~\eqref{eq:calibrated} reduces to the standard calibrated estimate of Wang and Lee\cite{wang2019simple}:
\begin{equation}
\tilde\theta_i = \hat\mu + \sqrt{\frac{\hat\tau^2}{\hat\tau^2+\hat\sigma_i^2}}\left(\hat\theta_i - \hat\mu\right).
\label{eq:calibrated-nomod}
\end{equation}

When studies share a common dataset or population (e.g., multiple publications from the same study population), we group them into $M$ clusters, stacking by cluster:
\[
\boldsymbol{Z} = \begin{bmatrix} \boldsymbol{Z}_1 \\ \vdots \\
\boldsymbol{Z}_M \end{bmatrix}, \qquad
\boldsymbol{\Theta} = \begin{bmatrix} \boldsymbol{\theta}_1 \\ \vdots \\
\boldsymbol{\theta}_M \end{bmatrix},
\]
where the $m$th cluster contributes $k_m$ studies with
$\sum_{m=1}^M k_m = k$. When referring to the clustered structure, we index study $i$ within cluster $m$ by the double subscript $(m,i)$, e.g., $\theta_{mi}$, $\hat\theta_{mi}$, $\hat\sigma_{mi}^2$; when clustering is not relevant we retain the single index $i$ used above.

\subsubsection{Publication Bias Correction}
\label{sec:pubbias_correction}

Let $A_{mi} \in \{0,1\}$ indicate whether study $i$ in cluster $m$ is \textit{affirmative}, defined as ``having a positive point estimate and a two-sided p-value below 0.05''. Following  Mathur and VanderWeele\cite{mathur2020sensitivity}, we refer to this as ``asymmetric selection process'', i.e. ``favoring studies that are both statistically significant and positive in direction'', as ``one-tailed selection'', in contrast to a ``two-tailed selection'' model in which publication favors significant results regardless of the direction. Note that this terminology describes the direction of the selection process itself, not that individual studies conducted one-tailed hypothesis tests; the p-value used to define $A_{mi}$ remains two-sided throughout.

Let $\eta \geq 1$ denote the ratio of publication probabilities for affirmative versus non-affirmative studies,
used to parameterize publication bias. Following Mathur and VanderWeele\cite{mathur2020sensitivity}, for publication-bias-corrected analyses, let
\begin{equation}
w_{mi} = \frac{\eta^{\,\mathbb{I}(A_{mi}=0)}}{\hat\sigma_{mi}^2 + \hat\tau_c^2}
\label{eq:omega_mi}
\end{equation}
denote the inverse probability of publication weight (IPPW) for study $i$ in cluster $m$, so that non-affirmative studies receive weight $\eta$ times that of affirmative studies. The per-cluster weight matrix is $W_m^{(\eta)} = \mathrm{diag}(w_{m1}, \ldots, w_{mk_m})$, and the full block-diagonal weight matrix is $\boldsymbol{W}^{(\eta)} = \mathrm{diag}(W_1^{(\eta)}, \ldots, W_M^{(\eta)})$. 

\noindent\textit{Remark.} We recommend obtaining $\hat\tau_c^2$ as a naive estimate from a standard meta-regression without correction for publication bias rather than jointly re-estimating $\tau_c^2$ under IPPW\cite{mathur2020sensitivity}. Although this naive estimate may itself be biased by publication bias, this does not compromise the consistency of $\hat{\boldsymbol\beta}_\eta$ or the validity of its robust sandwich variance estimator (later in Equation~\eqref{eq:sandwich}), as the robust specification of Hedges et al. \cite{hedges2010robust} provides unbiased point estimation and valid inference for any fixed choice of weight matrix $\boldsymbol{W}^{(\eta)}$, so an inaccurate $\hat\tau_c^2$ can only reduce estimation efficiency but not compromise validity.

\subsection{Prediction of Replication Studies Using Meta-Analysis Results}
\label{sec:prediction}

Wang and Lee\cite{wang2019simple} introduces nonparametric prediction intervals using calibrated estimates in standard meta-analysis, but their approach
does not accommodate study-level moderators. Mathur and VanderWeele\cite{mathur2021meta}
extends calibrated estimates to meta-regression by conditioning on
study-level covariates, but does not construct prediction intervals. To
fill both gaps simultaneously, we derive prediction intervals from the
meta-regression calibrated estimates of Mathur and VanderWeele\cite{mathur2021meta}, conditioning on each replication study's own covariate vector. We additionally incorporate IPPW to correct for publication bias.

\subsubsection{Assumptions}
The proposed method rests on four assumptions:
\begin{enumerate}[label=(A\arabic*)]
\item \textbf{Random-effects meta-regression:} In the population of all conducted studies, $\theta_i = \beta_0 + Z_i\boldsymbol\beta_1 + u_i$, where$(Z_i, \hat\sigma_i, u_i)$ are independent and identically distributed across studies, the moderators $Z_i$ have bounded support, and the residuals $u_i$ are independent of $(Z_i, \hat\sigma_i)$, with mean zero, positive variance $\tau_c^2$, and an absolutely continuous distribution. No parametric form is assumed for the distribution of $u_i$.
\item \textbf{Within-study normality:} $\hat\theta_i \mid \theta_i, Z_i \overset{\mathrm{ind}}{\sim} N(\theta_i, \hat\sigma_i^2)$, with the known within-study variances, bounded between zero and infinity.
\item \textbf{Selection model} (publication-bias-corrected models only): Affirmative studies are $\eta$ times more likely to be published than nonaffirmative studies, that is, $\Pr(\text{published} \mid A_i = 1) / \Pr(\text{published} \mid A_i = 0) = \eta$, where $\eta \ge 1$ is known and $\Pr(\text{published} \mid A_i = 0) > 0$. Publication depends on a study's point estimate only through whether a study is affirmative.
\item \textbf{Transportability and independence:} The population effect at replication site $j$, with moderator level $z_j$, follows the same conditional distribution as the population effects in the meta-analyzed literature, $\theta_j^{\mathrm{rep}} = \beta_0 + z_j\boldsymbol\beta_1 + u_j$ with $u_j$ having the same distribution as $u_i$; the replication estimate satisfies $\hat\theta_j^{\mathrm{rep}} \mid \theta_j^{\mathrm{rep}} \sim N\bigl(\theta_j^{\mathrm{rep}}, \hat\sigma_j^{2}\bigr)$ and is independent of the meta-analysis data.
\end{enumerate}

The following Proposition 1 shows that the meta-regression calibrated estimates $\tilde\theta_i(Z = z)$ (Equation~\eqref{eq:calibrated}) match the population effects at $Z = z$ in mean and variance, extending the key property of Wang and Lee~\cite{wang2019simple}, and match their full distribution when the population effects are normal. This justifies their use as the nonparametric basis for prediction intervals.

\bigskip
\noindent
\textbf{Proposition 1} (Calibrated estimates). Under (A1) and (A2), in the population of all conducted studies, the oracle calibrated estimates $\tilde\theta_i^*(Z = z)$, defined as in Equation~\eqref{eq:calibrated} but with the true parameters $(\beta_0, \beta_1, \tau_c^2)$, are independent and identically distributed with mean $\E(\theta \mid Z = z)$ and variance $\tau_c^2$ for every $k$. If $(\hat\beta_0, \hat\beta_1, \hat\tau_c^2)$ is consistent, the feasible calibrated estimates $\tilde\theta_i(Z = z)$, which use the estimated parameters as in Equation~\eqref{eq:calibrated}, satisfy $\tilde\theta_i(Z = z) - \tilde\theta_i^{*}(Z = z) \to 0$ in probability as $k \to \infty$.
If, in addition, the population effects are normally distributed, then $\tilde\theta_i^*(Z = z)$ has the same distribution as $\theta \mid Z = z$, and $\tilde\theta_i(Z = z)$ converges in distribution to $\theta \mid Z = z$.

Proposition 1 formalizes the moment-matching argument of Mathur and VanderWeele~\cite{mathur2021meta} for the shifted calibrated estimates and extends it to convergence in distribution when the population effects are normal. The proof is given in Appendix~\ref{app:prop1}.

\textit{Remarks.} Denote $\mu(z) = \beta_0 + z\beta_1$, $e_i = \hat\theta_i - \theta_i$, and $\lambda_i^{*} = \sqrt{\tau_c^2/(\tau_c^2 + \hat\sigma_i^2)}$, so that $\tilde\theta_i^{*}(Z = z) = \mu(z) + \lambda_i^{*}\{\hat\theta_i - \mu(Z_i)\}$.
Proposition 1 guarantees that the calibrated estimates match the population effects in mean and variance, but not in shape. When the population effects are not normal, the oracle estimate $\tilde\theta_i^{*}(Z = z) - \mu(z) = \lambda_i^{*}(u_i + e_i)$ mixes a shrunken residual with normal sampling noise, so its distribution generally differs from that of $\theta \mid Z = z$; the two are close when the population effects are approximately normal or when within-study variances are small relative to $\tau_c^2$. Proposition~1 also describes conducted studies; among published studies, the weights $\omega_i$ need to be used to correct for publication bias under (A3).

We now construct prediction intervals for the point estimate
$\hat{\theta}_j^{\mathrm{rep}}$ of a future replication study by using the
empirical distribution of calibrated estimates from the meta-analysis studies,
shifted to the replication study's covariate level $Z = z_j$.

\subsubsection{Prediction Interval}
\paragraph{Uncorrected for publication bias.}

The goal is to estimate the marginal predictive density
\[
f_{\hat{\theta}_j}(t) = \int f_{\hat{\theta}_j \mid \theta}(t \mid w)
\cdot f_\theta(w) \, dw,
\]
where $j$ indexes the replication study. We assume
$\hat{\theta}_j \mid \theta_j \sim N(\theta_j, \hat{\sigma}_j^2)$, i.e., the reported point estimate is normally distributed around the true population effect. Because the distribution of true population effects is unknown, we approximate it nonparametrically using the empirical distribution of calibrated estimates $\tilde{\theta}_i(Z = z_j)$ from
the meta-analysis studies shifted to $z_j$.

Following Wang and Lee~\cite{wang2019simple}, the predictive density via KDE is
\begin{equation}
    \hat f_j(t) = \frac{1}{k}\sum_{i=1}^k \phi\bigl(t;\ \tilde\theta_i(Z=z_j),\ h^2+\hat\sigma_j^2\bigr),
    \label{eq:kde-pi}
\end{equation}
where $\phi(\cdot;\mu,s^2)$ denotes the Gaussian density with mean $\mu$ and variance $s^2$, and $h$ is the bandwidth, selected using the Sheather--Jones method~\cite{sheather1991reliable}, as recommended by Wang and Lee~\cite{wang2019simple}. The $(1-\alpha)$ prediction interval consists of the $\alpha/2$ and $1-\alpha/2$ quantiles of $\hat f_j$; for example, the 95\% interval uses the 2.5th and 97.5th percentiles.

\paragraph{Corrected for publication bias.}
\label{sec:corrected}

Following Mathur and VanderWeele\cite{mathur2020sensitivity}, we assume that affirmative studies are $\eta$ times more likely to be published than non-affirmative studies. We use IPPW, with weights $w_{mi}$ and weight matrices $\boldsymbol{W}_m^{(\eta)}$ as defined in Section~\ref{sec:pubbias_correction}, to upweight non-affirmative studies.

The IPPW meta-regression coefficient estimate is
\begin{align}
    \hat{\boldsymbol{\beta}}_\eta = \left(\sum_{m=1}^M \boldsymbol{Z}_m
    \boldsymbol{W}_m^{(\eta)} \boldsymbol{Z}_m^T \right)^{-1}
    \left(\sum_{m=1}^M \boldsymbol{Z}_m \boldsymbol{W}_m^{(\eta)}
    \hat{\boldsymbol{\theta}}_m \right), \label{eq:beta_eta}
\end{align}
with sandwich variance estimator \cite{mathur2020sensitivity}:
\begin{align}
    \widehat{\Var}(\hat{\boldsymbol{\beta}}_\eta) = \frac{M}{M-1}
    \left(\sum_{m=1}^M \boldsymbol{Z}_m \boldsymbol{W}_m^{(\eta)}
    \boldsymbol{Z}_m^T\right)^{-1}
    \left[\sum_{m=1}^M \boldsymbol{Z}_m^T \boldsymbol{W}_m^{(\eta)}
    \boldsymbol{e}_m \boldsymbol{e}_m^T \boldsymbol{W}_m^{(\eta)}
    \boldsymbol{Z}_m\right]
    \left(\sum_{m=1}^M \boldsymbol{Z}_m \boldsymbol{W}_m^{(\eta)}
    \boldsymbol{Z}_m^T\right)^{-1}\!, \label{eq:sandwich}
\end{align}
where $\boldsymbol{e}_m = \hat{\boldsymbol{\theta}}_m -
\boldsymbol{Z}_m \hat{\boldsymbol{\beta}}_\eta$ is the residual vector for cluster $m$.

We then obtain IPPW-calibrated estimates $\tilde{\theta}_i(Z = z_j)$
using Equation~\eqref{eq:calibrated} with $\hat{\boldsymbol{\beta}}_\eta$ in place of the unweighted $\hat{\boldsymbol{\beta}}$, and construct prediction intervals as described above. 
The next proposition characterizes the asymptotic coverage of these intervals.

To state it, we need the limit of the heterogeneity estimate. For the corrected models, $\hat\tau_c^2$ is taken from the corresponding uncorrected model, so it need not be consistent for $\tau_c^2$ under publication bias. Let $\tau_\dagger^2 > 0$ denote its probability limit, i.e., $\hat\tau_c^2 \xrightarrow{p} \tau_\dagger^2$ as $k \to \infty$, with $\tau_\dagger^2 = \tau_c^2$ when $\hat\tau_c^2$ is consistent. Replacing $\tau_c^2$ with $\tau_\dagger^2$ in the oracle calibrated estimate of Proposition~1 gives the limiting calibrated estimate
\[
\tilde\theta_i^{\dagger}(Z = z) = \mu(z) + \lambda_i^{\dagger}\bigl\{\hat\theta_i - \mu(Z_i)\bigr\},
\qquad
\lambda_i^{\dagger} = \sqrt{\frac{\tau_\dagger^2}{\tau_\dagger^2 + \hat\sigma_i^2}} .
\]
Let $g_z^{\dagger}$ denote the density of $\tilde\theta_i^{\dagger}(Z = z)$ in the population of conducted studies, and $f_{\theta \mid z}$ the density of $\theta \mid Z = z$. When $\tau_\dagger^2 = \tau_c^2$, $\tilde\theta_i^{\dagger}(Z = z)$ is the oracle calibrated estimate $\tilde\theta_i^{*}(Z = z)$.

Proposition~2 below shows that the asymptotic coverage of the KDE-based prediction interval deviates from the nominal level by at most the $L^1$ distance between $g_{z_j}^{\dagger}$ and $f_{\theta \mid z_j}$, and attains the nominal level when $\hat\tau_c^2$ is consistent and the population effects are normal.

\textbf{Proposition 2} (Asymptotic coverage). Assume (A1), (A2), and (A4). For the corrected models, assume (A3); for the uncorrected models, assume that (A3) holds with $\eta = 1$,  

Suppose that as $k \to \infty$,
\begin{enumerate}[label=(\alph*)]
    \item $\|\hat{\boldsymbol\beta} - \boldsymbol\beta\| = O_p(k^{-1/2})$, with $\hat{\boldsymbol\beta}_\eta$ in place of $\hat{\boldsymbol\beta}$ for the corrected models;
    \item $|\hat\tau_c^2 - \tau_\dagger^2| = O_p(k^{-1/2})$;
    \item the bandwidth satisfies $h \to 0$ and $h\sqrt{k} \to \infty$.
\end{enumerate}
Then the KDE-based $(1-\alpha)$ prediction interval $\widehat{\mathrm{PI}}_j$ satisfies
\[
\lim_{k \to \infty} \Pr\bigl(\hat\theta_j^{\mathrm{rep}} \in \widehat{\mathrm{PI}}_j\bigr) = 1 - \alpha + R_j,
\qquad
|R_j| \le \bigl\| g_{z_j}^{\dagger} - f_{\theta \mid z_j} \bigr\|_1 .
\]
In particular, if $\tau_\dagger^2 = \tau_c^2$ and the population effects are normally distributed, then $R_j = 0$, and the interval PI achieves nominal coverage asymptotically.

Condition (a) above states that the meta-regression coefficients are root-$k$ consistent, which holds for the robust variance estimators used here \cite{hedges2010robust}. In the special case above, $g_{z_j}^{\dagger}$ is the density of the oracle calibrated estimates, which equals $f_{\theta \mid z_j}$ by Proposition~1. The proof is given in Appendix~\ref{app:prop2}.

\smallskip
\textit{Remarks.} Four conditions determine how closely the prediction interval approaches nominal coverage in practice. 

(i) Beyond the exact case in Proposition~2, $R_j$ is small when the population effects are approximately normal or when the meta-analyzed studies are precise relative to the heterogeneity, because the calibrated estimates then closely resemble the population effects themselves. 

(ii) The heterogeneity estimate determines the shrinkage applied to each calibrated estimate. For the uncorrected models, $\hat\tau_c^2$ is consistent under standard meta-regression theory when publication does not depend on the study results. For the corrected models, we use the estimate from the corresponding uncorrected model as a plug-in (Section~\ref{sec:pubbias_correction}), because the IPPW estimator does not directly yield a heterogeneity estimate. Under publication bias, this plug-in generally converges to some $\tau_\dagger^2 \neq \tau_c^2$, so the shrinkage factors $\lambda_i^{\dagger}$ are misspecified and the coverage error need not vanish even under normality. This error shrinks as $\eta \to 1$, and also when $\hat\sigma_i^2 \ll \tau_\dagger^2$, because then $\lambda_i^{\dagger} \approx 1$ regardless of $\tau_\dagger^2$. 

(iii) The IPPW weights in the KDE address a separate problem: they ensure that the estimated distribution represents all conducted studies rather than only the published ones. Without these weights, the uncorrected intervals target the distribution of published studies, and Proposition~2 provides no guarantee for them under publication bias. 

(iv) If, conditional on a given moderator level $z_j$, the population effects at the replication sites differ systematically from those in the meta-analyzed literature, for example, because of unmeasured moderators, Assumption~(A4) fails and nominal coverage is not guaranteed even asymptotically.

For comparison with the KDE-based intervals, one may also construct standard parametric prediction intervals, which assume that the replication estimate follows a $t$-distribution centered at the point forecast $\hat\mu(z_j)$:
\[
  \hat\mu(z_j) \pm t_{\mathrm{df},\,1-\alpha/2} \sqrt{\hat\tau_c^2 + \mathrm{SE}^2\bigl(\hat\mu(z_j)\bigr) + \hat\sigma^2_j},
\]
where $\mathrm{SE}(\hat\mu(z_j))$ is the robust standard error of the point forecast and $t_{\mathrm{df},\,1-\alpha/2}$ is the $1-\alpha/2$ quantile of the $t$-distribution with Satterthwaite degrees of freedom~\cite{jpt2009re, tipton2015small}. Unlike the KDE-based intervals, these intervals assume that the population effects are normally distributed around the forecast. Appendix~\ref{app:pi-kde-param} describes both types of intervals in detail.

\subsection{Forecasting models}
\label{sec:4models}
Throughout the paper, we compare four forecasting models, formed by crossing two settings: whether the model conditions on study-level moderators, and whether it corrects for publication bias.

\begin{enumerate}[(1)]
    \item[(1)] MA-U: intercept-only meta-analysis, uncorrected for publication bias;
  \item[(2)] MR-U: meta-regression, uncorrected for publication bias;
  \item[(3)] MA-C: intercept-only meta-analysis, corrected for publication bias;
  \item[(4)] MR-C: meta-regression, corrected for publication bias.
\end{enumerate}

The intercept-only meta-analysis models produce the same point forecast for every replication study, whereas the meta-regression models produce study-specific forecasts at each replication study's moderator levels. For each model, we construct both KDE-based prediction intervals and standard parametric prediction intervals for comparison.

All models are fitted in R using the \texttt{robumeta} package\cite{fisher2015robumeta}, which provides robust variance estimation with Satterthwaite small-sample degrees of freedom\cite{tipton2015small}. The uncorrected models used hierarchical weights where studies contribute multiple estimates, or correlated-effects weights where each study contributed a single estimate, and the corrected models used the IPPW weights, supplied via the \texttt{userweights} argument. Kernel density bandwidths were selected using the Sheather--Jones method\cite{sheather1991reliable}.

\subsection{Evaluation Metrics}
\label{sec:metrics}

We evaluate the four forecasting models using seven metrics that compare each model's prediction to the reported estimate in each replication study. Three are accuracy metrics: mean absolute error (MAE), root mean square error (RMSE), and mean prediction interval width. Four are calibration metrics~\cite{gneiting2007strictly}: raw difference, which quantifies the directional difference between the meta-analysis studies and replications; empirical coverage of the 95\% prediction interval, which tests calibration at the tails of the predictive distribution; percentage of overestimation, which tests calibration at the center by measuring the proportion of replication point estimates below the point prediction; and the mean $\bar P_{\mathrm{orig}}$, which measures how consistent each
replication study's estimate is with the distribution of effects predicted
by the forecasting model.

\begin{enumerate}[(1)]
    \item \textbf{Raw difference}:  the signed difference between the pooled meta-analytic effect estimate from the replication studies and the pooled estimate from the meta-analysis studies. Because this comparison is made at the pooled level, positive and negative study-level deviations can cancel, so raw difference captures systematic directional bias rather than the per-study error magnitude captured by MAE and RMSE below.

    \item \textbf{$\bar P_{\mathrm{orig}}$}~\cite{mathur2020new}: the probability that
    replication study $j$ would yield a point estimate at least as extreme as
    the one observed, if its true effect were drawn from the same distribution
    of true effects as the meta-analyzed studies (conditional on moderators for
    the meta-regression models). Mathur and VanderWeele~\cite{mathur2020new}
    proposed $\bar P_{\mathrm{orig}}$ to assess whether a single original study is
    consistent with a heterogeneous distribution of effects estimated from $k$
    replications. We reverse the roles: each replication study is the single
    estimate under assessment, and the $k$ meta-analyzed studies supply the
    reference distribution. Both setups share the same many-to-one structure,
    in which one estimate is compared with a heterogeneous distribution
    estimated from many others, so the statistic applies without modification.
    Specifically, for replication study $j$,
    \begin{align}
        P_{\mathrm{orig},j} = 2\left[1 - \Phi\!\left(
        \frac{\bigl|\hat{\theta}_j^{\mathrm{rep}} - \hat{\theta}_j^{\mathrm{pred}}\bigr|}
        {\sqrt{\hat{\tau}^2 + \bigl(\widehat{\SE}_j^{\mathrm{rep}}\bigr)^2 +
        \widehat{\SE}\bigl(\hat{\theta}_j^{\mathrm{pred}}\bigr)^2}}\right)\right],
        \label{eq:porig}
    \end{align}
    where $\hat{\theta}_j^{\mathrm{rep}}$ and $\widehat{\SE}_j^{\mathrm{rep}}$
    are the point estimate and standard error of replication study $j$, and
    $\hat{\theta}_j^{\mathrm{pred}}$ is the model-based prediction for that
    study. For the intercept-only models (MA-U and MA-C),
    $\hat{\theta}_j^{\mathrm{pred}} = \hat{\mu}$, the uncorrected or
    bias-corrected pooled estimate. For the meta-regression models (MR-U and
    MR-C), $\hat{\theta}_j^{\mathrm{pred}} = \mathbf{x}_j^{\top}\hat{\boldsymbol{\beta}}$,
    where $\mathbf{x}_j$ is the moderator vector of replication study $j$.
    $\widehat{\SE}\bigl(\hat{\theta}_j^{\mathrm{pred}}\bigr)$ is the robust
    standard error of this prediction, that is, $\widehat{\SE}(\hat{\mu})$ or
    $\bigl(\mathbf{x}_j^{\top}\widehat{\mathbf{V}}_{R}\,\mathbf{x}_j\bigr)^{1/2}$,
    with $\widehat{\mathbf{V}}_{R}$ the robust covariance matrix of
    $\hat{\boldsymbol{\beta}}$. $\hat{\tau}^2$ is the between-study variance, estimated from the uncorrected model (MA-U or MR-U); MA-C and MR-C use the $\hat{\tau}^2$ from MA-U and MR-U, respectively. A small
    $\bar P_{\mathrm{orig},j}$ indicates that replication study $j$ lies in the
    tails of the distribution predicted by the meta-analysis; a large value
    indicates relative consistency. For each model, we report the mean
    $\bar{P}_{\mathrm{orig}} = J^{-1}\sum_{j=1}^{J} P_{\mathrm{orig},j}$ across
    the $J$ replication studies to characterize average consistency between the replication study estimate and its predicted value. 

    \item \textbf{Mean absolute error (MAE)}: the mean of the absolute
    differences between the replication study estimate and the forecasting model's predicted value $\hat{\theta}_j^{\mathrm{pred}}$:
    \[
    \mathrm{MAE} = \frac{1}{n}\sum_{j=1}^n \bigl|\hat{\theta}_j^{
    \mathrm{rep}} - \hat{\theta}_j^{\mathrm{pred}}\bigr|.
    \]

    \item \textbf{Root mean square error (RMSE)}:
    \[
    \mathrm{RMSE} = \sqrt{\frac{1}{n}\sum_{j=1}^n
    \bigl(\hat{\theta}_j^{\mathrm{rep}} -
    \hat{\theta}_j^{\mathrm{pred}}\bigr)^2}.
    \]

    \item \textbf{Percentage of overestimation}: the percentage of replication
    studies for which the model's predicted estimate exceeded the
    reported estimate ($\hat{\theta}_j^{\mathrm{pred}} >
    \hat{\theta}_j^{\mathrm{rep}}$). The target is 50\%.

    \item \textbf{Coverage of prediction intervals}: the percentage of
    replication studies whose reported point estimate fell within its
    corresponding prediction interval.  Nominal target is 95\%.

    \item \textbf{Mean width of prediction intervals}: the mean
    difference between the upper and lower bounds of the prediction
    intervals across all replication studies.
\end{enumerate}

\section{Simulation Study}
\label{sec:simulation}

Proposition~2 characterizes the asymptotic coverage of the KDE-based prediction intervals, but several practically important questions can be addressed only through simulation. First, the proposition provides no guarantee about finite-sample performance when the meta-analysis includes few studies. Second, the IPPW-corrected models use a plug-in heterogeneity estimate from the corresponding uncorrected model, which under publication bias converges to some $\bar\tau^2 \neq \tau_c^2$; Proposition~2 implies that the resulting coverage error need not vanish asymptotically, but its practical magnitude is unknown without empirical evaluation. Third, Proposition~2 relies on transportability conditional on the measured moderators (Assumption~(A4)), which is unlikely to hold exactly in practice: the meta-analyzed studies may differ on moderators that are not recorded, and moderator effects may differ at the replication sites. Fourth, no single model is expected to dominate across all metrics, and the tradeoffs among point prediction accuracy, interval calibration, and robustness to publication bias depend on the combination of sample size, heterogeneity, moderator strength, and selection severity in ways that are difficult to characterize analytically. The simulation study in this section addresses these questions by comparing the four forecasting models across scenarios that vary the number of studies in the meta-analysis, the between-study variance, the severity of selection, and the strength of the moderators. To assess robustness to departures from (A4), the data generating process includes unmeasured moderators that are omitted from all fitted models, and moderator effects at the replication sites differ from those in the meta-analyzed literature.

\subsection{Simulation Design}
\label{sec:simdesign}

In each iteration of the simulation study, we generated a meta-analysis dataset subject to publication bias and a separate set of replication studies not subject to selection. In both, the population effects were drawn from a right-skewed distribution around their conditional means (Section~\ref{sec:DGP}), reflecting literatures in which a few large effects coexist with many small ones; results under normally distributed population effects are reported in the Supplementary Material. We fitted the four forecasting models in Section~\ref{sec:4models} to the meta-analysis dataset and used each fitted model to forecast the estimate of each replication study, yielding a point forecast and a 95\% prediction interval; for the meta-regression models, forecasts were evaluated at each replication study's measured moderator levels. We then compared these forecasts with the simulated replication estimates using the metrics in Section~\ref{sec:metrics}. Section~\ref{sec:scenarios} describes the four design factors varied across scenarios, and Section~\ref{sec:DGP} describes how the moderators, meta-analysis studies, and replication studies were generated.

\subsubsection{Simulation scenarios}
\label{sec:scenarios}

We generated 81 scenarios from the full factorial combination of four design factors: (a) the number of studies in the meta-analysis, $k \in \{20, 50, 100\}$; (b) the residual variance of the population effects, $\tau^2_c \in \{0.04, 0.09, 0.25\}$, corresponding to standard deviations of 0.2, 0.3, and 0.5 on the Cohen's $d$ scale; (c) the severity of publication bias, $\eta \in \{1, 5, 10\}$, where $\eta = 1$ denotes no selection; and (d) the overall moderator strength, $\beta_{\mathrm{mod}} \in \{0, 0.20, 0.40\}$ (Table~\ref{tab:sim_moderators}), where $\beta_{\mathrm{mod}} = 0$ denotes no moderation. All other parameters, including the skewed distribution of the population effects, were held fixed across scenarios. We ran $B = 1000$ iterations per scenario, which kept the Monte Carlo standard error of coverage below 0.5 percentage points in every scenario~\cite{morris2019using}.

\subsubsection{Data generating process}
\label{sec:DGP}

\paragraph{Moderators.}
Each study, whether in the meta-analysis or the replication, had five true moderators, generated independently from the distributions in Table~\ref{tab:sim_moderators}. All five were predictors of the population effects, but only $z_1$, $z_2$, and $z_3$ were measured and included in the meta-regression models; $z_4$ and $z_5$ were unmeasured and omitted from every fitted model, representing moderators that were not recorded in the literature. The mean slope of each moderator was $\beta_\ell = s_\ell\,\beta_{\mathrm{mod}}$ for $\ell = 1, \ldots, 5$, where $\beta_{\mathrm{mod}}$ is the overall moderator strength and $s_\ell$ are fixed scaling constants that allow the moderators to differ in their influence on the population effects. Varying $\beta_{\mathrm{mod}}$ therefore scaled all five slopes together while holding their relative influence fixed, so that differences in performance across levels of $\beta_{\mathrm{mod}}$ reflected overall moderation strength. The scaling constants were chosen for illustration.

\begin{table}[H]
  \centering
  \caption{Moderators in the simulation and their mean slopes in the meta-analysis studies under each level of the overall moderator strength $\beta_{\mathrm{mod}}$.}
  \label{tab:sim_moderators}
  \begin{tabular}{@{}llccccc@{}}
    \toprule
    & & & & \multicolumn{3}{c}{$\beta_\ell$} \\
    \cmidrule(l){5-7}
    Moderator & Distribution & Measured? & $s_\ell$ & $\beta_{\mathrm{mod}} = 0$ & $\beta_{\mathrm{mod}} = 0.20$ & $\beta_{\mathrm{mod}} = 0.40$ \\
    \midrule
    $z_1$ & $\mathrm{Bernoulli}(0.5)$ & Yes & 1.00 & 0 & 0.20 & 0.40 \\
    $z_2$ & $\mathcal{N}(0,1)$        & Yes & 0.50 & 0 & 0.10 & 0.20 \\
    $z_3$ & $\mathcal{N}(0,1)$        & Yes & 0.75 & 0 & 0.15 & 0.30 \\
    $z_4$ & $\mathrm{Bernoulli}(0.5)$ & No  & 0.50 & 0 & 0.10 & 0.20 \\
    $z_5$ & $\mathcal{N}(0,1)$        & No  & 0.25 & 0 & 0.05 & 0.10 \\
    \bottomrule
  \end{tabular}
\end{table}

\paragraph{Meta-analysis studies.}
To reflect variation in moderator effects across studies, each meta-analysis study $i$ drew its own slopes,
\[
  \beta_{\ell i} \sim \mathcal{N}(\beta_\ell, \omega^2), \quad \ell = 1, \ldots, 5, \quad \omega^2 = 0.01.
\]
We then generated the true population effect and the reported point estimate as
\[
  \theta_i = \beta_0 + \sum_{\ell=1}^{5} z_{\ell i} \beta_{\ell i} + u_i,
  \qquad
  u_i = \tau_c\,\frac{X_i - 4}{\sqrt{8}}, \quad X_i \sim \chi^2_4,
\]
\[
  \hat\theta_i \mid \theta_i \sim \mathcal{N}(\theta_i, \sigma^2),
\]
with intercept $\beta_0 = 0.20$ and fixed within-study variance $\sigma^2 = 0.04$. The residual $u_i$ has mean zero, variance $\tau^2_c$, and skewness $\sqrt{2} \approx 1.41$. Because the slopes varied across studies and two moderators were omitted from the model fitting procedure, the residual variance of the population effects conditional on the measured moderators exceeded $\tau^2_c$ and depended on the moderator levels, with the excess due to the omitted moderators increasing with $\beta_{\mathrm{mod}}$. The simulation therefore evaluated the forecasting models under a departure from the constant residual variance assumed in (A1), as is likely in practice.

\paragraph{Publication bias.}
A study was defined as affirmative if $\hat\theta_i > 0$ and its two-sided $p < 0.05$. Affirmative studies were published with probability 1 and nonaffirmative studies with probability $1/\eta$. We applied this selection process to a large pool of simulated studies and retained the first $k$ published studies as the meta-analysis dataset.

\paragraph{Replication studies.}
Replication studies $j = 1, \ldots, n_{\mathrm{rep}}$ were generated in the same way as the meta-analysis studies, including the same skewed residual distribution, with two differences. First, because replication populations may differ systematically from those in the published literature, each mean slope was shifted by $\delta s_\ell$, with $\delta = 0.05$, so that replication study $j$ drew its slopes from
\[
  \beta^{\mathrm{rep}}_{\ell j} \sim \mathcal{N}(\beta_\ell + \delta s_\ell, \omega^2), \quad \ell = 1, \ldots, 5.
\]
This shift applied at every level of $\beta_{\mathrm{mod}}$, including $\beta_{\mathrm{mod}} = 0$. Second, because replication results are reported regardless of outcome, replication studies were not subject to publication bias. We generated $n_{\mathrm{rep}} = 30$ replication studies per iteration.

\subsection{Estimation Procedure}
\label{sec:sim-estimation}
In each iteration, we fitted the four forecasting models in Section~\ref{sec:4models} to the simulated meta-analysis dataset only, using the measured moderators $z_1$, $z_2$, and $z_3$ in the meta-regression models, and used each fitted model to forecast the estimate of every replication study. Because the replication studies were simulated with a known within-study variance, $\sigma^2 = 0.04$, we treated each replication study's sampling variance as known, as in the application (Section~\ref{sec:estimation-procedure}). The selection ratio $\eta$ used for publication bias correction was set to its true value in each scenario.

\paragraph{MA-U and MR-U.} We fitted an intercept-only meta-analysis and a meta-regression on $z_1$, $z_2$, and $z_3$ using \texttt{robumeta} with correlated-effects weights and small-sample corrections\cite{fisher2015robumeta, tipton2015small}. The point forecast for every replication study was the pooled mean $\hat{\mu}$ under MA-U, and $\mathbf{z}_j\hat{\boldsymbol{\beta}}$ under MR-U, evaluated at replication study $j$'s moderator levels.

\paragraph{MA-C and MR-C.} We refitted the same two models with the IPPW weights of Equation~\ref{eq:omega_mi}, supplied to \texttt{robumeta} via the \texttt{userweights} argument. Following Section~\ref{sec:prediction}, the heterogeneity estimate in the weights and in the prediction intervals was taken from the corresponding uncorrected model (MA-U or MR-U). The point forecasts are the corrected pooled mean under MA-C and $\mathbf{z}_j\hat{\boldsymbol{\beta}}_\eta$ under MR-C.

\paragraph{Prediction intervals.} We constructed prediction intervals using two approaches: KDE-based and standard parametric intervals. For the KDE-based intervals, we computed calibrated estimates of the meta-analysis studies from each fitted model (Equation~\eqref{eq:calibrated}) and selected the kernel bandwidth $h$ from the unweighted calibrated estimates by the Sheather--Jones method~\cite{sheather1991reliable}. Convolving the kernel density with the replication study's sampling distribution yields the Gaussian mixture $\hat f(t) = \sum_{i=1}^{k}\omega_i\,\phi(t;\, \tilde\theta_i,\, h^2 + \sigma^2) / \sum_{i=1}^{k}\omega_i$ (Equation~\eqref{eq:kde-pi}), with $\omega_i = 1$ for the uncorrected models and $\omega_i = \eta^{\mathbb{I}\{A_i = 0\}}$ for the corrected models; we evaluated its cumulative distribution on a grid from $-3$ to $3$ in steps of $0.01$ and took the 2.5th and 97.5th percentiles. For the meta-regression models, the interval was computed at a reference moderator level and shifted to each replication study's moderator level, which is equivalent to calibrating at $\mathbf{z}_j$ because the bandwidth and sampling variance do not depend on the moderators.

For comparison, the parametric intervals are $\hat\theta_j \pm t_{\mathrm{df}}\sqrt{\hat\tau^2 + \mathrm{SE}^2(\hat\theta_j) + \sigma^2}$, with $\mathrm{SE}(\hat\theta_j)$ obtained from the robust variance estimate of the model coefficients and $t_{\mathrm{df}}$ based on Satterthwaite degrees of freedom (for the meta-regression models, the smallest across coefficients) (Appendix \ref{app:pi-kde-param}).

Both interval methods were applied to identical simulated datasets, so that differences between them reflect the interval method alone. For each model, the seven evaluation metrics (Section~\ref{sec:metrics}) were computed across the $n_{\mathrm{rep}} = 30$ replication studies within each iteration and then averaged across the $B = 1000$ iterations of each scenario. To summarize performance across scenarios, we calculated win rates: in each of the 81 scenarios, we identified the best-performing model on each of the seven metrics, defined as the smallest absolute raw difference; the lowest MAE, RMSE, and mean prediction interval width; the overestimation rate closest to 50\%; the coverage closest to 95\%; and the highest mean $\bar P_{\mathrm{orig}}$. We then counted the number of scenarios in which each model performed best, separately for the KDE-based and parametric intervals.

\subsection{Simulation Results}

Table~\ref{tab:winrate} summarizes, for each metric, the number of the 81 simulation scenarios in which each model performed best. MR-C led on MAE, RMSE, and mean $\bar P_{\mathrm{orig}}$, whereas MA-C led on raw difference and overestimation rate. For coverage, MA-C led with the KDE-based intervals and MR-U with the parametric intervals.

\paragraph{Publication bias ($\eta$).}
Without publication bias ($\eta = 1$), the corrected and uncorrected models coincide because IPPW weights equal 1, and the two meta-regression models were more accurate than the two intercept-only models. Overestimation rates were close to the 50\% benchmark, and mean $\bar P_{\mathrm{orig}}$ was close to 50\%. As $\eta$ increased, the uncorrected models increasingly overpredicted the replication results: overestimation rates increased well above 50\% to over 70\%, raw difference became increasingly negative, mean $\bar P_{\mathrm{orig}}$ declined substantially, and MAE and RMSE increased. Correcting for publication bias largely removed this directional bias. The overestimation rates of the corrected models stayed below 55\%, with MA-C remaining near the benchmark and MR-C rising moderately above it under strong selection; the raw difference of MA-C remained close to zero, whereas that of MR-C drifted moderately below zero. The corrected models' MAE and RMSE increased only slightly with selection, and their mean $\bar P_{\mathrm{orig}}$ declined much less than that of the uncorrected models. As a result, MR-C remained the most accurate model at every level of publication bias.

Correcting for publication bias also improved the coverage of the KDE-based intervals. KDE coverage was slightly below the nominal 95\% when $\eta = 1$ and declined for all four models as $\eta$ increased, but much less for the corrected models: under strong publication bias, the uncorrected models fell more than 10 percentage points below the nominal level, whereas MA-C and MR-C remained closer to it. MA-C's KDE intervals also narrowed less than those of MA-U as $\eta$ increased. In contrast, MR-C and MR-U had nearly identical widths, so MR-C's coverage advantage came from shifting its intervals toward the replication estimates, correcting the upward bias induced by publication bias.

The parametric intervals behaved differently. Coverage of the uncorrected models declined with $\eta$, while increased with $\eta$. MA-C maintained coverage close to the nominal 95\% level and MR-C increased above it. The corrected models achieved this by widening their parametric intervals as $\eta$ increased, whereas the parametric intervals of the uncorrected models narrowed.

\paragraph{Between-study heterogeneity ($\tau^2_c$).}
Point forecasts from all four models became less accurate as heterogeneity increased: MAE and RMSE increased with $\tau^2_c$, and both KDE and parametric intervals widened accordingly. MR-C remained the most accurate model at every level of $\tau^2_c$, but the accuracy advantage of the meta-regression models over the intercept-only models declined as $\tau^2_c$ increased. This pattern was consistent with the measured moderators explaining a smaller share of the variation in population effects as the residual heterogeneity increased. Heterogeneity also amplified the bias of the uncorrected models: their raw difference became more negative and their overestimation rates increased slightly as $\tau^2_c$ increased, whereas the raw difference of the corrected models changed little and their overestimation rates remained much closer to 50\%. 

In contrast, coverage was largely stable across $\tau^2_c$ levels. KDE interval coverage remained below the nominal 95\% for all four models, but the corrected models stayed consistently above the uncorrected models at every level of $\tau^2_c$. Mean $\bar P_{\mathrm{orig}}$ was also stable, with the corrected models consistently above the uncorrected models.

\paragraph{Moderator strength ($\beta_{\mathrm{mod}}$).}
Moderator strength showed a crossover in the accuracy of the point forecasts between the intercept-only and meta-regression models. Without moderation ($\beta_{\mathrm{mod}} = 0$), MA-C was the most accurate model, because the meta-regression models lost some precision by estimating slopes that were in fact zero.
As $\beta_{\mathrm{mod}}$ increased, the MAE and RMSE of the intercept-only models increased substantially, because they ignored variation in the population effects due to moderators, whereas those of the meta-regression models remained nearly flat; at nonzero levels of moderation, MR-C was the most accurate model and had the highest mean $\bar P_{\mathrm{orig}}$. The overestimation rates of the corrected models remained close to 50\% at all levels of moderator strength, whereas those of the uncorrected models declined somewhat as moderation increased but remained well above 50\%. However, the raw difference of MR-C became increasingly negative with $\beta_{\mathrm{mod}}$, while that of MA-C remained close to zero. 

Moderation also affected interval performance. As $\beta_{\mathrm{mod}}$ increased, the KDE intervals of the intercept-only models widened, yet their KDE coverage declined, whereas the KDE intervals and coverage of the meta-regression models changed little; MA-C had the KDE coverage closest to the nominal level at every level of moderation, followed by MR-C when moderation was present, and all four models remained below the nominal level.
Under the parametric method, MR-C stayed above the nominal level at all levels of moderation, whereas MA-C moved slightly below it as moderation increased.

\paragraph{Number of studies in the meta-analysis ($k$).}
The number of meta-analyzed studies had a smaller impact than $\eta$, $\tau^2_c$, or $\beta_{\mathrm{mod}}$ on most metrics, and its impact depended on the model. The accuracy of the point forecasts from the intercept-only models changed little with $k$, whereas the meta-regression models, which estimate more parameters, became more accurate as $k$ increased, MR-C most of all. The bias of MR-C also decreased with $k$: its raw difference moved toward zero, and its overestimation rate decreased toward 50\%. KDE coverage increased with $k$ for all four models, most for the meta-regression models, while KDE interval widths changed little; even at the largest $k$, however, KDE coverage remained below the nominal 95\%.
In contrast, the parametric intervals of the corrected models and of MR-U narrowed substantially as $k$ increased, and their coverage decreased accordingly: MA-C moved from the nominal level to slightly below it, and MR-C moved closer to, but remained above, the nominal level. At small $k$, the parametric intervals of MR-C were therefore conservative but wide.

\paragraph{Summary.}
Table~\ref{tab:winrate} summarizes, for each metric, the number of the 81 scenarios in which each model performed best. MR-C was the most accurate model, with the lowest MAE and RMSE in 32 scenarios each, and most often had the highest mean $\bar P_{\mathrm{orig}}$ (37 scenarios).
MA-C most often had a raw difference closest to zero (39 scenarios), an overestimation rate closest to 50\% (43 scenarios), and KDE coverage closest to the nominal level (36 scenarios), whereas MR-U most often had parametric coverage closest to the nominal level (35 scenarios). MR-U also produced the narrowest KDE intervals in 55 scenarios but rarely achieved the best KDE coverage (only 7 scenarios), indicating that its narrower intervals reflected undercoverage rather than greater precision.
When publication bias is present ($\eta > 1$), the corrected models performed best in all or nearly all of the 54 scenarios on every point-forecast metric and on KDE coverage, and in 34 of 54 scenarios on parametric coverage. Overall, the results supported correcting for publication bias in the meta-analysis dataset when forecasting replication results.

\begin{table}[H]
\centering
\caption{Win rates across 81 simulation scenarios. Each cell reports the number of scenarios in which a model achieved the best value for the given metric. Best is defined as: lowest MAE, RMSE, and PI width; closest to 50\% overestimation and 95\% coverage; highest mean $\bar P_\text{orig}$.}
\label{tab:winrate}
\begin{tabular}{lccccccccc}
\toprule
 & & & & & & \multicolumn{2}{c}{Coverage} & \multicolumn{2}{c}{PI Width} \\
\cmidrule(lr){7-8} \cmidrule(lr){9-10}
Model & Raw Diff & MAE & RMSE & Mean $\bar P_\text{orig}$ & Overestimation & KDE & Param. & KDE & Param. \\
\midrule
MA-U & 16 & 9 & 9 & 9 & 16 & 20 & 12 &  0 & 52 \\
MR-U & 11 & 18 & 19 & 18 & 11 & 7 & 35 & 55 & 29\ \\
MA-C & 39 & 22 & 21 & 17 & 43 & 36 & 26 & 0 &  0 \\
MR-C & 15 & 32 & 32 & 37 & 11 & 18 & 8 &  26 &  0 \\
\bottomrule
\multicolumn{9}{l}{\footnotesize MA-U: MA, uncorrected; MR-U: MR, uncorrected;} \\
\multicolumn{9}{l}{\footnotesize MA-C: MA, IPPW-corrected; MR-C: MR, IPPW-corrected.} \\
\multicolumn{9}{l}{\footnotesize PI: prediction interval.} \\
\multicolumn{9}{l}{\footnotesize Param.: standard parametric prediction interval assuming a $t$-distribution.} \\
\end{tabular}
\end{table}

\begin{figure}[H]
\centering
\begin{subfigure}[t]{0.48\textwidth}
  \centering
  \includegraphics[width=\textwidth]{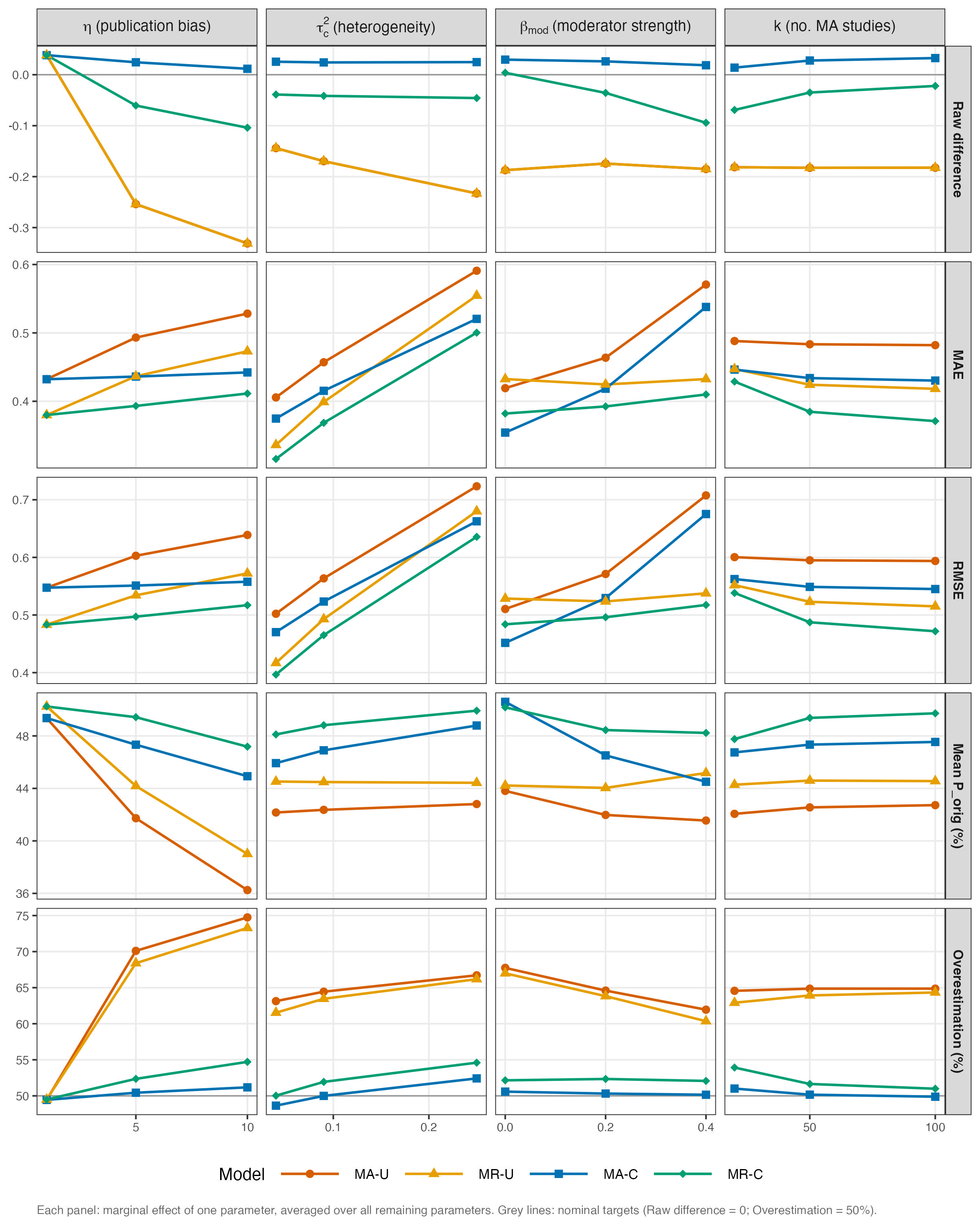}
  \caption{Shared metrics across KDE and parametric methods.}
  \label{fig:sim_metrics1}
\end{subfigure}
\hfill
\begin{subfigure}[t]{0.48\textwidth}
  \centering
  \includegraphics[width=\textwidth]{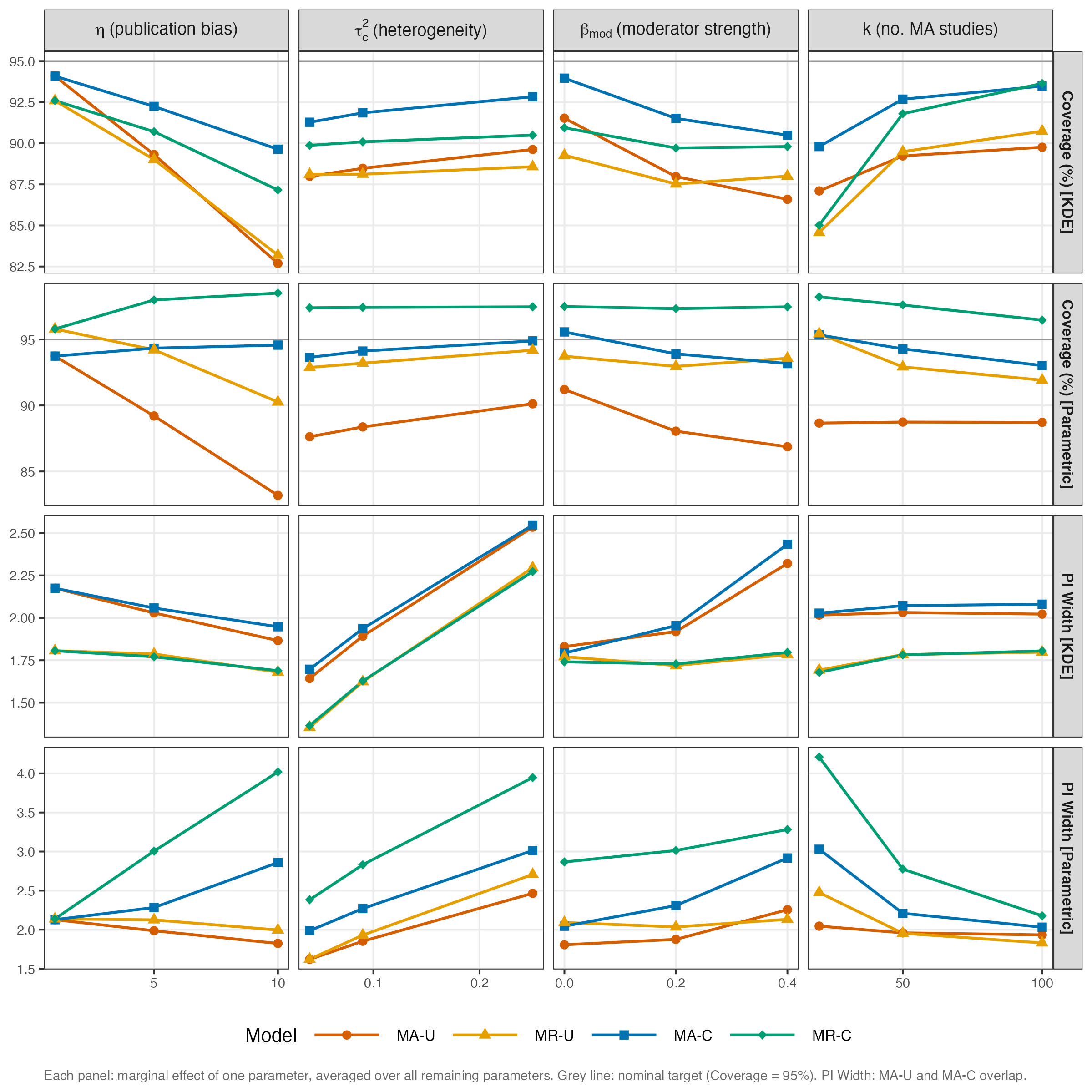}
  \caption{Coverage and prediction interval width by PI method (KDE vs.\ parametric).}
  \label{fig:sim_metrics2}
\end{subfigure}
\caption{Marginal performance of four forecasting models across simulation parameters. Each panel shows one metric (row) as a function of one design parameter (column), averaging over the remaining parameters.}
\label{fig:sim_metrics}
\end{figure}

\paragraph{Normally distributed population effects.}
As a benchmark, we repeated the simulation with normally distributed population effects, the setting in which Proposition~2 guarantees nominal asymptotic coverage of the KDE-based intervals; all other settings were unchanged (Appendix~\ref{app:sim-normal_mu}). The main conclusions were the same as under skewed population effects. MR-C was again the most accurate model on MAE and RMSE and most often had the highest mean $\bar P_{\mathrm{orig}}$, and MA-C again most often had a raw difference closest to zero and KDE coverage closest to the nominal level. Two leads changed with the distribution of the population effects. Under normality, MR-C most often had an overestimation rate closest to 50\% and MA-C most often had parametric coverage closest to the nominal level, whereas under skewness these leads shifted to MA-C and MR-U, respectively. MR-C's lead on mean $\bar P_{\mathrm{orig}}$ was also larger under normality, consistent with $\bar P_{\mathrm{orig}}$ being derived under a normal model for the population effects.


\section{Application to ManyBabies}
\label{sec:mbapplication}

Now we re-analyze the ManyBabies 1 and ManyBabies 4 meta-analysis and replication datasets using the four models and compared their performances. 

\subsection{Data Preprocessing}
\label{sec:mbmoderators}
\subsubsection{ManyBabies 1} 
We excluded studies with missing Cohen's $d$ or its variance from the corrected Dunst et al.\cite{dunst2012preference} dataset in \cite{zettersten2024evidence}. For the replication data, we used the same dataset as ManyBabies-Consortium \cite{manybabies2020quantifying} and Zettersten et al.\cite{zettersten2024evidence}. This resulted in 104 point estimates in ManyBabies 1 and 51 point estimates in the corresponding meta-analyses. The effect size was Cohen's $d$ of looking time. The measured covariates were mean age (days), test language and experimental method. 

Across the 51 point estimates from the meta-analyses, the mean infant age was 144.3 days. Most studies (86.3\%) used the infants' native language, with 7.8\% using a non-native language and 5.9\% an artificial language. Central fixation was the most common experimental method (66.7\%), followed by head-turn preference procedure (19.6\%). Among the 104 point estimates from the replication dataset, infants were older on average (285.7 days) and more likely to be tested in a non-native language (55.8\% vs. 44.2\% native). Central fixation remained the predominant method (66.3\%), with a larger share using head-turn preference procedure (33.7\%) compared to the meta-analytic samples.

The naive meta-analytic estimate of the 51 point estimates in the meta-analyses was 0.69. The funnel plot is shown in Figure~\ref{fig:funnel}(a). 
The selection model provided no evidence of selection favoring statistically significant results (selection ratio 2.33, 95\% CI 0.18 to 4.49; LRT P-value = 0.07). The selection-adjusted pooled estimate (0.92, 95\% CI 0.60 to 1.23) was larger than the unadjusted estimate (0.69, 95\% CI 0.50 to 0.88), suggesting that the main finding is robust to potential publication bias.

\subsubsection{ManyBabies 4} 
We used the dataset from \cite{tan2024prosocial} (available \href{https://docs.google.com/spreadsheets/d/1j2Af5jyMjqRC6FBQnqGeTPVJ5nacDVf2cu9a9D3Se4o/edit?pli=1&gid=1269834261#gid=1269834261}{here}), retaining only studies with the dependent measure of ``target selection'' and excluding one study \cite{van2018early} that used a between-condition comparison (prosocial vs.\ antisocial) rather than the standard design. For the replication dataset, we excluded studies with missing values in the number of exclusions due to infants ``misperforming'' during the choice task. This results in 33 point estimates in ManyBabies 4 and 138 point estimates in the corresponding meta-analyses.

The raw effect size was the odds ratio of infants who selected the more prosocial agent. The measured covariates are mean age (days), scenarios, stimuli, stimuli type, choice object, whether the study was from Hamlin lab (Hamlin lab tended to have larger point estimate than other labs \cite{tan2024prosocial}), intent valence, and outcome valence. Odds ratios were converted to Cohen's $d$ scale for analysis \cite{borenstein2009effect}. 

Across the 138 point estimates from the meta-analyses, the mean infant age was 401.3 days, with an average of 3.5 excluded infants per study. The majority of studies used a reach method (85.5\%), help/hinder scenarios (62.3\%), live show stimuli (61.6\%) with real events (82.6\%), and puppets as the choice object (50.0\%). Most studies measured opposing intent valence (84.1\%) and opposing outcome valence (75.4\%), and 34.1\% originated from the Hamlin lab. Among the 33 point estimates from ManyBabies 4, infants were younger on average (253.1 vs. 401.3 days) but had a higher number of excluded infants (6.2 vs. 3.5). ManyBabies 4 exclusively used reach methods, help/hinder scenarios, movie stimuli with real events, shapes as choice objects, opposing intent valence, and opposing outcome valence, reflecting a more homogeneous design across most covariates. Only one study was from the Hamlin lab.

Distributions of covariates are presented in Tables \ref{tab:cov_mb1} and \ref{tab:cov_mb4}.

\begin{table}[H]
\caption{Covariate distributions in ManyBabies 1.}
\label{tab:cov_mb1}
\centering
\small
\begin{tabular}{lrr}
\toprule
 & Meta-analysis dataset & Replication dataset \\
\midrule
Mean age, mean (sd) & 144.3 (79.4) & 285.7 (91.6) \\
Test language (\%) & & \\
\quad Native & 86.3 & 44.2 \\
\quad Non-native & 7.8 & 55.8 \\
\quad Artificial & 5.9 & 0 \\
Experimental method (\%) & & \\
\quad Central fixation & 66.7 & 66.3 \\
\quad Head-turn preference procedure & 19.6 & 33.7 \\
\quad Other & 13.7 & 0 \\
\bottomrule
\end{tabular}
\end{table}

\begin{table}[H]
\caption{Covariate distributions in ManyBabies 4.}
\label{tab:cov_mb4}
\centering
\small
\begin{tabular}{lrr}
\toprule
 & Meta-analysis dataset & Replication dataset\\
\midrule
Mean age, mean (sd) & 401.3 (199.0) & 253.1 (24.8) \\
Number of excluded infants, mean (sd) & 3.5 (4.0) & 6.2 (4.8) \\
Method (\%) & & \\
\quad Help & 14.5 & 0 \\
\quad Reach & 85.5 & 100 \\
Scenarios (\%) & & \\
\quad Fair/unfair & 19.6 & 0 \\
\quad Give/take & 10.1 & 0 \\
\quad Help/hinder & 62.3 & 100 \\
\quad Other & 8.0 & 0 \\
Stimuli (\%) & & \\
\quad Live show & 61.6 & 0 \\
\quad Movies & 38.4 & 100 \\
Stimuli type (\%) & & \\
\quad Cartoon & 17.4 & 0 \\
\quad Real events & 82.6 & 100 \\
Choice object (\%) & & \\
\quad Experimenters & 21.0 & 0 \\
\quad People & 1.4 & 0 \\
\quad Puppets & 50.0 & 0 \\
\quad Shapes & 27.5 & 100 \\
Hamlin lab (\%) & 34.1 & 3 \\
Intent valence (\%) & & \\
\quad Opposite (positive/negative) & 84.1 & 100 \\
\quad Positive & 9.4 & 0 \\
\quad Negative & 6.5 & 0 \\
Outcome valence (\%) & & \\
\quad Same (positive/positive, neutral/neutral, negative/negative) & 13.0 & 0 \\
\quad Opposite (positive/negative) & 75.4 & 100 \\
\quad Reversed (negative/positive) & 0.7 & 0 \\
\quad Neutral & 10.9 & 0 \\
\bottomrule
\end{tabular}
\end{table}

\begin{figure}[H]
    \centering
    \begin{subfigure}[b]{0.49\textwidth}
        \centering
        \includegraphics[width=\textwidth]{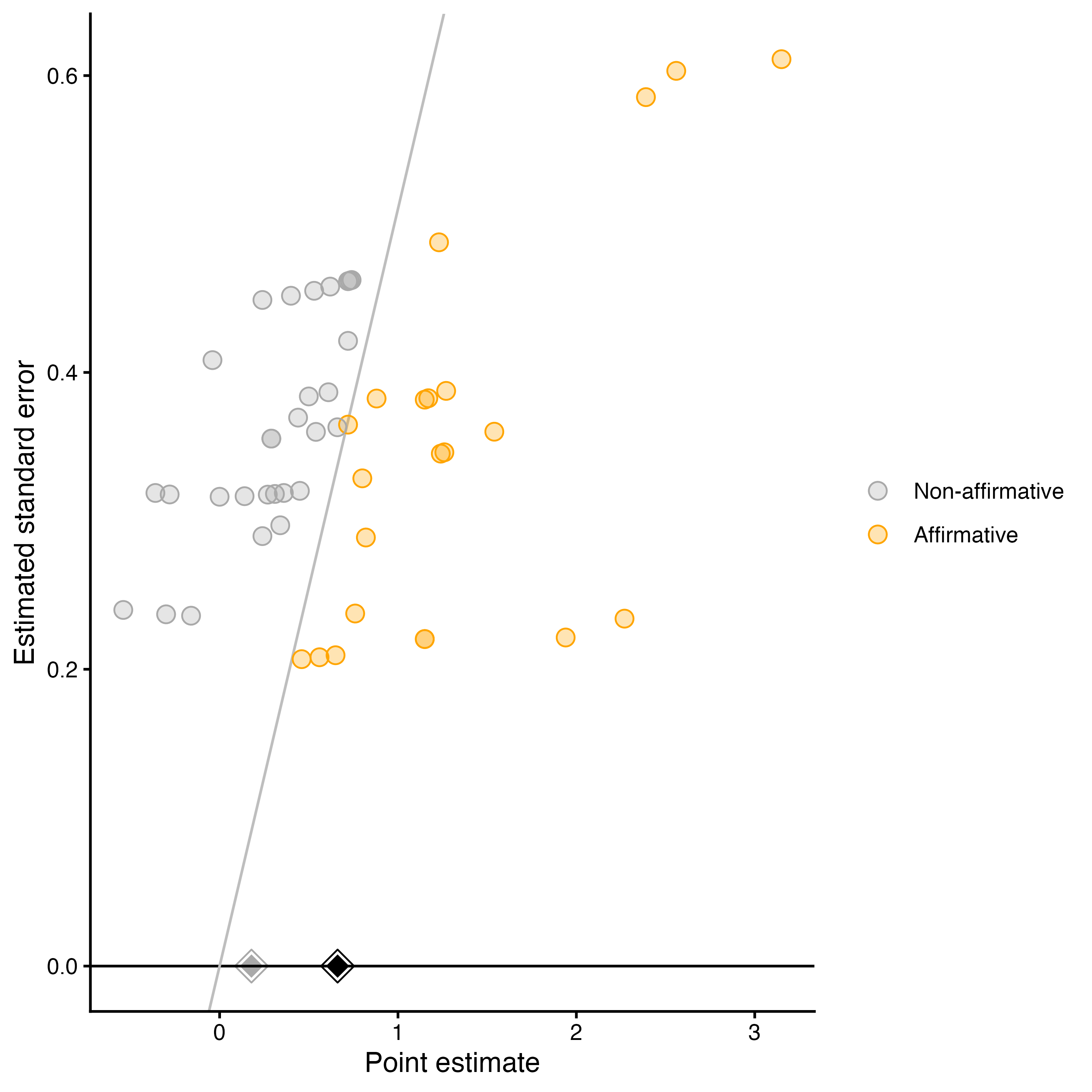}
        \caption{Meta-analysis studies in ManyBabies 1}
        \label{fig:pred-a}
    \end{subfigure}
    \hfill
    \begin{subfigure}[b]{0.49\textwidth}
        \centering
        \includegraphics[width=\textwidth]{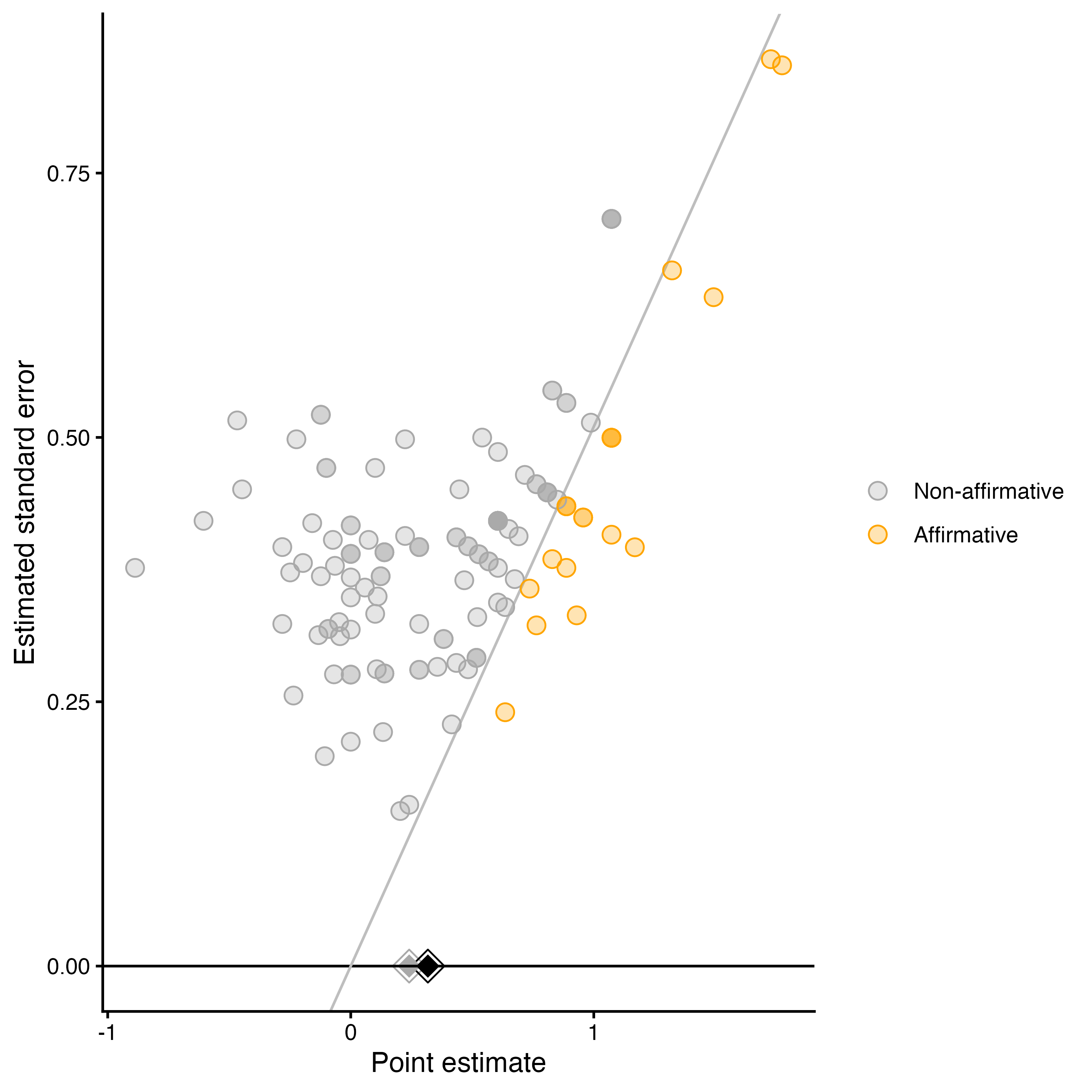}
        \caption{Meta-analysis studies in ManyBabies 4}
        \label{fig:pred-b}
    \end{subfigure}
    \caption{Significance funnel plots of the meta-analysis studies for ManyBabies 1 and ManyBabies 4. Gray line: $p = 0.05$. Publication bias was assumed to favor significant positive estimates. 
    Gray diamond: the pooled estimate using only the nonaffirmative studies; black diamond: the naive pooled estimate using all studies, not corrected for publication bias. The gap between the two pooled-estimate diamonds was noticeably wider in panel (a) versus a much smaller gap in panel (b), indicating that the meta-analysis studies for ManyBabies 1 had more severe publication bias than those for ManyBabies 4.}
    \label{fig:funnel}
\end{figure}

The naive meta-analytic estimate of the 138 point estimates in the meta-analysis was 0.33. In Figure~\ref{fig:funnel}(b), the funnel plot of the meta-analysis was less asymmetric than that in Figure~\ref{fig:funnel}(a). 
The selection model provided no evidence of selection favoring statistically significant results (selection ratio 1.29, 95\% CI 0.02 to 2.56; LRT P-value = 0.062). The selection-adjusted pooled estimate (0.36, 95\% CI 0.21 to 0.51) was similar to the unadjusted estimate (0.33, 95\% CI 0.26 to 0.39), suggesting that the main finding is robust to potential publication bias.

\subsection{Estimation Procedure}
\label{sec:estimation-procedure}
Because the ManyBabies~1 and ManyBabies~4 replication datasets were already available when we conducted this project, we forecasted the replication estimates retrospectively. The first author was blinded to the replication point estimates but had access to all other variables, including each replication study's moderator levels and sampling variance. For a not-yet-conducted replication, the sampling variance would instead be projected from the planned sample size; here, we treated it as known.

We followed the estimation procedure of the simulation study (Section~\ref{sec:sim-estimation}), fitting the four forecasting models to the meta-analysis dataset, constructing KDE-based and parametric prediction intervals, and evaluating the forecasts using the seven metrics in Section~\ref{sec:metrics}, with two differences. First, because the true severity of publication bias was unknown, the corrected models used a selection ratio of $\eta = 4.70$, the estimated selection ratio for meta-analyses in Metalab, a database of developmental psychology meta-analyses, where affirmative results were an estimated 4.70 times more likely to be published than non-affirmative results\cite{mathur2021estimating}. Second, prediction intervals used each replication study's own sampling variance, $\hat\sigma^2_j$, in place of the common variance used in the simulation. The meta-regression models conditioned on the moderators described in Section~\ref{sec:mbmoderators}.

\subsection{Results}
\label{sec:mbresults}
Prediction intervals for ManyBabies 1 and ManyBabies 4 are shown in Figures ~\ref{fig:PI-mb1} and ~\ref{fig:PI-mb4}.
The reported point estimates (black dots in both figures) in ManyBabies 1's replication dataset were more heterogeneous than those of ManyBabies 4. Tables~\ref{tab:metrics_mb1} and ~\ref{tab:metrics_mb4} show the seven evaluation metrics for the four forecasting models applied to the ManyBabies 1 and ManyBabies 4 replication datasets.

\begin{table}[H]
\centering
\footnotesize
\begin{threeparttable}
\caption{Performance of four forecasting models on the ManyBabies 1 replication dataset. Point-forecast metrics are identical under the KDE and parametric methods because both use the same point forecasts; only coverage and interval width differ.}
\label{tab:metrics_mb1}
\setlength{\tabcolsep}{4pt}
\begin{tabular}{@{}l rrrrr rr rr@{}}
\toprule
 & & & & & &
  \multicolumn{2}{c}{Coverage} &
  \multicolumn{2}{c}{PI Width} \\
\cmidrule(lr){7-8} \cmidrule(lr){9-10}
Model &
  \shortstack{Raw\\diff.} &
  \shortstack{Mean\\$\bar P_{\mathrm{orig}}$} &
  MAE & RMSE &
  \shortstack{Over-\\estimation} &
  KDE & Param. &
  KDE & Param. \\
\midrule
MA--U & -0.36 & 0.53 & 0.42 & 0.51 & 89.4\% & 100\%  & 100\% & 2.71 & 2.56 \\
MR--U & -0.41 & 0.44 & 0.61 & 0.71 & 87.5\% & 88.5\% & 100\% & 2.65 & 5.27 \\
MA--C & -0.06 & 0.65 & 0.28 & 0.37 & 52.9\% & 99.0\% & 100\% & 2.32 & 2.77 \\
MR--C & -0.17 & 0.50 & 0.49 & 0.58 & 76.9\% & 95.2\% & 100\% & 2.15 & 4.47 \\
\bottomrule
\end{tabular}
\begin{tablenotes}[flushleft]
\footnotesize
\item MA-U: MA, uncorrected; MR-U: MR, uncorrected; MA-C: MA, IPPW-corrected; MR-C: MR, IPPW-corrected; PI: prediction interval; Param.: standard parametric prediction interval assuming a $t$-distribution for the predictive density.
\end{tablenotes}
\end{threeparttable}
\end{table}

\paragraph{MB1.}
For the intercept-only models, moving from MA-U to MA-C reduced the overestimation rate from 89.4\% to 52.9\%, close to the 50\% benchmark, shrank the raw difference from $-0.36$ to $-0.06$ on the Cohen's $d$ scale, reduced the MAE from 0.42 to 0.28 and the RMSE from 0.51 to 0.37, and increased the mean $\bar P_{\mathrm{orig}}$ from 0.53 to 0.65. Correction improved the meta-regression models in the same direction: moving from MR-U to MR-C reduced the overestimation rate by 10.6 percentage points (from 87.5\% to 76.9\%), shrank the raw difference from $-0.41$ to $-0.17$, reduced the MAE from 0.61 to 0.49 and the RMSE from 0.71 to 0.58, increased the mean $\bar P_{\mathrm{orig}}$ from 0.44 to 0.50, and increased KDE coverage from 88.5\% to 95.2\%, the nominal level, while narrowing the KDE interval width.

In contrast, moving from meta-analysis to meta-regression did not improve point forecasts in ManyBabies~1. Among the uncorrected models, MR-U slightly reduced the overestimation rate relative to MA-U (87.5\% versus 89.4\%) but had a higher MAE (0.61 versus 0.42); among the corrected models, MR-C had a higher MAE (0.49 versus 0.28), a higher overestimation rate (76.9\% versus 52.9\%), and a lower mean $\bar P_{\mathrm{orig}}$ (0.50 versus 0.65) than MA-C. 
The main benefit of meta-regression was in interval calibration: the KDE intervals of the intercept-only models covered all or nearly all replication estimates (100\% for MA-U and 99.0\% for MA-C), exceeding the nominal level, whereas MR-C attained the nominal coverage (95.2\%) with the narrowest KDE interval of the four models (2.15).
The parametric intervals of the meta-regression models also attained 100\% coverage, but only because they were considerably wider than their KDE intervals.

\begin{table}[H]
\centering
\footnotesize
\begin{threeparttable}
\caption{Performance of four forecasting models on the ManyBabies 4 replication dataset. Point-forecast metrics are identical under the KDE and parametric methods because both use the same point forecasts; only coverage and interval width differ.}
\label{tab:metrics_mb4}
\setlength{\tabcolsep}{4pt}
\begin{tabular}{@{}l rrrrr rr rr@{}}
\toprule
 & & & & & &
  \multicolumn{2}{c}{Coverage} &
  \multicolumn{2}{c}{PI Width} \\
\cmidrule(lr){7-8} \cmidrule(lr){9-10}
Model &
  \shortstack{Raw\\diff.} &
  \shortstack{Mean\\$\bar P_{\mathrm{orig}}$} &
  MAE & RMSE &
  \shortstack{Over-\\estimation} &
  KDE & Param. &
  KDE & Param. \\
\midrule
MA--U & -0.30 & 0.36 & 0.82 & 1.07 & 60.6\% & 84.8\% & 84.8\% & 2.68 & 2.79 \\
MR--U & -0.32 & 0.42 & 0.79 & 1.05 & 57.6\% & 84.8\% & 100\% & 2.62 & 6.30 \\
MA--C & -0.24 & 0.36 & 0.81 & 1.05 & 54.5\% & 84.8\% & 87.9\% & 2.67 & 2.83 \\
MR--C & -0.26 & 0.41 & 0.79 & 1.03 & 51.5\% & 87.9\% & 100\% & 2.62 & 9.63 \\
\bottomrule
\end{tabular}
\begin{tablenotes}[flushleft]
\footnotesize
\item MA-U: MA, uncorrected; MR-U: MR, uncorrected; MA-C: MA, IPPW-corrected; MR-C: MR, IPPW-corrected; PI: prediction interval; Param.: standard parametric prediction interval assuming a $t$-distribution for the predictive density.
\end{tablenotes}
\end{threeparttable}
\end{table}

\paragraph{MB4.}
In ManyBabies~4, publication bias correction and meta-regression each improved the forecasts, but both improvements were modest, and the four models differed much less than in ManyBabies~1. Correcting for publication bias reduced the overestimation rate by 6.1 percentage points for both model types, from 60.6\% (MA-U) to 54.5\% (MA-C) and from 57.6\% (MR-U) to 51.5\% (MR-C), and shrank the raw difference ($-0.30$ to $-0.24$ for the intercept-only models; $-0.32$ to $-0.26$ for the meta-regression models). Its effect on point accuracy was small (MAE lower by at most 0.01; RMSE lower by 0.02).

Moving from meta-analysis to meta-regression produced comparably small gains. The meta-regression models had slightly lower MAE (0.79 versus 0.81 to 0.82) and RMSE (1.03 to 1.05 versus 1.05 to 1.07), overestimation rates about 3 percentage points lower, and higher mean $\bar P_{\mathrm{orig}}$ (0.41 to 0.42 versus 0.36) than the corresponding intercept-only models. Combining both adjustments, MR-C performed best on most point-forecast metrics, with the lowest MAE and RMSE and an overestimation rate closest to the 50\% benchmark (51.5\%). However, mean $\bar P_{\mathrm{orig}}$ remained well below 0.5 for all four models (0.36 to 0.42), indicating that the replication estimates were more dispersed around the forecasts than the meta-analyses predicted. Because the shared protocol held most moderators constant across replication sites, the meta-regression models could shift their forecasts toward the ManyBabies~4 design but could not capture variation among sites, which likely aincreased from unmeasured factors.

Neither adjustment substantially improved interval calibration. KDE coverage was 84.8\% for MA-U, MR-U, and MA-C and 87.9\% for MR-C, so correction raised the coverage of the meta-regression model by about 3 percentage points but left all four models 7 to 10 percentage points below the nominal 95\%, with similar KDE interval widths.
Under the parametric method, correction raised the coverage of the intercept-only model from 84.8\% to 87.9\%, and the meta-regression models attained 100\% coverage, but only because their intervals were much wider than the corresponding KDE intervals (6.30 for MR-U and 9.63 for MR-C, compared with 2.62), making them largely uninformative. The especially wide parametric intervals of MR-C may reflect the interaction of IPPW reweighting with the robust variance estimator, which inflates the estimated uncertainty of the coefficients.

The minimal improvement from publication bias correction in ManyBabies 4 was consistent with the less severe asymmetry as we see in Figure~\ref{fig:funnel}(b). The minimal improvement from conditioning on moderators likely reflects the fact that, although the main moderators in ManyBabies 4 were homogeneous across replication sites due to the shared protocol \cite{lucca2025infants}, the replication point estimates were nevertheless heterogeneous as shown in Figure~\ref{fig:PI-mb4}. Unmeasured moderators likely drove the between-study variability in replication outcomes in this specific dataset.

\begin{figure}[H]
    \centering
    \begin{subfigure}[b]{\textwidth}
        \centering      \includegraphics[height=0.85\textheight]{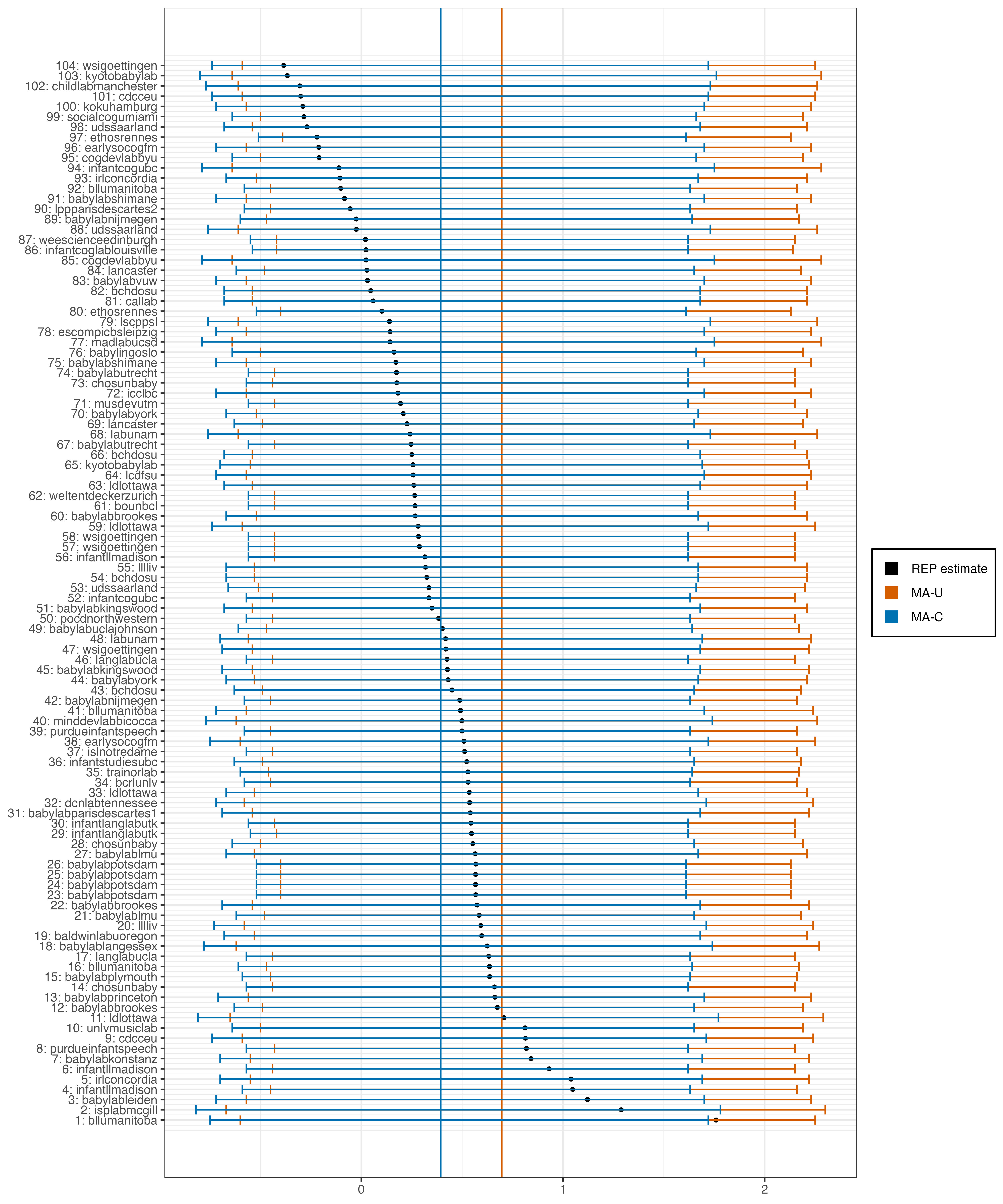}
        \caption{MB1: Prediction intervals under MA-U and MA-C models}
        \label{fig:pred-a}
    \end{subfigure}
\end{figure}

\begin{figure}[p]
    \ContinuedFloat
    \centering
    \begin{subfigure}[b]{\textwidth}
        \centering        \includegraphics[height=0.85\textheight]{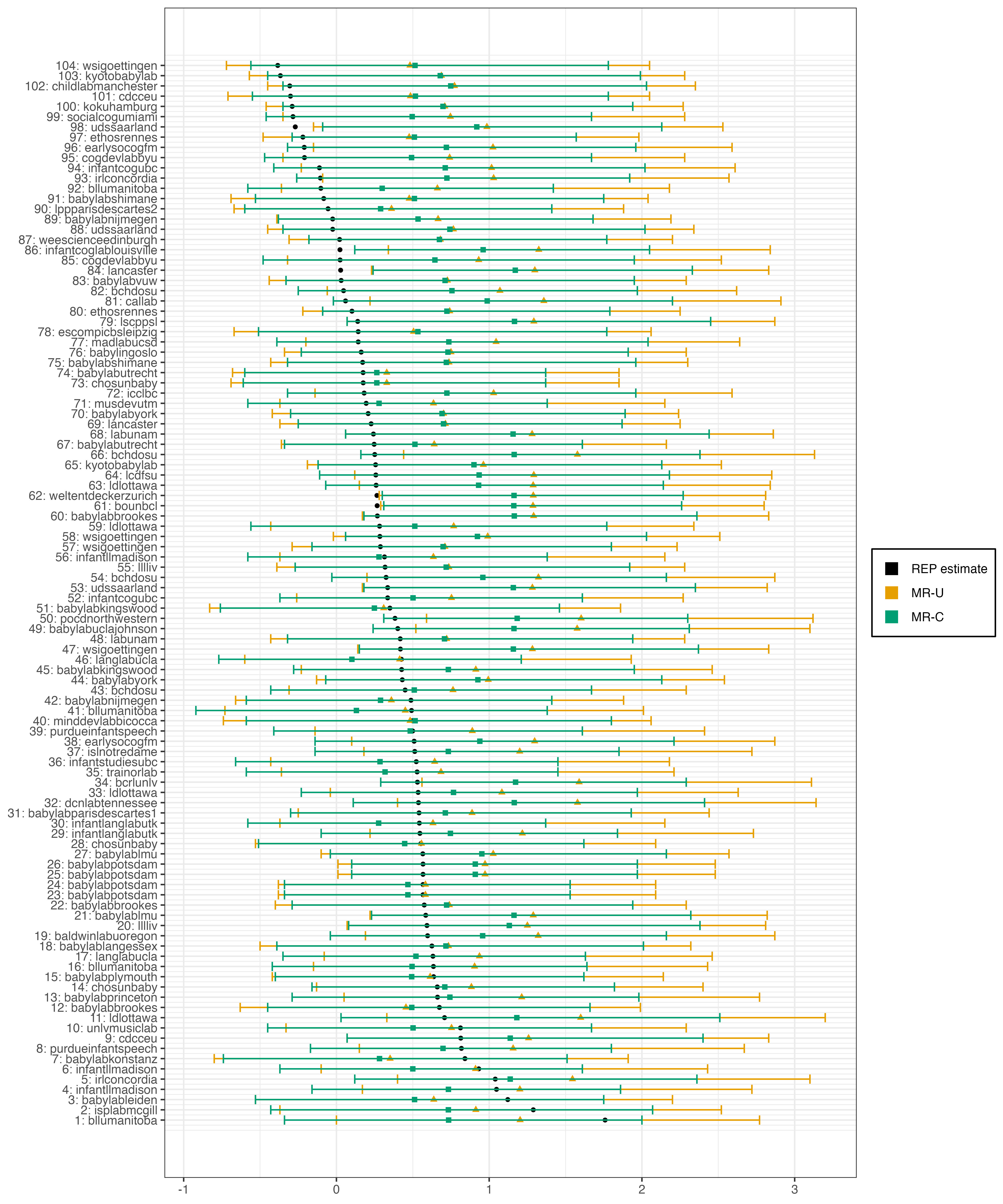}
        \caption{MB1: Prediction intervals under MR-U and MR-C models}
        \label{fig:pred-b}
    \end{subfigure}
    \caption{KDE-based 95\% prediction intervals for ManyBabies 1 replication studies. 
    MA-U: uncorrected intercept-only meta-analysis; MA-C: publication-bias-corrected intercept-only meta-analysis; MR-U: uncorrected meta-regression; MR-C: publication-bias-corrected meta-regression;
    Black dot: reported point estimate; Vertical orange line: uncorrected meta-analyzed estimate; Vertical blue line: corrected meta-analyzed estimate; 
    Yellow triangle: uncorrected point forecast by meta-regression; Green square: corrected point forecast by meta-regression.}
    \label{fig:PI-mb1}
\end{figure}

\begin{figure}[H]
    \centering
    \begin{subfigure}[b]{\textwidth}
        \centering        \includegraphics[height=0.85\textheight]{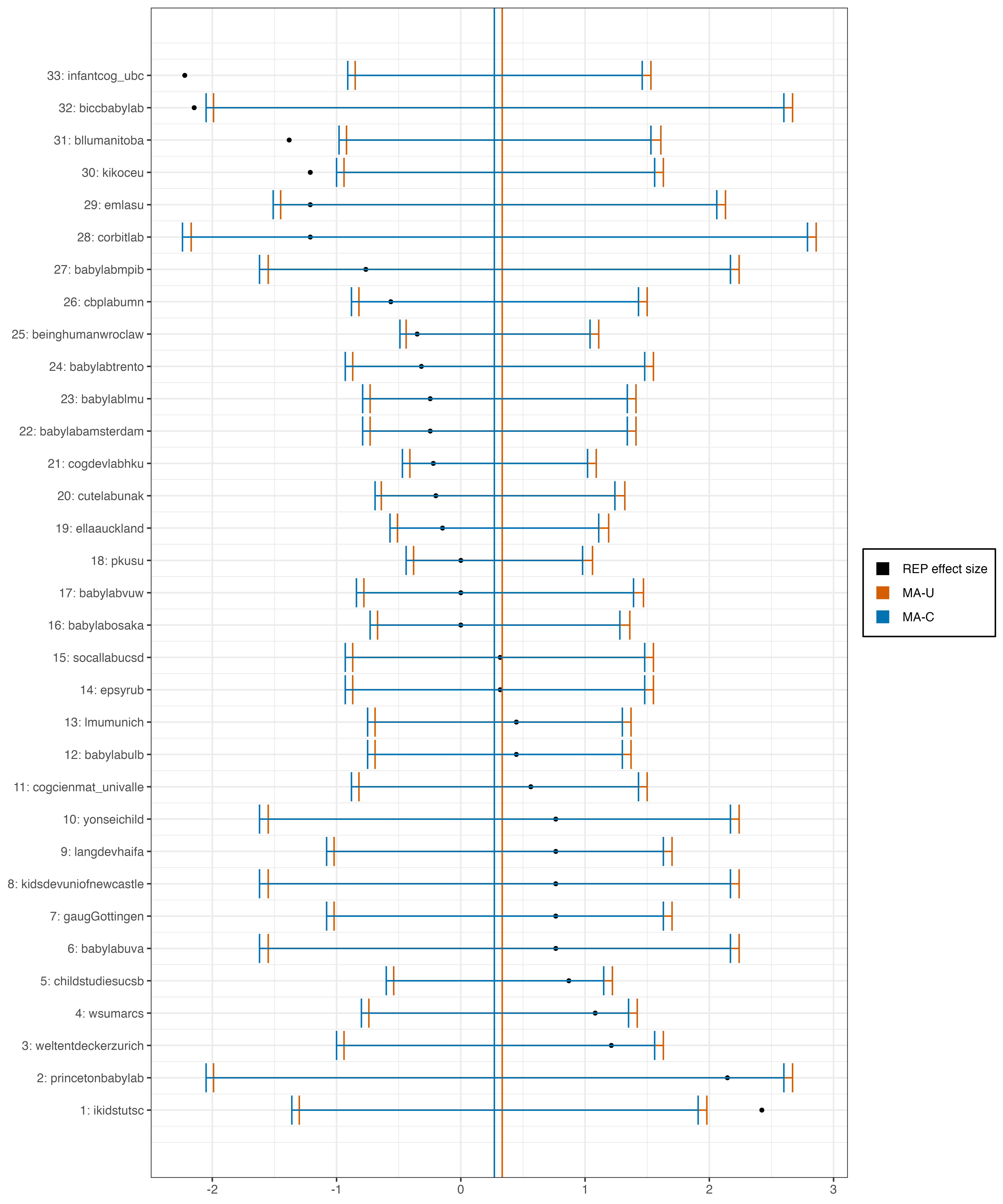}
        \caption{MB4: Prediction intervals under MA-U and MA-C models}
        \label{fig:pred-mb4-a}
    \end{subfigure}
\end{figure}

\begin{figure}[p]
    \ContinuedFloat
    \centering
    \begin{subfigure}[b]{\textwidth}
        \centering       \includegraphics[height=0.85\textheight]{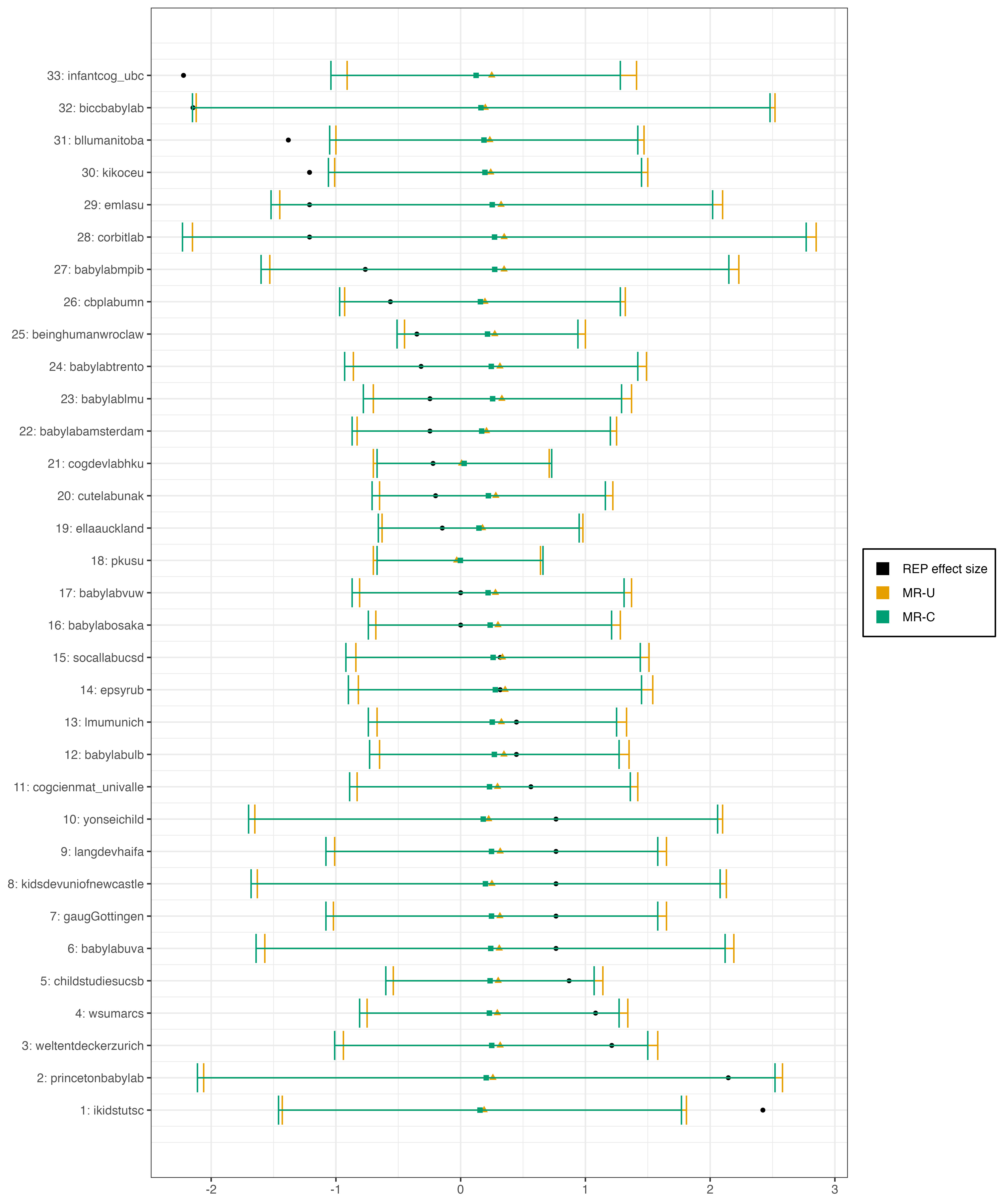}
        \caption{MB4: Prediction intervals under MR-U and MR-C models}
        \label{fig:pred-mb4-b}
    \end{subfigure}
    \caption{KDE-based 95\% prediction intervals for ManyBabies 4 replication studies. 
    MA-U: uncorrected intercept-only meta-analysis; MA-C: publication-bias-corrected intercept-only meta-analysis; MR-U: uncorrected meta-regression; MR-C: publication-bias-corrected meta-regression;
    Black dots=: reported point estimate; Vertical orange line: uncorrected meta-analyzed estimate; Vertical blue line: corrected meta-analyzed estimate; 
    Yellow triangle: uncorrected point forecast by meta-regression; Green square: corrected point forecast by meta-regression.}
    \label{fig:PI-mb4}
\end{figure}

\section{Discussion}
\label{sec:discussion}
\subsection{Principal findings}
We developed and evaluated methods for forecasting replication study results from published meta-analyses, combining meta-regression calibrated estimates, IPPW correction for publication bias, and KDE-based prediction intervals. Taken together, the simulation and the two ManyBabies applications clarified when each component of the approach improves forecasts.

Publication bias correction improved forecasts roughly in proportion to the severity of selection. When selection was strong, as in the simulation with large $\eta$ (severe publication bias), the uncorrected models systematically overpredicted the replication results, and correction largely removed this bias and substantially improved accuracy; when selection was mild, correction still moved the forecasts in the right direction, although the gain was small. Correction also improved the calibration of the prediction intervals when the IPPW weights were applied to the KDE as well as to the model fit. In the simulation, the corrected models' KDE coverage declined much less with increasing selection than that of the uncorrected models. Under strong selection, however, coverage remained below the nominal level, likely because the plug-in heterogeneity estimate from the uncorrected model was biased downward under selection, and because the meta-analysis and replication studies may differ in their underlying populations.

Meta-regression, in turn, improved forecasts only when two conditions held: the measured moderators explained variation in the population effects, and the replication sites differed on those moderators. The simulation met both conditions whenever moderation was present, and the corrected meta-regression was then the most accurate model. ManyBabies~1 met only the second condition, and meta-regression did not improve its forecasts; ManyBabies~4 met neither, and the gains were slight. In ManyBabies~4, the replication estimates varied considerably more across sites than any model predicted, driven by sources of heterogeneity that none of the measured moderators captured. These results suggest a practical guideline: publication bias correction, applied to both the model and the prediction interval, should be the default when forecasting replication results from meta-analyses, whereas meta-regression is worthwhile only when the meta-analysis provides evidence that the moderators matter and the replication sites vary on them.

\subsection{Accuracy-calibration tradeoff}
No single model performed best on both point accuracy and interval calibration. In the simulation, MR-C most often had the best point accuracy, overestimation rate, and mean $\bar P_{\mathrm{orig}}$, whereas MA-C most often had the raw difference closest to zero and the coverage closest to the nominal level, clearly so for the KDE-based intervals and narrowly for the parametric intervals. ManyBabies~1 showed a similar tradeoff: MA-C gave the most accurate forecasts, but had the full coverage , exceeding the nominal level, whereas MR-C attained the nominal coverage with the narrowest KDE-based interval of the four models, at the cost of a higher MAE and overestimation rate.

The tradeoff also appeared between the two interval methods. With IPPW weights applied to the KDE, the corrected models' KDE-based intervals were better calibrated than those of the uncorrected models and remained informative, but under strong selection they still fell short of the nominal level. The parametric intervals of the corrected models, especially MR-C, reached or exceeded nominal coverage, but only by becoming considerably wider, to the point of being largely uninformative in ManyBabies~4. Neither method produced intervals that were both fully calibrated and informative under strong selection. 
MR-U illustrated the opposite failure in the simulation: it produced the narrowest KDE intervals in nearly all scenarios but rarely achieved the best coverage, but intervals are of little value when the point forecasts are biased.

\subsection{Unexplained heterogeneity and unmeasured moderators}
In ManyBabies 4, the improvement from meta-regression and IPPW correction was minimal. The replication dataset exhibited substantial unexplained heterogeneity in point estimates despite a shared protocol that held key moderator levels constant across sites. Because the meta-regression and IPPW-corrected models improve predictions by conditioning on moderator levels and correcting for publication bias, their advantage diminishes when replication-study moderators are homogeneous and residual heterogeneity is driven by unmeasured factors. This pattern highlights a broader tension in applying replication forecasting methods: meta-analysis datasets are typically heterogeneous across study populations and designs, which is precisely the variation that meta-regression exploits, whereas large-scale replication projects deliberately standardize their protocols to isolate the effect of interest. When this standardization compresses moderator variability in the replication dataset, the forecasting models lose their primary source of predictive capability. In this respect, ManyBabies 1, where replication sites varied meaningfully on study-level covariates, provided a more informative demonstration of the four models' comparative strengths than ManyBabies 4.

This limitation is not unique to our application; it mirrors a broader pattern documented in meta-analyses and replication studies across developmental psychology. Using data from Metalab, Cao et al.\cite{cao2025estimating}
unexpectedly found no evidence of age-related growth in infants' linguistic and cognitive skills, linear or nonlinear, across all 25 datasets examined, contrary to the field's longstanding assumption that such skills refine with age. They further found that an age-related meta-analytic effect was present in only a subset of studies, identified either by a specific feature of the experimental paradigm (e.g., paradigms depicting ``giving and taking'' actions to infants) \cite{margoni2018infants} or by particular participant characteristics (e.g., bilingual infants)\cite{tsui2019associative}. This subset-specific pattern indicates that moderation effects are not uniform across studies, challenging the common assumption in developmental psychology that such effects are generalizable.

Cao et al.\cite{cao2025estimating} offered two explanations for this pattern, both relevant to replication forecasting. First, the psychometric properties of many infant paradigms, such as their reliability and validity across ages, are rarely examined\cite{byers2022six}, so measurement artifacts may distort effect size estimates. Second, studies pooled in a meta-analysis differ in participants and procedures, such as infant age, stimuli, and testing paradigm, largely by the choice of individual laboratories rather than by design, so differences across studies may reflect true heterogeneity, measurement artifacts, or both. Because primary studies are not designed to be pooled, many potentially relevant moderators are neither measured nor reported, which we view as an inherent limitation of meta-analysis. Cao et al. therefore called for more standardized methods and more detailed reporting, which would reduce avoidable heterogeneity and allow moderators to be studied more precisely in future meta-analyses.

Our ManyBabies applications illustrated this same pattern. The meta-analysis datasets underlying both ManyBabies 1 and ManyBabies 4 showed markedly different distributions of covariates across studies (mean age, test
language, and experimental method in ManyBabies 1; method, scenario, stimuli, choice object, intent valence, and outcome valence in ManyBabies 4), consistent with the paradigm-driven moderation described by Cao et al.\cite{cao2025estimating}. In the ManyBabies 1 meta-analysis dataset, most studies tested infants in their native language, whereas most replication studies tested infants in a non-native language; this difference could plausibly alter infants' responses, though the mechanism cannot be confirmed from the available data. More generally, because most experiment-level moderators were fixed at a single level across ManyBabies replication sites, distinguishing true moderators from unmeasured sources of heterogeneity remains difficult.

\subsection{Future directions}
Given the findings and ongoing replication projects, we do not advocate for homogenizing moderators across all replication studies. Doing so is unnecessary for drawing valid conclusions about replicability, and it forecloses the opportunity to characterize real-world heterogeneity, in which study participants exhibit both genuinely heterogeneous responses to the outcome of interest and study-level variation attributable to differences in experimental setting and methodology. Building on this perspective, we outline directions for strengthening the use of meta-analytic results in forecasting replication studies, particularly in settings where replication protocols allow moderator levels to vary in a controlled manner rather than being fully standardized.

Replication projects could adopt designs that deliberately assign different moderator levels across sites. In ManyBabies 4, the shared protocol held key moderators constant, which compressed the moderator variation that meta-regression relies on for conditional prediction. By contrast, a design in which sites vary a prespecified set of moderators (e.g., participant age groups, stimulus types, or experimental methods) under careful data collection would preserve the meta-regression's predictive leverage while maintaining the measurement quality and internal validity characteristic of large-scale replications. The central methodological question is which moderators to vary and at what levels, a problem that connects replication planning to optimal experimental design.

Given a fitted meta-regression from the existing literature, researchers could allocate moderator configurations across replication sites to maximize the informativeness of the resulting predictions. Sites could be assigned to fill gaps in the moderator space covered by the meta-analysis, targeting the covariate combinations where conditional predictions are most uncertain or where the corrected and uncorrected models diverge most sharply. This would improve the precision of forecasts by meta-regression while avoiding extrapolation beyond the range of the meta-analytic evidence.

Historically, large-scale replication projects have standardized moderator levels across sites so that differences in observed point estimates can be attributed to replication failure rather than methodological variation, simplifying the assessment of whether a finding replicates. However, this simplification is not strictly necessary when appropriate statistical methods are available. The meta-regression and calibrated estimation framework developed in this paper, for example, can generate moderator-conditional predictions that account for design differences across sites, allowing researchers to evaluate replicability even when study protocols vary on key moderators. As such methods continue to develop, replication projects can move toward designs that deliberately vary moderators while still drawing valid conclusions about whether the underlying effect replicates.

\begin{Backmatter}

\paragraph{Original Publication}
The manuscript contains original unpublished work and is not being submitted for publication elsewhere at the same time.

\paragraph{Funding Statement}
This research was supported by the National Institutes of Health grants R01 LM013866, UL1TR003142, P30CA124435, and P30DK116074.

\paragraph{Competing Interests}
None.

\paragraph{Data Availability Statement}
All code and data required to reproduce the simulation study and applied examples are publicly available and documented at
\url{https://github.com/einsley1993/rfmb14}. 

\paragraph{Ethical Standards}
The research meets all ethical guidelines, including adherence to the legal requirements of the study country.

\paragraph{Author Contributions}
Conceptualization: Y.L; M.B.M. Methodology: Y.L; M.B.M. Data curation: M.C.F; A.W.T; F.M. Data visualisation: Y.L. Writing original draft: Y.L; M.B.M. All authors approved the final submitted draft.

\newpage
\printbibliography

@article{aczel2026investigating,
  title={Investigating the analytical robustness of the social and behavioural sciences},
  author={Aczel, Balazs and Szaszi, Barnabas and Clelland, Harry T and Kovacs, Marton and Holzmeister, Felix and Van Ravenzwaaij, Don and Schulz-K{\"u}mpel, Hannah and Hoffmann, Sabine and Nilsonne, Gustav and Kosa, Livia and others},
  journal={Nature},
  volume={652},
  number={8108},
  pages={135--142},
  year={2026},
  publisher={Nature Publishing Group UK London}
}

@article{tyner2026investigating,
  title={Investigating the replicability of the social and behavioural sciences},
  author={Tyner, Andrew H and Abatayo, Anna Lou and Daley, Mason and Field, Samuel and Fox, Nicholas and Haber, Noah A and Hahn, Krystal M and Struhl, Melissa Kline and Mawhinney, Brinna and Miske, Olivia and others},
  journal={Nature},
  volume={652},
  number={8108},
  pages={143--150},
  year={2026},
  publisher={Nature Publishing Group UK London}
}

@article{miske2026investigating,
  title={Investigating the reproducibility of the social and behavioural sciences},
  author={Miske, Olivia and Abatayo, Anna Lou and Daley, Mason and Dirzo, Mirka and Fox, Nicholas and Haber, Noah and Hahn, Krystal M and Struhl, Melissa Kline and Mawhinney, Brinna and Silverstein, Priya and others},
  journal={Nature},
  volume={652},
  number={8108},
  pages={126--134},
  year={2026},
  publisher={Nature Publishing Group UK London}
}

@article{ebersole2016many,
  title={Many Labs 3: Evaluating participant pool quality across the academic semester via replication},
  author={Ebersole, Charles R and Atherton, Olivia E and Belanger, Aimee L and Skulborstad, Hayley M and Allen, Jill M and Banks, Jonathan B and Baranski, Erica and Bernstein, Michael J and Bonfiglio, Diane BV and Boucher, Leanne and others},
  journal={Journal of Experimental Social Psychology},
  volume={67},
  pages={68--82},
  year={2016},
  publisher={Elsevier}
}

@article{simons2014introduction,
  title={An introduction to registered replication reports at perspectives on psychological science},
  author={Simons, Daniel J and Holcombe, Alex O and Spellman, Barbara A},
  journal={Perspectives on Psychological Science},
  volume={9},
  number={5},
  pages={552--555},
  year={2014},
  publisher={Sage Publications Sage CA: Los Angeles, CA}
}

@article{cao2025estimating,
  title={Estimating Age-Related Change in Infants' Linguistic and Cognitive Development Using (Meta-) Meta-Analysis},
  author={Cao, Anjie and Lewis, Molly and Tsuji, Sho and Bergmann, Christina and Cristia, Alejandrina and Frank, Michael C},
  journal={Developmental Science},
  volume={28},
  number={4},
  pages={e70028},
  year={2025},
  publisher={Wiley Online Library}
}

@article{tsui2019associative,
  title={Associative word learning in infancy: A meta-analysis of the switch task.},
  author={Tsui, Angeline Sin Mei and Byers-Heinlein, Krista and Fennell, Christopher T},
  journal={Developmental psychology},
  volume={55},
  number={5},
  pages={934},
  year={2019},
  publisher={American Psychological Association}
}

@article{choi2023understanding,
  title={Understanding preferences in infancy},
  author={Choi, Youjung and Luo, Yuyan},
  journal={Wiley Interdisciplinary Reviews: Cognitive Science},
  volume={14},
  number={4},
  pages={e1643},
  year={2023},
  publisher={Wiley Online Library}
}

@article{margoni2024violation,
  title={The violation-of-expectation paradigm: A conceptual overview.},
  author={Margoni, Francesco and Surian, Luca and Baillargeon, Ren{\'e}e},
  journal={Psychological Review},
  volume={131},
  number={3},
  pages={716},
  year={2024},
  publisher={American Psychological Association}
}

@article{lee2015development,
  title={The development of infants’ sensitivity to behavioral intentions when inferring others’ social preferences},
  author={Lee, Young-eun and Yun, Jung-eun Ellie and Kim, Eun Young and Song, Hyun-joo},
  journal={PLoS One},
  volume={10},
  number={9},
  pages={e0135588},
  year={2015},
  publisher={Public Library of Science San Francisco, CA USA}
}

@article{byers2022six,
  title={Six solutions for more reliable infant research},
  author={Byers-Heinlein, Krista and Bergmann, Christina and Savalei, Victoria},
  journal={Infant and Child Development},
  volume={31},
  number={5},
  pages={e2296},
  year={2022},
  publisher={Wiley Online Library}
}

@article{kvarven2020comparing,
  title={Comparing meta-analyses and preregistered multiple-laboratory replication projects},
  author={Kvarven, Amanda and Str{\o}mland, Eirik and Johannesson, Magnus},
  journal={Nature Human Behaviour},
  volume={4},
  number={4},
  pages={423--434},
  year={2020},
  publisher={Nature Publishing Group UK London}
}

@article{mathur2021meta,
  title={Meta-regression methods to characterize evidence strength using meaningful-effect percentages conditional on study characteristics},
  author={Mathur, Maya B and VanderWeele, Tyler J},
  journal={Research synthesis methods},
  volume={12},
  number={6},
  pages={731--749},
  year={2021},
  publisher={Wiley Online Library}
}

@article{wang2019simple,
  title={A simple method to estimate prediction intervals and predictive distributions: summarizing meta-analyses beyond means and confidence intervals},
  author={Wang, Chia-Chun and Lee, Wen-Chung},
  journal={Research Synthesis Methods},
  volume={10},
  number={2},
  pages={255--266},
  year={2019},
  publisher={Wiley Online Library}
}

@article{mathur2020new,
  title={New statistical metrics for multisite replication projects},
  author={Mathur, Maya B and VanderWeele, Tyler J},
  journal={Journal of the Royal Statistical Society Series A: Statistics in Society},
  volume={183},
  number={3},
  pages={1145--1166},
  year={2020},
  publisher={Oxford University Press}
}

@article{morris2019using,
  title={Using simulation studies to evaluate statistical methods},
  author={Morris, Tim P and White, Ian R and Crowther, Michael J},
  journal={Statistics in medicine},
  volume={38},
  number={11},
  pages={2074--2102},
  year={2019},
  publisher={Wiley Online Library}
}

@article{mathur2020sensitivity,
  title={Sensitivity analysis for publication bias in meta-analyses},
  author={Mathur, Maya B and VanderWeele, Tyler J},
  journal={Journal of the Royal Statistical Society Series C: Applied Statistics},
  volume={69},
  number={5},
  pages={1091--1119},
  year={2020},
  publisher={Oxford University Press}
}

@article{hedges2010robust,
  title={Robust variance estimation in meta-regression with dependent effect size estimates},
  author={Hedges, Larry V and Tipton, Elizabeth and Johnson, Matthew C},
  journal={Research synthesis methods},
  volume={1},
  number={1},
  pages={39--65},
  year={2010},
  publisher={Wiley Online Library}
}

@article{hamlin2007social,
  title={Social evaluation by preverbal infants},
  author={Hamlin, Kiley and Wynn, Karen and Bloom, Paul},
  journal={Nature},
  volume={450},
  number={7169},
  pages={557--559},
  year={2007},
  publisher={Nature Publishing Group UK London}
}

@article{bloom2016develops,
  title={What develops in moral development},
  author={Bloom, Paul and Wynn, Karen},
  journal={Core knowledge and conceptual change},
  pages={347--364},
  year={2016},
  publisher={Oxford University Press New York, NY}
}

@article{van2016evidence,
  title={Evidence for intuitive morality: Preverbal infants make sociomoral evaluations},
  author={Van de Vondervoort, Julia W and Hamlin, Kiley},
  journal={Child Development Perspectives},
  volume={10},
  number={3},
  pages={143--148},
  year={2016},
  publisher={Wiley Online Library}
}

@article{holvoet2016infants,
  title={Infants’ preference for prosocial behaviors: A literature review},
  author={Holvoet, Claire and Scola, C{\'e}line and Arciszewski, Thomas and Picard, Delphine},
  journal={Infant Behavior and Development},
  volume={45},
  pages={125--139},
  year={2016},
  publisher={Elsevier}
}

@article{buon2014friend,
  title={Friend or foe? Early social evaluation of human interactions},
  author={Buon, Marine and Jacob, Pierre and Margules, Sylvie and Brunet, Isabelle and Dutat, Michel and Cabrol, Dominique and Dupoux, Emmanuel},
  journal={PloS one},
  volume={9},
  number={2},
  pages={e88612},
  year={2014},
  publisher={Public Library of Science San Francisco, USA}
}

@article{kiley2010three,
  title={Three-month-olds show a negativity bias in their social evaluations},
  author={Hamlin, Kiley and Wynn, Karen and Bloom, Paul},
  journal={Developmental science},
  volume={13},
  number={6},
  pages={923--929},
  year={2010},
  publisher={Wiley Online Library}
}

@article{hamlin2011young,
  title={Young infants prefer prosocial to antisocial others},
  author={Hamlin, Kiley and Wynn, Karen},
  journal={Cognitive development},
  volume={26},
  number={1},
  pages={30--39},
  year={2011},
  publisher={Elsevier}
}

@article{loheide2021toddlers,
  title={Toddlers' preference for prosocial versus antisocial agents: No associations with empathy or attachment security},
  author={Loheide-Niesmann, Lisa and de Lijster, Jasmijn and Hall, Ruby and van Bakel, Hedwig and Cima, Maaike},
  journal={Social Development},
  volume={30},
  number={2},
  pages={410--427},
  year={2021},
  publisher={Wiley Online Library}
}

@article{scola2015further,
  title={Further evidence for infants' preference for prosocial over antisocial behaviors},
  author={Scola, Celine and Holvoet, Claire and Arciszewski, Thomas and Picard, Delphine},
  journal={Infancy},
  volume={20},
  number={6},
  pages={684--692},
  year={2015},
  publisher={Wiley Online Library}
}

@article{shimizu2021cultural,
  title={Cultural similarities and differences in the development of sociomoral judgments: An eye-tracking study},
  author={Shimizu, Yuki and Senzaki, Sawa and Cowell, Jason M},
  journal={Cognitive development},
  volume={57},
  pages={100974},
  year={2021},
  publisher={Elsevier}
}

@article{salvadori2015probing,
  title={Probing the strength of infants' preference for helpers over hinderers: Two replication attempts of Hamlin and Wynn (2011)},
  author={Salvadori, Eliala and Blazsekova, Tatiana and Volein, Agnes and Karap, Zsuzsanna and Tatone, Denis and Mascaro, Olivier and Csibra, Gergely},
  journal={PloS one},
  volume={10},
  number={11},
  pages={e0140570},
  year={2015},
  publisher={Public Library of Science San Francisco, CA USA}
}

@article{scarf2012social,
  title={Social evaluation or simple association? Simple associations may explain moral reasoning in infants},
  author={Scarf, Damian and Imuta, Kana and Colombo, Michael and Hayne, Harlene},
  year={2012},
  publisher={Public Library of Science San Francisco, USA}
}

@article{cowell2015precursors,
  title={Precursors to morality in development as a complex interplay between neural, socioenvironmental, and behavioral facets},
  author={Cowell, Jason M and Decety, Jean},
  journal={Proceedings of the National Academy of Sciences},
  volume={112},
  number={41},
  pages={12657--12662},
  year={2015},
  publisher={National Acad Sciences}
}

@article{nighbor2017stability,
  title={Stability of infants’ preference for prosocial others: Implications for research based on single-choice paradigms},
  author={Nighbor, Tyler and Kohn, Carolynn and Normand, Matthew and Schlinger, Henry},
  journal={PloS one},
  volume={12},
  number={6},
  pages={e0178818},
  year={2017},
  publisher={Public Library of Science San Francisco, CA USA}
}

@article{vaporova2020social,
  title={Social evaluation and imitation of prosocial and antisocial agents in infants, children, and adults},
  author={Vaporova, Elena and Zmyj, Norbert},
  journal={Plos one},
  volume={15},
  number={9},
  pages={e0235595},
  year={2020},
  publisher={Public Library of Science San Francisco, CA USA}
}

@article{abramson2016preferences,
  title={Preferences for helpers and hinders in 9-and 18-month-old infants},
  author={Abramson, L and Dar, M and Te’eni, A and Knafo-Noam, A},
  journal={Unpublished raw data},
  year={2016}
}

@article{hamlin2016box,
  title={A box show replication attempt},
  author={Hamlin, Kiley},
  journal={Unpublished raw data},
  year={2016}
}

@article{woo2016older,
  title={Older, but not younger, babies prefer helpers in a cartoon version of the hill paradigm},
  author={Woo, BM and Hamlin, Kiley},
  journal={Manuscript in preparation},
  year={2016}
}

@article{schlingloff202015,
  title={Do 15-month-old infants prefer helpers? A replication of Hamlin et al.(2007)},
  author={Schlingloff, Laura and Csibra, Gergely and Tatone, Denis},
  journal={Royal Society open science},
  volume={7},
  number={4},
  pages={191795},
  year={2020},
  publisher={The Royal Society}
}

@article{margoni2018infants,
  title={Infants’ evaluation of prosocial and antisocial agents: A meta-analysis.},
  author={Margoni, Francesco and Surian, Luca},
  journal={Developmental psychology},
  volume={54},
  number={8},
  pages={1445},
  year={2018},
  publisher={American Psychological Association}
}

@article{byers2020building,
  title={Building a collaborative psychological science: Lessons learned from ManyBabies 1.},
  author={Byers-Heinlein, Krista and Bergmann, Christina and Davies, Catherine and Frank, Michael C and Hamlin, J Kiley and Kline, Melissa and Kominsky, Jonathan F and Kosie, Jessica E and Lew-Williams, Casey and Liu, Liquan and others},
  journal={Canadian Psychology/Psychologie Canadienne},
  volume={61},
  number={4},
  pages={349},
  year={2020},
  publisher={Educational Publishing Foundation}
}

@article{manybabies2020quantifying,
  title={Quantifying sources of variability in infancy research using the infant-directed-speech preference},
  author={ManyBabies-Consortium},
  journal={Advances in Methods and Practices in Psychological Science},
  volume={3},
  number={1},
  pages={24--52},
  year={2020},
  publisher={Sage Publications Sage CA: Los Angeles, CA}
}

@article{klein2014data,
  title={Data from investigating variation in replicability: A" many labs" replication project},
  author={Klein, Richard and Ratliff, Kate and Vianello, Michelangelo and Adams Jr, Reginald B and Bahn{\'\i}k, St{\u{e}}p{\'a}n and Bernstein, Michael J and Brandt, MJ and Ijzerman, H and Bocian, Konrad and Brooks, Beach and others},
  journal={Journal of Open Psychology Data},
  volume={2},
  number={1},
  year={2014}
}

@article{chae2018negativity,
  title={Negativity bias in infants’ expectations about agents’ dispositions},
  author={Chae, Joanna Joo Kyung and Song, Hyun-joo},
  journal={British Journal of Developmental Psychology},
  volume={36},
  number={4},
  pages={620--633},
  year={2018},
  publisher={Wiley Online Library}
}

@article{tan2024prosocial,
  title={Prosocial evaluation: A new meta-analysis},
  author={Tan, Alvin WM},
  year={2024},
  publisher={OSF}
}

@article{borenstein2009effect,
  title={Effect sizes for continuous data},
  author={Borenstein, Michael and Cooper, Harris and Hedges, L and Valentine, J},
  journal={The handbook of research synthesis and meta-analysis},
  volume={2},
  pages={221--235},
  year={2009}
}

@article{dunst2012preference,
  title={Preference for infant-directed speech in preverbal young children},
  author={Dunst, Carl and Gorman, Ellen and Hamby, Deborah},
  journal={Center for Early Literacy Learning},
  volume={5},
  number={1},
  pages={1--13},
  year={2012}
}

@article{sheather1991reliable,
  title={A reliable data-based bandwidth selection method for kernel density estimation},
  author={Sheather, Simon J and Jones, Michael C},
  journal={Journal of the Royal Statistical Society Series B: Statistical Methodology},
  volume={53},
  number={3},
  pages={683--690},
  year={1991},
  publisher={Wiley Online Library}
}

@article{edgington1972additive,
  title={An additive method for combining probability values from independent experiments},
  author={Edgington, Eugene S},
  journal={The Journal of Psychology},
  volume={80},
  number={2},
  pages={351--363},
  year={1972},
  publisher={Taylor \& Francis}
}

@article{edgington1972normal,
  title={A normal curve method for combining probability values from independent experiments},
  author={Edgington, Eugene S},
  journal={The Journal of psychology},
  volume={82},
  number={1},
  pages={85--89},
  year={1972},
  publisher={Taylor \& Francis}
}

@article{held2025comparison,
  title={A comparison of combined p-value functions for meta-analysis},
  author={Held, Leonhard and Hofmann, Felix and Pawel, Samuel},
  journal={Research Synthesis Methods},
  volume={16},
  number={5},
  pages={758--785},
  year={2025},
  publisher={Cambridge University Press}
}

@article{held2024assessment,
  title={The assessment of replicability using the sum of p-values},
  author={Held, Leonhard and Pawel, Samuel and Micheloud, Charlotte},
  journal={Royal Society Open Science},
  volume={11},
  number={8},
  year={2024},
  publisher={The Royal Society}
}

@article{held2022ReplicationSuccess,
  title={Design and analysis of replication studies with ReplicationSuccess},
  author={Held, Leonhard and Micheloud, Charlotte and Pawel, Samuel and Hofmann, Felix and Gerber, Florian},
  journal={URL https://crsuzh. github. io/ReplicationSuccess/index. html},
  year={2022}
}

@article{held2020sceptical,
  title={A new standard for the analysis and design of replication studies},
  author={Held, Leonhard},
  journal={Journal of the Royal Statistical Society Series A: Statistics in Society},
  volume={183},
  number={2},
  pages={431--448},
  year={2020},
  publisher={Oxford University Press}
}

@article{cochran1954combination,
  title={The combination of estimates from different experiments},
  author={Cochran, William G},
  journal={Biometrics},
  volume={10},
  number={1},
  pages={101--129},
  year={1954},
  publisher={JSTOR}
}

@article{fisher2015robumeta,
  title={robumeta: An R-package for robust variance estimation in meta-analysis},
  author={Fisher, Zachary and Tipton, Elizabeth},
  journal={arXiv preprint arXiv:1503.02220},
  year={2015}
}

@article{mathur2021estimating,
  title={Estimating publication bias in meta-analyses of peer-reviewed studies: A meta-meta-analysis across disciplines and journal tiers},
  author={Mathur, Maya B and VanderWeele, Tyler J},
  journal={Research Synthesis Methods},
  volume={12},
  number={2},
  pages={176--191},
  year={2021},
  publisher={Wiley Online Library}
}

@article{devroye1985nonparametric,
  title={Nonparametric density estimation},
  author={Devroye, Luc},
  journal={The L\_1 View},
  year={1985},
  publisher={John Wiley}
}

@book{van2000asymptotic,
  title={Asymptotic statistics},
  author={Van der Vaart, Aad W},
  volume={3},
  year={2000},
  publisher={Cambridge university press}
}

@article{gneiting2007strictly,
  title={Strictly proper scoring rules, prediction, and estimation},
  author={Gneiting, Tilmann and Raftery, Adrian E},
  journal={Journal of the American statistical Association},
  volume={102},
  number={477},
  pages={359--378},
  year={2007},
  publisher={Taylor \& Francis}
}

@book{sutton2000methods,
  title={Methods for meta-analysis in medical research},
  author={Sutton, Alex J and Abrams, Keith R and Jones, David R and Sheldon, Trevor A and Song, Fujian},
  volume={348},
  year={2000},
  publisher={Wiley Chichester}
}

@article{mathur2019new,
  title={New metrics for meta-analyses of heterogeneous effects},
  author={Mathur, Maya B and VanderWeele, Tyler J},
  journal={Statistics in Medicine},
  volume={38},
  number={8},
  pages={1336--1342},
  year={2019},
  publisher={Wiley Online Library}
}

@article{jpt2009re,
  title={A re-evaluation of random-effects meta-analysis},
  author={JPT, HIGGINS},
  journal={JR Stat. Soc. Ser. A Stat. Soc.},
  volume={172},
  pages={137--159},
  year={2009}
}

@article{graham2012robust,
  title={Robust meta-analytic conclusions mandate the provision of prediction intervals in meta-analysis summaries},
  author={Graham, Petra L and Moran, John L},
  journal={Journal of clinical epidemiology},
  volume={65},
  number={5},
  pages={503--510},
  year={2012},
  publisher={Elsevier}
}

@article{inthout2016plea,
  title={Plea for routinely presenting prediction intervals in meta-analysis},
  author={IntHout, Joanna and Ioannidis, John PA and Rovers, Maroeska M and Goeman, Jelle J},
  journal={BMJ open},
  volume={6},
  number={7},
  pages={e010247},
  year={2016},
  publisher={British Medical Journal Publishing Group}
}

@article{kontopantelis2012performance,
  title={Performance of statistical methods for meta-analysis when true study effects are non-normally distributed: a simulation study},
  author={Kontopantelis, Evangelos and Reeves, David},
  journal={Statistical methods in medical research},
  volume={21},
  number={4},
  pages={409--426},
  year={2012},
  publisher={SAGE Publications Sage UK: London, England}
}

@article{lee2008flexible,
  title={Flexible parametric models for random-effects distributions},
  author={Lee, Katherine J and Thompson, Simon G},
  journal={Statistics in medicine},
  volume={27},
  number={3},
  pages={418--434},
  year={2008},
  publisher={Wiley Online Library}
}

@article{yamaguchi2017random,
  title={A random effects meta-analysis model with Box-Cox transformation},
  author={Yamaguchi, Yusuke and Maruo, Kazushi and Partlett, Christopher and Riley, Richard D},
  journal={BMC medical research methodology},
  volume={17},
  number={1},
  pages={109},
  year={2017},
  publisher={Springer}
}

@article{burr2005bayesian,
  title={A Bayesian semiparametric model for random-effects meta-analysis},
  author={Burr, Deborah and Doss, Hani},
  journal={Journal of the American Statistical Association},
  volume={100},
  number={469},
  pages={242--251},
  year={2005},
  publisher={Taylor \& Francis}
}

@article{ohlssen2007flexible,
  title={Flexible random-effects models using Bayesian semi-parametric models: applications to institutional comparisons},
  author={Ohlssen, David I and Sharples, Linda D and Spiegelhalter, David J},
  journal={Statistics in medicine},
  volume={26},
  number={9},
  pages={2088--2112},
  year={2007},
  publisher={Wiley Online Library}
}

@article{fernald1989intonation,
  title={Intonation and communicative intent in mothers' speech to infants: Is the melody the message?},
  author={Fernald, Anne},
  journal={Child development},
  pages={1497--1510},
  year={1989},
  publisher={JSTOR}
}

@article{cooper1997development,
  title={The development of infants' preference for motherese},
  author={Cooper, Robin Panneton and Abraham, Jane and Berman, Sheryl and Staska, Margaret},
  journal={Infant behavior and development},
  volume={20},
  number={4},
  pages={477--488},
  year={1997},
  publisher={Elsevier}
}

@article{cooper1994developmental,
  title={Developmental differences in infant attention to the spectral properties of infant-directed speech},
  author={Cooper, Robin Panneton and Aslin, Richard N},
  journal={Child development},
  volume={65},
  number={6},
  pages={1663--1677},
  year={1994},
  publisher={Oxford University Press}
}

@article{fernald1985four,
  title={Four-month-old infants prefer to listen to motherese},
  author={Fernald, Anne},
  journal={Infant behavior and development},
  volume={8},
  number={2},
  pages={181--195},
  year={1985},
  publisher={Elsevier}
}

@article{hayashi2001developmental,
  title={Developmental change in auditory preferences for speech stimuli in Japanese infants},
  author={Hayashi, Akiko and Tamekawa, Yuji and Kiritani, Shigeru},
  journal={Journal of Speech, Language, and Hearing Research},
  volume={44},
  number={6},
  pages={1189--1200},
  year={2001}
}

@article{kitamura2009age,
  title={Age-specific preferences for infant-directed affective intent},
  author={Kitamura, Christine and Lam, Christa},
  journal={Infancy},
  volume={14},
  number={1},
  pages={77--100},
  year={2009},
  publisher={Wiley Online Library}
}

@article{newman2006changes,
  title={Changes in preference for infant-directed speech in low and moderate noise by 4.5-to 13-month-olds},
  author={Newman, Rochelle S and Hussain, Isma},
  journal={Infancy},
  volume={10},
  number={1},
  pages={61--76},
  year={2006},
  publisher={Wiley Online Library}
}

@article{pegg1992preference,
  title={Preference for infant-directed over adult-directed speech: Evidence from 7-week-old infants},
  author={Pegg, Judith E and Werker, Janet F and McLeod, Peter J},
  journal={Infant behavior and development},
  volume={15},
  number={3},
  pages={325--345},
  year={1992},
  publisher={Elsevier}
}

@article{santesso2007frontal,
  title={Frontal brain electrical activity (EEG) and heart rate in response to affective infant-directed (ID) speech in 9-month-old infants},
  author={Santesso, Diane L and Schmidt, Louis A and Trainor, Laurel J},
  journal={Brain and cognition},
  volume={65},
  number={1},
  pages={14--21},
  year={2007},
  publisher={Elsevier}
}

@article{singh2002infants,
  title={Infants' listening preferences: Baby talk or happy talk?},
  author={Singh, Leher and Morgan, James L and Best, Catherine T},
  journal={Infancy},
  volume={3},
  number={3},
  pages={365--394},
  year={2002},
  publisher={Wiley Online Library}
}

@article{werker1989infant,
  title={Infant preference for both male and female infant-directed talk: a developmental study of attentional and affective responsiveness.},
  author={Werker, Janet F and McLeod, Peter J},
  journal={Canadian Journal of Psychology/Revue canadienne de psychologie},
  volume={43},
  number={2},
  pages={230},
  year={1989},
  publisher={Canadian Psychological Association}
}

@article{frank2017collaborative,
  title={A collaborative approach to infant research: Promoting reproducibility, best practices, and theory-building},
  author={Frank, Michael C and Bergelson, Elika and Bergmann, Christina and Cristia, Alejandrina and Floccia, Caroline and Gervain, Judit and Hamlin, J Kiley and Hannon, Erin E and Kline, Melissa and Levelt, Claartje and others},
  journal={Infancy},
  volume={22},
  number={4},
  pages={421--435},
  year={2017},
  publisher={Wiley Online Library}
}

@article{zettersten2024evidence,
  title={Evidence for infant-directed speech preference is consistent across large-scale, multi-site replication and meta-analysis},
  author={Zettersten, Martin and Cox, Christopher and Bergmann, Christina and Tsui, Angeline Sin Mei and Soderstrom, Melanie and Mayor, Julien and Lundwall, Rebecca A and Lewis, Molly and Kosie, Jessica E and Kartushina, Natalia and others},
  journal={Open Mind},
  volume={8},
  pages={439--461},
  year={2024},
  publisher={MIT Press One Broadway, 12th Floor, Cambridge, Massachusetts 02142, USA~…}
}

@article{lucca2025infants,
  title={Infants’ Social Evaluation of Helpers and Hinderers: A Large-Scale, Multi-Lab, Coordinated Replication Study},
  author={Lucca, Kelsey and Yuen, Francis and Wang, Yiyi and Alessandroni, Nicol{\'a}s and Allison, Olivia and Alvarez, Mario and Axelsson, Emma L and Baumer, Janina and Baumgartner, Heidi A and Bertels, Julie and others},
  journal={Developmental Science},
  volume={28},
  number={1},
  pages={e13581},
  year={2025},
  publisher={Wiley Online Library}
}

@article{van2018early,
  title={The early emergence of sociomoral evaluation: Infants prefer prosocial others},
  author={Van de Vondervoort, Julia W and Hamlin, J Kiley},
  journal={Current Opinion in Psychology},
  volume={20},
  pages={77--81},
  year={2018},
  publisher={Elsevier}
}

@article{tipton2015small,
  title={Small sample adjustments for robust variance estimation with meta-regression.},
  author={Tipton, Elizabeth},
  journal={Psychological methods},
  volume={20},
  number={3},
  pages={375},
  year={2015},
  publisher={American Psychological Association}
}

\end{Backmatter}

\newpage
\pagenumbering{arabic}
\renewcommand{\thepage}{page ~\arabic{page}}
\begin{appendix}
\section{Appendix for ``Forecasting Replication Study Results from Meta-Analyses''}\label{appendixA}
\markboth{Appendix}{Appendix}

\subsection{Setup and assumptions}\label{app:setup}

For notational simplicity, we index the $k$ published meta-analysis studies by $i$ and treat their estimates as independent; the arguments extend to independent clusters of bounded size, as in the main text, in which case the number of clusters and $k$ are of the same order. Let $\mu(z) = \beta_0 + z\bbeta_1$ denote the conditional mean of the population effects at moderator level $z$, and let $A_i = \mathbb{I}\{\hat\theta_i > 0 \text{ and two-sided } p_i < 0.05\}$ indicate an affirmative study. We use the following assumptions.

\begin{enumerate}[label=(A\arabic*)]
\item \textbf{Random-effects meta-regression:} In the population of all conducted studies, $\theta_i = \beta_0 + Z_i\boldsymbol\beta_1 + u_i$, where$(Z_i, \hat\sigma_i, u_i)$ are independent and identically distributed across studies, the moderators $Z_i$ have bounded support, and the residuals $u_i$ are independent of $(Z_i, \hat\sigma_i)$, with mean zero, positive variance $\tau_c^2$, and an absolutely continuous distribution. No parametric form is assumed for the distribution of $u_i$.
\item \textbf{Within-study normality:} $\hat\theta_i \mid \theta_i, Z_i \overset{\mathrm{ind}}{\sim} N(\theta_i, \hat\sigma_i^2)$, with the known within-study variances, bounded between zero and infinity.
\item \textbf{Selection model} (publication-bias-corrected models only): Affirmative studies are $\eta$ times more likely to be published than nonaffirmative studies, that is, $\Pr(\text{published} \mid A_i = 1) / \Pr(\text{published} \mid A_i = 0) = \eta$, where $\eta \ge 1$ is known and $\Pr(\text{published} \mid A_i = 0) > 0$. Publication depends on a study's point estimate only through whether a study is affirmative.
\item \textbf{Transportability and independence:} The population effect at replication site $j$, with moderator level $z_j$, follows the same conditional distribution as the population effects in the meta-analyzed literature, $\theta_j^{\mathrm{rep}} = \beta_0 + z_j\boldsymbol\beta_1 + u_j$ with $u_j$ having the same distribution as $u_i$; the replication estimate satisfies $\hat\theta_j^{\mathrm{rep}} \mid \theta_j^{\mathrm{rep}} \sim N\bigl(\theta_j^{\mathrm{rep}}, \hat\sigma_j^{2}\bigr)$ and is independent of the meta-analysis data.

\end{enumerate}

For the uncorrected models, we assume instead that (A3) holds with $\eta = 1$, so that publication does not depend on the study data and the published studies are a random sample of the conducted studies.

Let $\omega_i = \eta^{\mathbb{I}\{A_i = 0\}}$ denote the study weight used in the KDE, so that $\omega_i \equiv 1$ for the uncorrected models. Under (A3), these weights invert the selection probabilities. Write $p = \P(\text{published} \mid A_i = 1)$, so that $\P(\text{published} \mid A_i = 0) = p/\eta$. If $f^{\mathrm{pub}}$ and $f$ denote the densities of $(\hat\theta_i, Z_i, \hat\sigma_i)$ among published and conducted studies, then $f^{\mathrm{pub}} = p\,\eta^{-\mathbb{I}\{A=0\}} f / \P(\text{published})$, and the published studies are independent and identically distributed draws from $f^{\mathrm{published}}$. Hence, for any integrable $g$,
\begin{equation}
\E_{\mathrm{pub}}\bigl[\omega_i\, g(\hat\theta_i, Z_i, \hat\sigma_i)\bigr]
= c\,\E\bigl[g(\hat\theta_i, Z_i, \hat\sigma_i)\bigr],
\qquad
c = \frac{p}{\P(\text{published})} = \frac{1}{\E\bigl[\eta^{-\mathbb{I}\{A_i = 0\}}\bigr]} \in [1, \eta].
\label{eq:ipw}
\end{equation}
The constant $c$ does not depend on the unknown $p$, and $c = 1$ when $\eta = 1$. Because every estimator below normalizes by $\sum_i \omega_i$, the constant cancels. That is, the $\omega$-weighted empirical distribution of the published studies estimates the distribution of all conducted studies.

\subsection{Proof of Proposition 1. Calibrated estimates}\label{app:prop1}
For fixed $z$ and study $i$,  denote $\mu(z) = \beta_0 + z\beta_1 = \E(\theta \mid Z = z)$, $e_i = \hat\theta_i - \theta_i$, and
\[
\tilde\theta_i^{*}(Z = z) = \mu(z) + \lambda_i^{*} r_i^{*},
\qquad
r_i^{*} = \hat\theta_i - \mu(Z_i) = u_i + e_i,
\qquad
\lambda_i^{*} = \sqrt{\frac{\tau_c^2}{\tau_c^2 + \hat\sigma_i^2}} .
\]

\textit{(i) Moments of the oracle estimates.} By (A1), $u_i$ is independent of $(Z_i, \hat\sigma_i)$ with mean zero and variance $\tau_c^2$. By (A2), $e_i \mid (u_i, Z_i, \hat\sigma_i) \sim N(0, \hat\sigma_i^2)$. Hence
\[
\E\bigl[r_i^{*} \mid Z_i, \hat\sigma_i\bigr] = 0,
\qquad
\Var\bigl(r_i^{*} \mid Z_i, \hat\sigma_i\bigr) = \tau_c^2 + \hat\sigma_i^2 .
\]
Because $\lambda_i^{*}$ is a function of $\hat\sigma_i$ only,
\[
\E\bigl[\tilde\theta_i^{*}(Z = z) \mid Z_i, \hat\sigma_i\bigr] = \mu(z) + \lambda_i^{*}\,\E\bigl[r_i^{*} \mid Z_i, \hat\sigma_i\bigr] = \mu(z),
\]
\[
\Var\bigl(\tilde\theta_i^{*}(Z = z) \mid Z_i, \hat\sigma_i\bigr) = \lambda_i^{*2}\,\Var\bigl(r_i^{*} \mid Z_i, \hat\sigma_i\bigr) = \frac{\tau_c^2}{\tau_c^2 + \hat\sigma_i^2}\,(\tau_c^2 + \hat\sigma_i^2) = \tau_c^2 .
\]
Neither depends on $(Z_i, \hat\sigma_i)$, so by the laws of total expectation and total variance, $\tilde\theta_i^{*}(Z = z)$ has mean $\E(\theta \mid Z = z)$ and variance $\tau_c^2$. 

Finally, because $(Z_i, \hat\sigma_i, u_i, e_i)$ are independent and identically distributed across studies by (A1)--(A2), so are the $\tilde\theta_i^{*}(Z = z)$, which are a fixed function of them.

\textit{(ii) Feasible versus oracle estimates.} Let $\hat\mu(z) = \hat\beta_0 + z\hat\beta_1$ and $\hat\lambda_i = \sqrt{\hat\tau_c^2/(\hat\tau_c^2 + \hat\sigma_i^2)}$, so that the feasible estimate is $\tilde\theta_i(Z = z) = \hat\mu(z) + \hat\lambda_i\bigl(\hat\theta_i - \hat\mu(Z_i)\bigr)$. Since $\hat\theta_i - \hat\mu(Z_i) = r_i^{*} - \bigl(\hat\mu(Z_i) - \mu(Z_i)\bigr)$,
\[
\tilde\theta_i(Z = z) - \tilde\theta_i^{*}(Z = z)
= \underbrace{\bigl(\hat\mu(z) - \mu(z)\bigr) - \hat\lambda_i\bigl(\hat\mu(Z_i) - \mu(Z_i)\bigr)}_{\text{(a): error in } \hat\beta_0,\, \hat\beta_1}
+ \underbrace{\bigl(\hat\lambda_i - \lambda_i^{*}\bigr)\, r_i^{*}}_{\text{(b): error in } \hat\tau_c^2} .
\]

For term (a), $|\hat\mu(x) - \mu(x)| \le |\hat\beta_0 - \beta_0| + \|x\|\,\|\hat\beta_1 - \beta_1\|$ for any $x$. Because $z$ and $Z_i$ lie in a bounded set and $0 < \hat\lambda_i \le 1$,
\[
|\text{(a)}| \le C\,\bigl(|\hat\beta_0 - \beta_0| + \|\hat\beta_1 - \beta_1\|\bigr) \xrightarrow{p} 0
\]
for some constant $C$.

For term (b), $\lambda_i^{*}$ is a continuous function of $\tau_c^2 > 0$, so $\hat\tau_c^2 \xrightarrow{p} \tau_c^2$ implies $\hat\lambda_i \xrightarrow{p} \lambda_i^{*}$ by the continuous mapping theorem. Because $r_i^{*}$ has finite variance, $r_i^{*} = O_p(1)$, and hence $\text{(b)} \xrightarrow{p} 0$.

\textit{(iii) Normal population effects.} Suppose the population effects are normal. Because $\theta \mid Z = z$ has the distribution of $\mu(z) + u_i$ under (A1), this means $u_i \sim N(0, \tau_c^2)$.

Given $(Z_i, \hat\sigma_i)$, $u_i \sim N(0, \tau_c^2)$ by (A1), and by (A2), $e_i \mid (u_i, Z_i, \hat\sigma_i) \sim N(0, \hat\sigma_i^2)$, which does not depend on $u_i$; hence $e_i$ is independent of $u_i$. Their sum is therefore normal:
\[
r_i^{*} \mid (Z_i, \hat\sigma_i) \sim N\bigl(0,\, \tau_c^2 + \hat\sigma_i^2\bigr),
\qquad\text{so}\qquad
\lambda_i^{*} r_i^{*} \mid (Z_i, \hat\sigma_i) \sim N\bigl(0,\, \lambda_i^{*2}(\tau_c^2 + \hat\sigma_i^2)\bigr) = N(0, \tau_c^2).
\]
Because this conditional distribution does not depend on $(Z_i, \hat\sigma_i)$, $\lambda_i^{*} r_i^{*}$ is independent of $(Z_i, \hat\sigma_i)$, and its marginal distribution is also $N(0, \tau_c^2)$. Adding the constant $\mu(z)$ gives
\[
\tilde\theta_i^{*}(Z = z) = \mu(z) + \lambda_i^{*} r_i^{*} \sim N\bigl(\mu(z), \tau_c^2\bigr),
\]
which is exactly the distribution of $\theta \mid Z = z$.

Finally, by (ii), $\tilde\theta_i(Z = z) = \tilde\theta_i^{*}(Z = z) + o_p(1)$. Since $\tilde\theta_i^{*}(Z = z)$ has the distribution of $\theta \mid Z = z$ for every $k$, Slutsky's theorem implies that $\tilde\theta_i(Z = z)$ converges in distribution to $\theta \mid Z = z$. \qquad $\square$

\textit{Remarks.} Let $g_z$ denote the density of $\tilde\theta_i^{*}(Z = z)$ among conducted studies, and $f_{\theta \mid z}$ the density of $\theta \mid Z = z$. (i) The shrinkage factor $\lambda_i^{*}$ removes the extra dispersion that sampling error adds to each estimate \cite{wang2019simple}. (ii) Without normality, $\tilde\theta_i^{*}(Z = z) - \mu(z) = \lambda_i^{*}(u_i + e_i)$ is a shrunken residual plus scaled normal noise. Its distribution matches that of $u_i$ in mean and variance but generally differs in shape, so $g_z \neq f_{\theta \mid z}$. The two are close when $u_i$ is approximately normal, or when $\hat\sigma_i^2 \ll \tau_c^2$; in the latter case, $\lambda_i^{*} \approx 1$, $e_i$ is negligible relative to $u_i$, and $\tilde\theta_i^{*}(Z = z) \approx \mu(z) + u_i$. (iii) Proposition~1 focuses on describing conducted studies. Among published studies, the distribution of $\tilde\theta_i^{*}(Z = z)$ is distorted by publication bias; under (A3), Equation~\eqref{eq:ipw} implies that the $\omega$-weighted distribution among published studies recovers $g_z$.

\subsection{Proof of Proposition 2. Asymptotic coverage}\label{app:prop2}
\textit{Notation.} For the replication study $j$, denote $\tilde\theta_i = \tilde\theta_i(Z = z_j)$, $\tilde\theta_i^{\dagger} = \tilde\theta_i^{\dagger}(Z = z_j)$, and $g^{\dagger} = g_{z_j}^{\dagger}$, and recall $r_i^{*} = \hat\theta_i - \mu(Z_i)$ from Appendix~\ref{app:prop1}. Let $K_h(x) = \phi(x; 0, h^2)$ be the Gaussian kernel. The weighted KDE of the population effects and the predictive density of the replication estimate are
\[
\hat f_\theta(w) = \frac{\sum_{i=1}^k \omega_i\, K_h(w - \tilde\theta_i)}{\sum_{i=1}^k \omega_i},
\]
\begin{equation}
\hat f_j(t) = \bigl(\hat f_\theta * \phi(\cdot\,; 0, \hat\sigma_j^2)\bigr)(t)
= \frac{\sum_{i=1}^k \omega_i\, \phi\bigl(t; \tilde\theta_i, h^2 + \hat\sigma_j^2\bigr)}{\sum_{i=1}^k \omega_i},
\label{eq:pred_density}
\end{equation}
which reduces to Equation~\ref{eq:kde-pi} when $\omega_i \equiv 1$. The interval is $\widehat{\mathrm{PI}}_j = [\hat q_{\alpha/2}, \hat q_{1-\alpha/2}]$, where $\hat q_p$ is the $p$-quantile of $\hat f_j$ with CDF $\hat F_j$. The corresponding targets are
\begin{align*}
f_j &= f_{\theta \mid z_j} * \phi(\cdot\,; 0, \hat\sigma_j^2),\\
g_j^{\dagger} &= g^{\dagger} * \phi(\cdot\,; 0, \hat\sigma_j^2),
\end{align*}
where $f_j$ is the density of $\hat\theta_j^{\mathrm{rep}}$ under Assumption (A4), with CDF $F_j$, and $g_j^{\dagger}$ has CDF $G_j^{\dagger}$ and $p$-quantile $q_p^{\dagger} = (G_j^{\dagger})^{-1}(p)$.

\textit{From published to all conducted studies (IPPW).} Under Assumption (A3), for any integrable function $\psi$ of $(\hat\theta_i, Z_i, \hat\sigma_i)$, Equation~\eqref{eq:ipw} gives
\begin{equation}
\E_{\mathrm{pub}}\bigl[\omega_i\, \psi(\hat\theta_i, Z_i, \hat\sigma_i)\bigr] = c\, \E\bigl[\psi(\hat\theta_i, Z_i, \hat\sigma_i)\bigr],
\label{eq:reweight}
\end{equation}
where $\E_{\mathrm{pub}}$ is the expectation over published studies, $\E$ is the expectation over all conducted studies, and $c = P(\text{published} \mid A = 1)/P(\text{published}) > 0$. For the uncorrected models, $\eta = 1$ and $\omega_i \equiv 1$, so $c = 1$. The value of $c$ is not needed, because it either multiplies a zero expectation (Step 1) or cancels in a ratio with $\sum_i \omega_i$ in Steps 2--3.

The proof proceeds in 5 steps. Step 1 verifies that condition (a) in Proposition 2 is attainable. Steps 2--3 show $\hat f_\theta \to g^{\dagger}$ in $L^1$, Step 4 carries this to $\hat f_j \to g_j^{\dagger}$, and Step 5 converts density convergence into coverage.

\textit{Step 1: Root-$k$ consistency of $\hat{\boldsymbol\beta}$.} 

With $Z_i$ including the intercept, $\hat{\boldsymbol\beta}$ solves
\[
\sum_{i=1}^k \omega_i v_i Z_i^{\top}(\hat\theta_i - Z_i\boldsymbol\beta) = 0,
\qquad  \text{where} \
v_i = \frac{1}{\hat\sigma_i^2 + \hat\tau_c^2},
\]
so that $\omega_i v_i$ is the weight $w_{mi}$ in Equation~\ref{eq:omega_mi}, with $\omega_i \equiv 1$ for the uncorrected models. By (A1)--(A2),
\[
\E\bigl[\hat\theta_i - Z_i\boldsymbol\beta \mid Z_i, \hat\sigma_i\bigr] = \E\bigl[u_i + e_i \mid Z_i, \hat\sigma_i\bigr] = 0 .
\]
Because $v_i$ depends on $\hat\tau_c^2$, which is estimated from all studies, we work with its limit $v_i^{\dagger} = 1/(\hat\sigma_i^2 + \tau_\dagger^2)$, a function of study $i$'s data only; replacing $v_i^{\dagger}$ with $v_i$ does not affect the limit of $\hat{\boldsymbol\beta}$, because $\hat\tau_c^2 \xrightarrow{p} \tau_\dagger^2$. 

Applying Equation~\eqref{eq:reweight} with $\psi = v_i^{\dagger} Z_i^{\top}(\hat\theta_i - Z_i\boldsymbol\beta)$, and then iterated expectations,
\[
\begin{aligned}
\E_{\mathrm{pub}}\bigl[\omega_i v_i^{\dagger} Z_i^{\top}(\hat\theta_i - Z_i\boldsymbol\beta)\bigr]
&= c\, \E\bigl[v_i^{\dagger} Z_i^{\top}(\hat\theta_i - Z_i\boldsymbol\beta)\bigr]
&& \text{by Equation~\eqref{eq:reweight}} \\
&= c\, \E\Bigl[v_i^{\dagger} Z_i^{\top}\, \E\bigl(u_i + e_i \mid Z_i, \hat\sigma_i\bigr)\Bigr]
&& \text{by iterated expectations} \\
&= 0 .
\end{aligned}
\]
The estimating equation is therefore unbiased at the true $\boldsymbol\beta$, and $\hat{\boldsymbol\beta}$ is root-$k$ consistent by standard estimating-equation arguments \cite{hedges2010robust, mathur2020sensitivity}. The same argument applies to any weights that depend only on the within-study variances and $\hat\tau_c^2$, including the hierarchical and correlated-effects weights of Hedges et al.~\cite{hedges2010robust}. In particular, it does not require $\tau_\dagger^2 = \tau_c^2$.

\textit{Step 2: Consistency of the KDE based on $\tilde\theta_i^{\dagger}$.} 

Let $\tilde f_\theta^{\dagger}$ denote $\hat f_\theta$ computed from $\tilde\theta_i^{\dagger}$. The published studies are independent and identically distributed, with $1 \le \omega_i \le \eta$. By Equation~\eqref{eq:reweight} with $\psi = K_h(w - \tilde\theta_i^{\dagger})$ and $\psi = 1$,
\[
\E_{\mathrm{pub}}\Bigl[\frac{1}{k}\sum_{i=1}^k \omega_i K_h(w - \tilde\theta_i^{\dagger})\Bigr] = c\,(K_h * g^{\dagger})(w),
\qquad
\frac{1}{k}\sum_{i=1}^k \omega_i \xrightarrow{\text{a.s.}} c .
\]
Split the error into a stochastic and a bias term:
\[
\|\tilde f_\theta^{\dagger} - g^{\dagger}\|_1
\le \underbrace{\|\tilde f_\theta^{\dagger} - K_h * g^{\dagger}\|_1}_{\to_p\, 0 \text{ as } kh \to \infty}
+ \underbrace{\|K_h * g^{\dagger} - g^{\dagger}\|_1}_{\to\, 0 \text{ as } h \to 0},
\]
by the $L^1$ consistency of KDE, which holds for any density when $h \to 0$ and $kh \to \infty$ and extends to bounded weights \cite{devroye1985nonparametric}.

\textit{Step 3: Replacing $\tilde\theta_i^{\dagger}$ with the feasible $\tilde\theta_i$.} 

As in Appendix~\ref{app:prop1}(ii),
\[
\tilde\theta_i - \tilde\theta_i^{\dagger}
= \bigl(\hat\mu(z_j) - \mu(z_j)\bigr) - \hat\lambda_i\bigl(\hat\mu(Z_i) - \mu(Z_i)\bigr) + \bigl(\hat\lambda_i - \lambda_i^{\dagger}\bigr) r_i^{*}.
\]
For the first two terms, $|\hat\mu(x) - \mu(x)| \le |\hat\beta_0 - \beta_0| + \|x\|\,\|\hat\beta_1 - \beta_1\|$ for any $x$. Because $z_j$ and $Z_i$ lie in a bounded set and $0 < \hat\lambda_i \le 1$, both terms are at most $C\|\hat{\boldsymbol\beta} - \boldsymbol\beta\|$.

For the third term, write the shrinkage factor as a function of the heterogeneity, i.e., $\lambda(t) = \sqrt{t/(t + \hat\sigma_i^2)}$, so that $\hat\lambda_i = \lambda(\hat\tau_c^2)$ and $\lambda_i^{\dagger} = \lambda(\tau_\dagger^2)$. Its derivative is bounded:
\[
\lambda'(t) = \frac{\hat\sigma_i^2}{2\sqrt{t}\,(t + \hat\sigma_i^2)^{3/2}}
\le \frac{1}{2\sqrt{t}\,(t + \hat\sigma_i^2)^{1/2}}
\le \frac{1}{2t},
\]
where the first inequality follows from $\hat\sigma_i^2/(t + \hat\sigma_i^2) \le 1$ and the second from $(t + \hat\sigma_i^2)^{1/2} \ge \sqrt{t}$. By the mean value theorem,
\[
\hat\lambda_i - \lambda_i^{\dagger} = \lambda'(\bar t)\,(\hat\tau_c^2 - \tau_\dagger^2)
\quad\text{for some } \bar t \text{ between } \hat\tau_c^2 \text{ and } \tau_\dagger^2 .
\]
Because $\hat\tau_c^2 \xrightarrow{p} \tau_\dagger^2 > 0$, $\Pr(\hat\tau_c^2 \ge \tau_\dagger^2/2) \to 1$ as $k \to \infty$. When $\hat\tau_c^2 \ge \tau_\dagger^2/2$, and $\bar t \ge \tau_\dagger^2/2$, so $|\lambda'(\bar t)| \le 1/(2\bar t) \le 1/\tau_\dagger^2$, and hence
\[
|\hat\lambda_i - \lambda_i^{\dagger}| \le \frac{|\hat\tau_c^2 - \tau_\dagger^2|}{\tau_\dagger^2} .
\]
Combining the bounds on all three terms, with probability tending to one, for every $i$,
\[
\bigl|\tilde\theta_i - \tilde\theta_i^{\dagger}\bigr| \le C\bigl(\|\hat{\boldsymbol\beta} - \boldsymbol\beta\| + |\hat\tau_c^2 - \tau_\dagger^2|\,|r_i^{*}|\bigr).
\]

We now carry this bound over to the KDEs. Shifting a Gaussian kernel changes it by at most a multiple of the shift: by the mean value theorem,
\[
\bigl\|K_h(\cdot - a) - K_h(\cdot - b)\bigr\|_1 \le |a - b|\,\|K_h'\|_1 = \sqrt{\frac{2}{\pi}}\,\frac{|a - b|}{h} .
\]
Since both KDEs use the same weights,
\[
\hat f_\theta - \tilde f_\theta^{\dagger}
= \frac{\sum_i \omega_i \bigl\{K_h(\cdot - \tilde\theta_i) - K_h(\cdot - \tilde\theta_i^{\dagger})\bigr\}}{\sum_i \omega_i},
\]
so, by the triangle inequality and the bound above,
\[
\|\hat f_\theta - \tilde f_\theta^{\dagger}\|_1
\le \frac{C}{h} \cdot \frac{\sum_i \omega_i |\tilde\theta_i - \tilde\theta_i^{\dagger}|}{\sum_i \omega_i}
\le \frac{C}{h}\Bigl(\|\hat{\boldsymbol\beta} - \boldsymbol\beta\| + |\hat\tau_c^2 - \tau_\dagger^2| \cdot \frac{\sum_i \omega_i |r_i^{*}|}{\sum_i \omega_i}\Bigr).
\]
By Equation~\eqref{eq:reweight} and the law of large numbers, $k^{-1}\sum_i \omega_i |r_i^{*}| \xrightarrow{p} c\,\E|r_i^{*}|$ and $k^{-1}\sum_i \omega_i \xrightarrow{p} c$, so the weighted average of $|r_i^{*}|$ converges to $\E|r_i^{*}| < \infty$ and is $O_p(1)$. 

By conditions (a) and (b) in Proposition 2, the right-hand side is therefore
\[
\frac{C}{h}\Bigl(O_p(k^{-1/2}) + O_p(k^{-1/2}) \cdot O_p(1)\Bigr) = O_p\Bigl(\frac{1}{h\sqrt{k}}\Bigr) \to 0
\]
by condition (c) in Proposition 2. 

Combining with Step 2,
\[
\|\hat f_\theta - g^{\dagger}\|_1 \le \|\hat f_\theta - \tilde f_\theta^{\dagger}\|_1 + \|\tilde f_\theta^{\dagger} - g^{\dagger}\|_1 \xrightarrow{p} 0 .
\]

\textit{Step 4: Convergence of the predictive density.} 

By Young's inequality, $\|f * g\|_1 \le \|f\|_1 \|g\|_1$ for integrable $f$ and $g$. Since convolution is linear and $\|\phi(\cdot\,; 0, \hat\sigma_j^2)\|_1 = 1$,
\[
\|\hat f_j - g_j^{\dagger}\|_1
= \bigl\|(\hat f_\theta - g^{\dagger}) * \phi(\cdot\,; 0, \hat\sigma_j^2)\bigr\|_1
\le \|\hat f_\theta - g^{\dagger}\|_1 \xrightarrow{p} 0 ,
\]
where the limit follows from Step 3. In words, adding the replication study's sampling noise to both densities does not increase the distance between them.

\textit{Step 5: Quantiles and coverage.} 

Let $f$ and $g$ be any two densities with CDFs $F$ and $G$. Because $\int (f - g) = 0$, we have $F(t) - G(t) = \int_{-\infty}^{t} (f - g) = -\int_{t}^{\infty} (f - g)$, so
\[
|F(t) - G(t)|
\le \min\Bigl\{\int_{-\infty}^{t} |f - g|,\ \int_{t}^{\infty} |f - g|\Bigr\}
\le \tfrac{1}{2}\Bigl(\int_{-\infty}^{t} |f - g| + \int_{t}^{\infty} |f - g|\Bigr)
= \tfrac{1}{2}\|f - g\|_1 .
\]
Taking the supremum over $t$,
\begin{equation}
\sup_t |F(t) - G(t)| \le \tfrac{1}{2}\|f - g\|_1 .
\label{eq:cdf_l1}
\end{equation}

By Equation~\eqref{eq:cdf_l1} and Step 4, $\sup_t |\hat F_j(t) - G_j^{\dagger}(t)| \le \tfrac{1}{2}\|\hat f_j - g_j^{\dagger}\|_1 \xrightarrow{p} 0$. Because $g_j^{\dagger}$ is a convolution with a normal density, it is strictly positive, so $G_j^{\dagger}$ is continuous and strictly increasing. Hence, for each $p \in (0,1)$ and any $\epsilon > 0$,
\[
G_j^{\dagger}(q_p^{\dagger} - \epsilon) < p < G_j^{\dagger}(q_p^{\dagger} + \epsilon).
\]
By the uniform convergence of $\hat F_j$, the same strict inequalities hold for $\hat F_j$ with probability tending to one, and then $|\hat q_p - q_p^{\dagger}| \le \epsilon$. Thus $\hat q_p \xrightarrow{p} q_p^{\dagger}$ (\cite{van2000asymptotic}, Lemma~21.2).

By Assumption (A4), $\hat\theta_j^{\mathrm{rep}}$ has CDF $F_j$, which is continuous because $f_j$ is a density, and $\hat\theta_j^{\mathrm{rep}}$ is independent of the meta-analysis data, hence of $(\hat q_{\alpha/2}, \hat q_{1-\alpha/2})$. Conditioning on the quantiles,
\[
\Pr\bigl(\hat\theta_j^{\mathrm{rep}} \in \widehat{\mathrm{PI}}_j\bigr)
= \E\bigl[F_j(\hat q_{1-\alpha/2}) - F_j(\hat q_{\alpha/2})\bigr] .
\]
By the continuous mapping theorem, $F_j(\hat q_p) \xrightarrow{p} F_j(q_p^{\dagger})$; since $F_j$ is bounded by 1, the bounded convergence theorem gives
\[
\Pr\bigl(\hat\theta_j^{\mathrm{rep}} \in \widehat{\mathrm{PI}}_j\bigr)
\;\longrightarrow\;
F_j(q_{1-\alpha/2}^{\dagger}) - F_j(q_{\alpha/2}^{\dagger}) =: 1 - \alpha + R_j .
\]
Since $G_j^{\dagger}(q_{1-\alpha/2}^{\dagger}) - G_j^{\dagger}(q_{\alpha/2}^{\dagger}) = 1 - \alpha$,
\[
R_j = \bigl[F_j(q_{1-\alpha/2}^{\dagger}) - G_j^{\dagger}(q_{1-\alpha/2}^{\dagger})\bigr] - \bigl[F_j(q_{\alpha/2}^{\dagger}) - G_j^{\dagger}(q_{\alpha/2}^{\dagger})\bigr].
\]
Each bracket is at most $\sup_t |F_j(t) - G_j^{\dagger}(t)|$ in absolute value. By Equation~\eqref{eq:cdf_l1}, and by Young's inequality as in Step 4 (since $f_j - g_j^{\dagger} = (f_{\theta \mid z_j} - g^{\dagger}) * \phi(\cdot\,; 0, \hat\sigma_j^2)$),
\[
|R_j| \le 2\sup_t |F_j(t) - G_j^{\dagger}(t)| \le \|f_j - g_j^{\dagger}\|_1 \le \|f_{\theta \mid z_j} - g^{\dagger}\|_1 .
\]
If $\tau_\dagger^2 = \tau_c^2$, then $\tilde\theta_i^{\dagger}$ is the oracle estimate, and if the population effects are normal, its density equals $f_{\theta \mid z_j}$ by Proposition~1, so $R_j = 0$. \qquad $\square$

\textit{Remarks.}
(i) \textit{Misspecified shrinkage.} If $\tau_\dagger^2 \neq \tau_c^2$, then even if the population effects are normal,
\[
\Var\bigl(\tilde\theta_i^{\dagger} - \mu(z_j)\bigr) = \E\bigl[\lambda_i^{\dagger 2}(\tau_c^2 + \hat\sigma_i^2)\bigr] \neq \tau_c^2,
\]
so $g^{\dagger} \neq f_{\theta \mid z_j}$ and $R_j \neq 0$ in general. For the uncorrected models with $\eta = 1$, $\hat\tau_c^2$ is consistent by standard method-of-moments theory \cite{hedges2010robust}, so $\tau_\dagger^2 = \tau_c^2$.

(ii) \textit{Uncorrected models under publication bias.} If $\eta > 1$ but $\omega_i \equiv 1$, Equation~\eqref{eq:reweight} no longer holds: the estimating equation in Step 1 is biased for $\boldsymbol\beta$, and the KDE in Step 2 targets the distribution of calibrated estimates among published studies rather than all conducted studies. Proposition~2 then provides no guarantee.

(iii) \textit{Bandwidth.} Condition (c) is slightly stronger than the usual $kh \to \infty$. It ensures that the $O_p(k^{-1/2})$ error from estimating the calibration parameters, which Step 3 magnifies by $1/h$, still vanishes. It holds for bandwidths of the form $h = C k^{-1/5}$, such as the Sheather--Jones bandwidth \cite{sheather1991reliable}, because then $h\sqrt{k} = C k^{-1/5} k^{1/2} = C k^{3/10} \to \infty$.
 In practice, $h$ is selected from the unweighted calibrated estimates and is therefore dependent on data. Our proof treats $h$ as a fixed sequence. We expect the result to hold also for a data-dependent bandwidth that behaves like such a sequence in large samples, as the Sheather--Jones bandwidth does.

\subsection{Standard parametric prediction interval versus KDE-based prediction interval}\label{app:pi-kde-param}

\textit{Standard parametric interval.} The point forecast is $\hat\mu(z_j) = \hat\beta_0 + z_j\hat\beta_1$ for the meta-regression models and the pooled estimate $\hat\mu$ for the intercept-only models, with the IPPW coefficients $\hat{\boldsymbol\beta}_\eta$ for the corrected models. The total prediction variance combines three components: (1) the estimated between-study variance $\hat\tau_c^2$ (residual heterogeneity for the meta-regression models), taken from the corresponding uncorrected model (MA-U for the intercept-only models and MR-U for the meta-regression models); (2) the squared robust standard error of the point forecast, $\SE^2(\hat\mu(z_j))$; and (3) the within-study sampling variance $\hat\sigma_j^2$ of the replication study. The $(1-\alpha)$ prediction interval is
\[
\hat\mu(z_j) \pm t_{\mathrm{df},\,1-\alpha/2} \sqrt{\hat\tau_c^2 + \SE^2\bigl(\hat\mu(z_j)\bigr) + \hat\sigma_j^2},
\]
where $t_{\mathrm{df},\,1-\alpha/2}$ is the $1-\alpha/2$ quantile of the $t$-distribution with Satterthwaite degrees of freedom, which account for the small-sample properties of the robust variance estimator \cite{tipton2015small}. For the meta-regression models, we used the smallest degrees of freedom among the coefficients as a conservative approximation. This interval assumes that the population effects are normally distributed around the forecast.

\textit{KDE-based interval.} We first computed the calibrated estimates of the meta-analysis studies by adding each study's shrunken residual to the forecast at the replication study's moderator level,
\[
\tilde\theta_i(Z = z_j) = \hat\mu(z_j) + \hat\lambda_i\bigl\{\hat\theta_i - \hat\mu(Z_i)\bigr\},
\qquad
\hat\lambda_i = \sqrt{\frac{\hat\tau_c^2}{\hat\tau_c^2 + \hat\sigma_i^2}},
\]
where $\hat\sigma_i^2$ is the meta-analysis study's own sampling variance. We then estimated the distribution of the population effects by a Gaussian KDE of these calibrated estimates, weighted by $\omega_i$ (Appendix~\ref{app:setup}; $\omega_i \equiv 1$ for the uncorrected models), with bandwidth $h$ selected by the Sheather--Jones method from the unweighted calibrated estimates. Convolving this density with the replication study's sampling distribution $N(0, \hat\sigma_j^2)$ yields the predictive density in Equation~\eqref{eq:pred_density}, a weighted mixture of normal densities. We evaluated its CDF on a grid with step size $0.01$ (from $-3$ to $3$ in the simulation and from $-5$ to $5$ in the ManyBabies applications) and took the $\alpha/2$ and $1-\alpha/2$ quantiles as the $(1-\alpha)$ prediction interval. Unlike the parametric interval, the KDE-based interval does not impose a normal shape on the estimated distribution of population effects.

\subsection{Simulation results under normal distribution of $\mu_i$}\label{app:sim-normal_mu}
\begin{table}[H]
\centering
\caption{Win rates across 81 simulation scenarios (under normal distribution of $\mu_i$). Each cell reports the number of scenarios in which a model achieved the best value for the given metric. Best is defined as: lowest MAE, RMSE, and PI width; closest to 50\% overestimation and 95\% coverage; highest mean $\bar P_\text{orig}$.}
\label{tab:winrate}
\begin{tabular}{lccccccccc}
\toprule
 & & & & & & \multicolumn{2}{c}{Coverage} & \multicolumn{2}{c}{PI Width} \\
\cmidrule(lr){7-8} \cmidrule(lr){9-10}
Model & Raw Diff & MAE & RMSE & Mean $\bar P_\text{orig}$ & Overestimation & KDE & Param. & KDE & Param. \\
\midrule
MA-U & 14 & 9 & 9 & 4 & 17 & 21 & 14 & 0 & 49 \\
MR-U & 13 & 24 & 25 & 22 & 10 & 6 & 21 & 80 & 31 \\
MA-C & 39 & 22 & 21 & 4 & 22 & 45 & 26 & 0 & 1 \\
MR-C & 15 & 26 & 26 & 51 & 32 & 9 & 20 & 1 & 0 \\
\bottomrule
\multicolumn{9}{l}{\footnotesize MA-U: MA, uncorrected; MR-U: MR, uncorrected;} \\
\multicolumn{9}{l}{\footnotesize MA-C: MA, IPPW-corrected; MR-C: MR, IPPW-corrected.} \\
\multicolumn{9}{l}{\footnotesize PI: prediction interval.} \\
\multicolumn{9}{l}{\footnotesize Param.: standard parametric prediction interval assuming a $t$-distribution.} \\
\end{tabular}
\end{table}

\begin{figure}[H]
\centering
\begin{subfigure}[t]{0.48\textwidth}
  \centering
  \includegraphics[width=\textwidth]{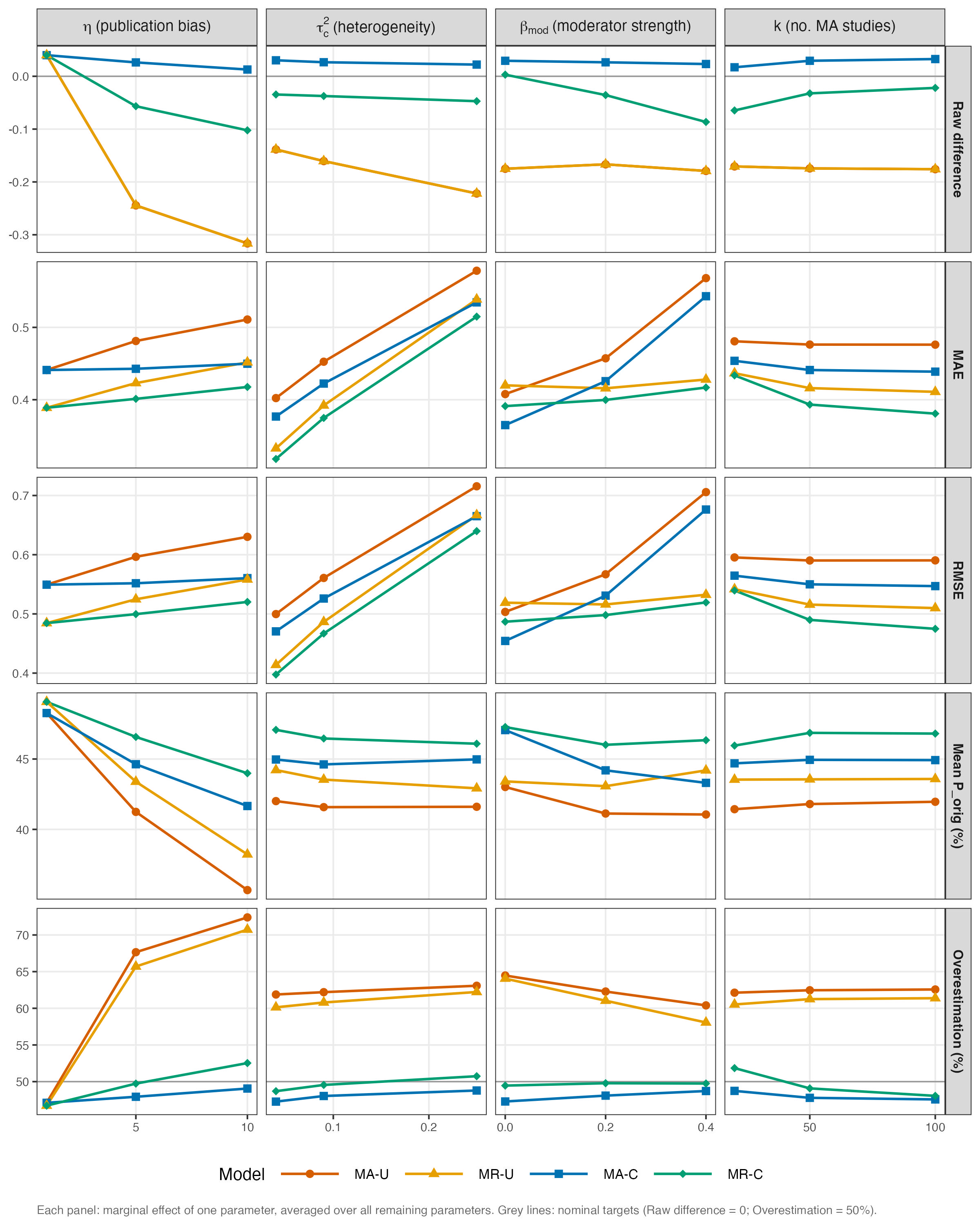}
  \caption{Shared metrics across KDE and parametric methods.}
  \label{fig:sim_metrics1}
\end{subfigure}
\hfill
\begin{subfigure}[t]{0.48\textwidth}
  \centering
  \includegraphics[width=\textwidth]{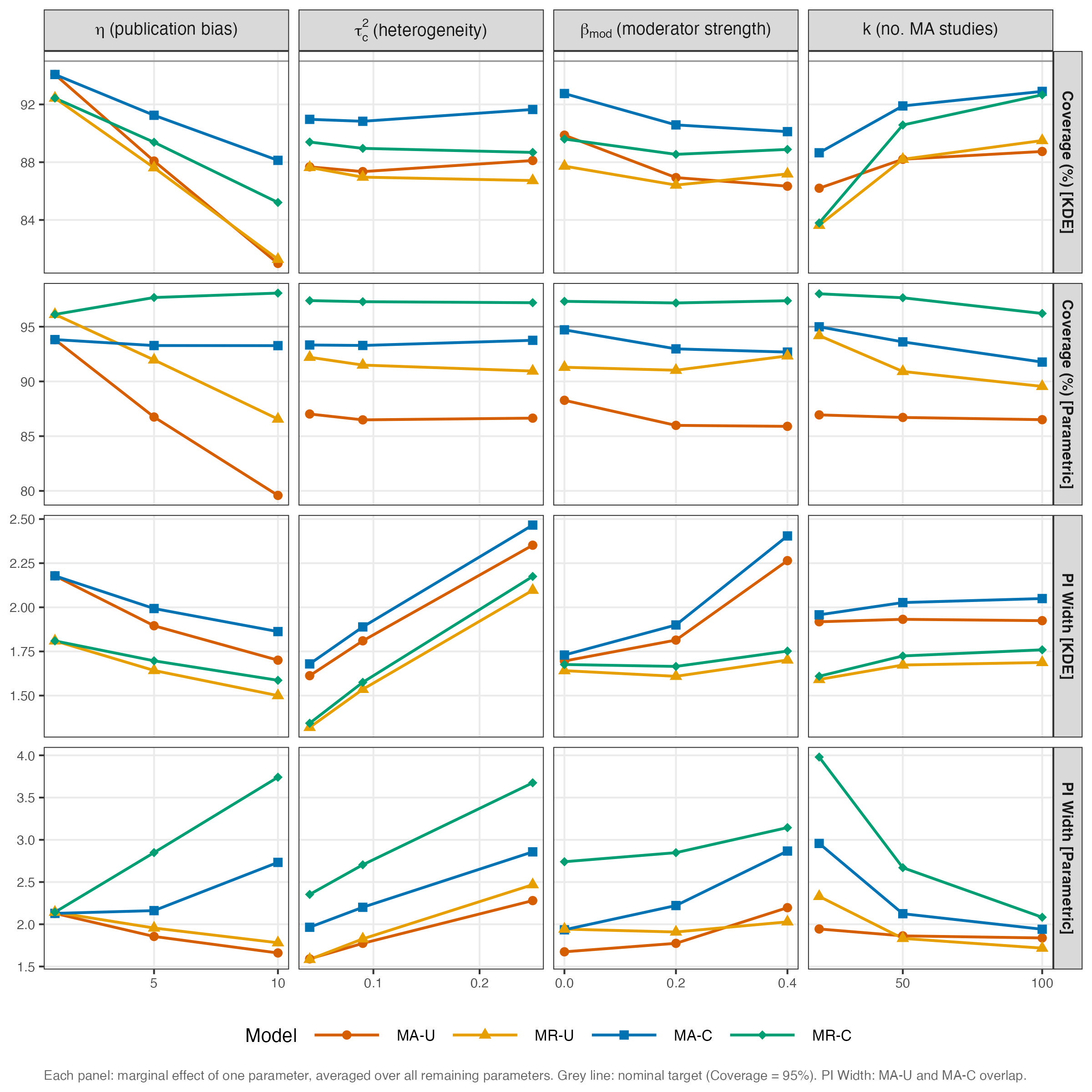}
  \caption{Coverage and prediction interval width by PI method (KDE vs.\ parametric).}
  \label{fig:sim_metrics2}
\end{subfigure}
\caption{Marginal performance of four forecasting models across simulation parameters (under normal distribution of $\mu_i$). Each panel shows one metric (row) as a function of one design parameter (column), averaging over the remaining parameters.}
\label{fig:sim_metrics}
\end{figure}

\end{appendix}



\end{document}